\documentclass[a4paper,11pt]{article}
\pdfoutput=1 

\usepackage{jheppub} 

\usepackage[T1]{fontenc} 

\title{\begin{center}
		\LARGE{\textbf{Heavy-Lifting of Gauge Theories \\ By Cosmic Inflation}}
	\end{center}
}
\author{Soubhik Kumar}
\author{and Raman Sundrum}
\affiliation{Maryland Center for Fundamental Physics, Department of Physics,\\University of Maryland, College Park, MD 20742}
\emailAdd{soubhik@terpmail.umd.edu}
\emailAdd{raman@umd.edu}
\abstract{Future measurements of primordial non-Gaussianity can reveal cosmologically produced particles with masses of order the inflationary Hubble scale and their interactions with the inflaton, 
	giving us crucial insights into the structure of fundamental physics at extremely high energies. We study gauge-Higgs theories that may be accessible in this regime, carefully imposing the constraints  of gauge symmetry and its (partial) Higgsing. 
	We distinguish two types of Higgs mechanisms: (i) a standard one in which the Higgs scale is constant before and after inflation, where the particles observable in non-Gaussianities are far heavier than can be accessed by laboratory experiments,  perhaps associated with gauge unification, 
	and (ii) a ``heavy-lifting'' mechanism in which couplings to curvature can 
	result in Higgs scales of order the 
	Hubble scale during inflation  while reducing to far lower scales in the current era, where they may now be accessible to collider and other laboratory experiments. 
	In the heavy-lifting option, renormalization-group running of terrestrial measurements yield predictions for 
	cosmological non-Gaussianities.
	If the heavy-lifted gauge theory suffers a hierarchy problem, such as does the Standard Model, confirming such predictions would demonstrate a striking violation of the Naturalness Principle.  
	While observing gauge-Higgs sectors in non-Gaussianities will be challenging given the constraints of cosmic variance, we show that it may be possible with reasonable precision given favorable couplings to the inflationary dynamics. }

\begin{document}
	\hspace{30em} UMD-PP-017-31
	\maketitle
	\flushbottom
	\section{Introduction}
	Cosmic Inflation (see \cite{Baumann:2009ds} for a review),  originally invoked to help explain the homogeneity and flatness of the universe on large scales, provides an attractive framework for understanding inhomogeneities on smaller scales, such as
	the spectrum of temperature fluctuations in the Cosmic Microwave Background (CMB) radiation. These fluctuations are consistent with an almost scale-invariant, adiabatic and Gaussian spectrum of primordial curvature perturbations $\mathcal{R}$ \cite{Ade:2015lrj}. The approximate scale invariance of these fluctuations can be naturally modeled as quantum oscillations of the inflaton field in a quasi-de Sitter (dS) spacetime. The adiabaticity property implies that 
	among the fields driving inflation, there is a single ``clock'', the inflaton, which governs the duration of inflation and the subsequent reheating process. Finally, Gaussianity of the present data \cite{Ade:2015ava} reflects very weak couplings among inflationary and gravitational fields. While these features point to successes of the inflationary paradigm, few details of the fundamental physics at play during inflation have emerged. Observing small non-Gaussianity (NG) of the fluctuations could change this situation radically, giving critical insights not only into the inflationary dynamics itself but also into the particle physics structure of that era. 
	
	Interactions of the inflaton with itself or other fields during or immediately after inflation can lead to a non-Gaussian spectrum of $\mathcal{R}$. However, NG can also be developed after fluctuation modes re-enter the horizon at the end of inflation. This can happen for various reasons, including, nonlinear growth of perturbations under gravity during structure formation (see \cite{2010AdAst2010E..73L,Bartolo:2010qu} for reviews in the context of CMB and Large-Scale Structure). Therefore it is crucial to understand and distinguish this latter type of NG which can ``contaminate'' the invaluable primordial NG. In this paper we will assume that this separation can be achieved in future experiments involving Large-Scale Structure (LSS) surveys \cite{Alvarez:2014vva} and 21-cm cosmology \cite{Loeb:2003ya,Munoz:2015eqa}, to reach close to a cosmic-variance-only limited precision. With this qualifier, a future measurement of NG can reveal important clues as to the underlying inflationary dynamics. For example, for the case single-field slow-roll inflation, there is a minimal amount of NG mediated by gravitational interactions \cite{Maldacena:2002vr,Acquaviva:2002ud}, while lying several orders of magnitude below the current limit on NG, can be achievable in the future.
	
	There also exist a variety of models which predict a larger than minimal NG (see \cite{Bartolo:2004if,Chen:2010xka} for reviews and references to original papers). A common feature among some of these models is the presence of additional fields beyond the inflaton itself. Such non-minimal structure can be motivated by the need to capture
	inflationary dynamics within a fully theoretically controlled and natural framework. If those additional fields are light with mass, $m\ll H$ (where $H$ denotes the Hubble scale during inflation), they can oscillate and co-evolve along with the inflaton during inflation. These fields can generate significant NG after inflation, with a functional form approximated by the ``local'' shape \cite{Salopek:1990jq,Komatsu:2001rj,Ade:2015ava}. 
	
	On the other hand, the additional fields can be heavy with masses $m\gtrsim H$. Such fields can be part of ``quasi-single-field inflation'' which was introduced in \cite{Chen:2009we} and further developed in \cite{Chen:2009zp,Baumann:2011nk,Chen:2012ge,Assassi:2012zq,Noumi:2012vr,Arkani-Hamed:2015bza,Dimastrogiovanni:2015pla}. In the presence of these massive fields, the three-point correlation function of the curvature perturbation $\mathcal{R}$ has a distinctive \textit{non-analytic} dependence on momenta,
	\begin{equation}\label{qsfi}
	\langle \mathcal{R}(\vec{k}_1) \mathcal{R}(\vec{k}_2) \mathcal{R}(\vec{k}_3)\rangle \propto \frac{1}{k_3^3}\frac{1}{k_1^3}\left(\frac{k_3}{k_1}\right)^{\Delta(m)}+\cdots, \text{ for } k_3\ll k_1,
	\end{equation}
	in the ``squeezed'' limit where one of 3-momenta becomes smaller than the other two. In the above,
	\begin{equation}
	\Delta(m)=\frac{3}{2}+i\sqrt{\frac{m^2}{H^2}-\frac{9}{4}},
	\end{equation}
	where $m$ is the mass of the new particle. The non-analyticity reflects the fact that the massive particles are not merely virtual within these correlators, but rather are physically present ``on-shell'' due to cosmological particle production, driven by the inflationary background time-dependence. Such production is naturally suppressed for 
	$m\gg H$, which is reflected by a ``Boltzmann-like suppression'' factor in the proportionality constant in \eqref{qsfi}.  The only effect of $m\gg H$ particles is then virtual-mediation of interactions among the remaining light fields \cite{Creminelli:2003iq}. At the other extreme, for $m\ll H$ the distinctive non-analyticity is lost.  Hence, we are led to a window of opportunity around $H$, where the non-analytic dependence of the three-point function is both non-trivial and observable, and can be used to do   spectroscopy of masses. Furthermore, if a massive particle has nontrivial spin \cite{Arkani-Hamed:2015bza,Lee:2016vti}, there will be an angle-dependent prefactor in \eqref{qsfi}, which can enable us to determine the spin as well \cite{MoradinezhadDizgah:2017szk}. These observations point to a program of ``Cosmological Collider Physics'' \cite{Arkani-Hamed:2015bza}, which has an unprecedented reach into the structure of fundamental physics at much higher energy scales than we can expect to probe at  colliders. The sensitivity of the measurements is ultimately constrained by cosmic variance, very roughly in the ball park of
	\begin{equation}\label{cosmicvar}
	\frac{\langle\mathcal{R}\mathcal{R}\mathcal{R}\rangle}{\langle\mathcal{R}\mathcal{R}\rangle^{\frac{3}{2}}}\sim \frac{1}{\sqrt{N_{\text{21-cm}}}}\sim 10^{-8},
	\end{equation}
	where we have assumed the number of modes accessible by a cosmic variance limited 21-cm experiment is $N_{\text{21-cm}}\sim 10^{16}$ \cite{Loeb:2003ya}. Achieving such a precision is  very important for realizing the full potential of the program.
	
\begin{figure}[h]
		\centering
		\includegraphics[width=0.6\linewidth]{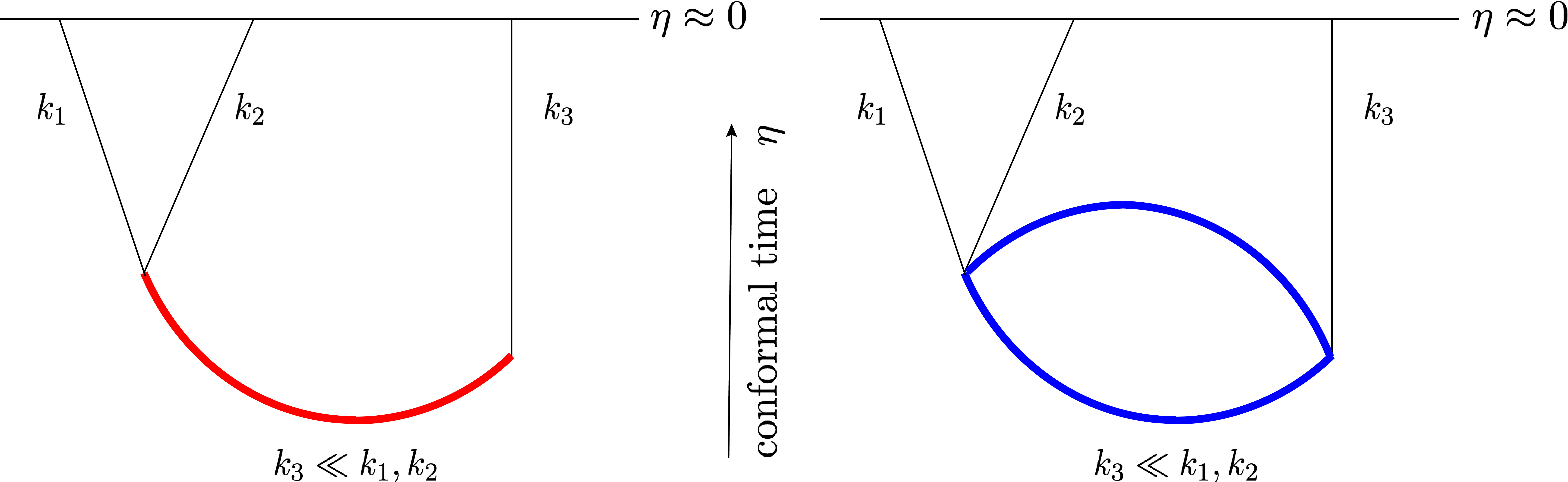}
		\caption{From left to right: (a) Tree level exchange of neutral massive scalar (in red) between inflatons (in black); (b) Loop level exchange of charged massive fields (in blue) between inflatons (in black). The external lines are taken to end at the end of inflation, conformal time, $\eta\approx 0$.}
		\label{fig:treeandloop}
	\end{figure}

	In this paper, we couple gauge-Higgs theories with $m\sim H$ to inflationary dynamics and ask to what extent the associated states can be seen via the cosmological collider physics approach.
	The contributions of massive particle to the three point function $\langle \mathcal{R}(\vec{k}_1) \mathcal{R}(\vec{k}_2) \mathcal{R}(\vec{k}_3)\rangle$ can be represented via ``in-in'' diagrams in (quasi-)dS space such as in Fig. \ref{fig:treeandloop}.
	From Fig. \ref{fig:treeandloop} (a), we see that since the inflaton has to have the internal quantum numbers of the vacuum, \footnote{In the context of Higgs inflation \cite{Bezrukov:2007ep} however, inflaton is the physical charge neutral Higgs field.} the same has to be true for the massive particles. The particles must therefore be gauge singlets. Keeping this fact in mind, let us analyze the two scenarios that can arise during inflation. 
	
	The gauge theory may be unbroken during inflation. Gauge singlet 1-particle states then can mediate NG via tree diagrams as shown in Fig. \ref{fig:treeandloop} (a). This is also the case that has been analyzed extensively in the literature. On the other hand, gauge charged states can contribute via loops, as shown in Fig. \ref{fig:treeandloop} (b), but are expected to be small. Alternatively, the gauge theory may be (partially) Higgsed during inflation. Then the massive particle in Fig. \ref{fig:treeandloop} (a)  need only be a gauge singlet of a residual gauge symmetry, but may be charged under the full gauge group. This possibility, which has received less attention in the literature  (however, see \cite{Chen:2016uwp,Chen:2016hrz} for a related scenario), will be our primary focus. There are two ways in which such a Higgsing can happen, as we discuss now.
	
	First, such a breaking can be due to a fixed tachyonic mass term for the Higgs $\mathcal{H}$, $\mu_{\mathcal{H}}^2\mathcal{H}^\dagger \mathcal{H}$ with $\mu_{\mathcal{H}}\sim H$. In this case, the gauge-Higgs theory remains Higgsed after inflation ends and its massive states can annihilate away as universe cools giving rise to standard cosmology. Grand unified theories are examples of gauge extensions of Standard Model (SM) containing very massive new particles and which are strongly motivated by existing lower energy experimental data. For example, non-supersymmetric unification is suggested by the near renormalization-group convergence of SM gauge couplings in the $10^{13} - 10^{14}$ GeV range, right in the high-scale inflation window \cite{Ade:2015lrj} of opportunity for cosmological collider physics!\footnote{Unification at such scales is disfavored in minimal unification schemes by proton decay constraints, but viable in non-minimal schemes such as that of Refs. \cite{Weiner:2001pv,Csaki:2001qm}.}  
	NG detection of some subset of these massive states could give invaluable clues to the structure and reality of our most ambitious theories. It is also possible that $H$-mass states revealed in NG are not connected to specific preconceived theories, but even this might provide us with valuable clues about the far UV.

	Another very interesting and testable option is a tachyonic ``mass'' term of the form $\mathcal{L}\supset c R \mathcal{H}^\dagger \mathcal{H}$, where  $R$ is the Ricci scalar and $c>0$ parametrizes the non-minimal coupling of Higgs to gravity. The effects of non-tachyonic terms for this form with $c<0$ have been considered before (see e.g. \cite{Espinosa:2007qp,Herranen:2014cua}). Note, spontaneous breaking triggered by $c>0$ is completely negligible at low temperatures, say below $100\text{ GeV}$. Whereas in the scenario above we needed the gauge-Higgs theory to fortuitously have states with $m\sim H$,  here we naturally get the Higgs particle at $H$ for $c\sim\mathcal{O}(1)$. Furthermore, if $(\text{gauge coupling}\times\text{Higgs VEV})\sim H$, we also get massive gauge bosons at $H$. In this way such a nonminimal coupling can lift up a gauge theory with a relatively low Higgs scale today, which we can access via collider or other probes, to the window of opportunity of cosmological collider physics during the inflationary era. We will call this the ``heavy-lifting'' mechanism. To make this idea concrete, we consider the example of heavy lifting the SM. 
	
	During the inflationary era the SM weak scale $v$ can be lifted to be very high, but we do not know where precisely because of the unknown parameter $c$ (even if we knew $H$). However, this uncertainty drops out in mass ratios,
	\begin{equation}\label{massratios}
	\begin{aligned}
	\frac{m_h}{m_Z}&=\frac{2\sqrt{2\lambda_h}}{\sqrt{g^2+g'^2}}\\
	\frac{m_h}{m_W}&=\frac{2\sqrt{2\lambda_h}}{g}\\
	\frac{m_h}{m_t}&=\frac{2\sqrt{\lambda_h}}{y_t}
	\end{aligned}
	\end{equation}
	where, $\lambda_h,g^\prime,g,y_t$ are Higgs quartic, $U(1)_Y$, $SU(2)_L$ and top Yukawa couplings of SM. While the top $t$ and $W$ boson can only appear in loops Fig. \ref{fig:treeandloop} (b), the physical Higgs $h$ and the $Z$ can appear in Fig. \ref{fig:treeandloop} (a) giving us one prediction in this case. However, an important subtlety of the couplings on the R.H.S. of the ratios above is that they are not those measured at the weak scale but rather are the results of running to $\sim H$. But it is well known that the SM effective potential develops an instability around $10^{10}-10^{12} \text{GeV}$ because of the Higgs quartic coupling running negative (see \cite{Buttazzo:2013uya} and references therein for older works). Since the inflationary $H$ can be higher, the Higgs field can sample values in its potential beyond the instability scale. Whether this is potentially dangerous for our universe has been considered before (see e.g. \cite{Espinosa:2007qp,Herranen:2014cua,Buttazzo:2013uya,Fairbairn:2014zia,Kobakhidze:2013tn,Kearney:2015vba,Espinosa:2015qea}). But
	it is possible that this instability is straightforwardly cured once dark matter (DM) is coupled to the SM. A simple example \cite{Burgess:2000yq,Gonderinger:2009jp,Cheung:2012nb,Kadastik:2011aa} would be if future experiments determine that DM is a SM gauge singlet scalar $S$ stabilized by a $Z_2$ symmetry, $S\rightarrow -S$. Then the most general renormalizable  new couplings are given by the Higgs portal coupling and scalar self-interaction,
	\begin{equation}
	\frac{k}{2}S^2|\mathcal{H}|^2+\frac{\lambda_S}{24}S^4.
	\end{equation}
	Since the coupling $k$ contributes positively to the Higgs quartic running, for appropriate choice of $k$ (and less sensitively to $\lambda_S$) the Higgs quartic never becomes negative. This solves the vacuum instability problem of the SM and we can reliably trust our effective theory up to even high scale inflation energies.
	
	Imagine a discovery of such a DM ($S$) is made in the coming years, along with a measurement of $k$ and its mass $m_S$ (and possibly a measurement of or at least a bound on, $\lambda_S$) . Also, imagine a measurement of $H$ is obtained via detecting the primordial tensor power spectrum. Then we can use the Renormalization Group (RG) to run all the measured couplings to the high scale $H$. These would then allow us to compute the run-up couplings needed to make a cosmological verification of \eqref{massratios}. Such a verification of this Next-to-Minimal SM (NMSM) would give strong evidence that no new physics intervenes between TeV and $H$. Since this NMSM clearly suffers from a hierarchy problem (worse than the SM), the precision NG measurements would therefore provide us with a test of ``\textit{un-naturalness}'' in Nature, perhaps explained by the anthropic principle \cite{Weinberg:1987dv,Agrawal:1997gf}.
	Whether the naturalness principle is undercut by the anthropic principle or by other considerations is one of the most burning questions in fundamental physics.

	Of course, the heavy-lifting mechanism may also apply to non-SM ``dark'' gauge-Higgs sectors, which we may uncover by lower energy experiments and observations in the coming years, or to gauge-Higgs extensions of the SM which may emerge from collider experiments. In this way, there may be more than one mass ratio of spin-0 and spin-1 particles that might appear in NG which we will be able to predict. As we will show, such new gauge structure may be more easily detectable in NG than the (NM)SM, depending on  details of its couplings.  It is important to note that different gauge theory sectors in the current era, with perhaps very different Higgsing scales, can be heavy-lifted to the same rough scale $~ H$ during inflation, with their contributions to NG being superposed.

	The heavy-lifting mechanism may not be confined to unnatural gauge-Higgs theories. For example, if low energy supersymmetry (SUSY) plays a role in stabilizing the electro-weak hierarchy, a suitable structure of SUSY breaking may permit the heavy-lifting mechanism to work. Heavy-lifting can then provide us with a new test of naturalness! Possibly non-tachyonic squarks and sleptons in the current era were tachyonic during inflation, higgsing QCD or electro-magnetism back then. We leave a study of the requisite SUSY-breaking structure for future work. Cosmological collider physics studies incorporating SUSY but restricted to gauge singlet fields have appeared in \cite{Baumann:2011nk,Craig:2014rta}.
	
NG potentially provide us with the boon of an ultra-high energy ``cosmological collider'', but cosmic variance implies it operates at frustratingly low ``luminosity''!  We will see that this constrains how much we can hope to measure, even under the best experimental/observational circumstances. For example, a pair of spin-1 particles appearing in the NG will be more difficult to decipher than only one of them appearing, due to the 
more complicated functional form of the pair that must be captured in the limited squeezed regime under cosmic variance. And yet,  we would ideally like to see a rich spectrum of particles at  $H$. 
The key to visibility of new physics under these harsh conditions is then determined by the strength of couplings to the inflaton. This is the central technical consideration of this paper, taking into account the significant suppressions imposed by   (spontaneously broken) gauge invariance. We study this within two effective field theory frameworks, one more conservative but less optimistic than the other. Single-field slow-roll inflation gives the most explicit known construction of inflationary dynamics, but we will see that minimal models under effective field theory control give relatively weak NG signals, although still potentially observable. We also consider the more agnostic approach in which the dynamics of inflation itself is parametrized as a given background process \cite{Cheung:2007st}, but in which the interactions of the  gauge-Higgs sector and inflaton fluctuations are explicitly described. This will allow for larger NG signals, capable in principle of allowing even multiple particles to be discerned.

	This paper is organized as follows. We start in Section \ref{prelim} by reviewing the in-in formalism and its use in calculation of the relevant non-Gaussian observables. We also include a discussion of different gauges and conventions used for characterizing NG. Then in Section \ref{squeezed} we review the significance of the squeezed limit of cosmological correlators, both in the absence and presence of new fields beyond the inflaton. In particular, we review the derivation of \eqref{qsfi}. In Section \ref{gaugeccp} we discuss some general aspects of gauge-Higgs theory dynamics during inflation and elaborate upon the two alternatives for Higgs mechanism discussed above. We then specialize in Section \ref{slowroll} to slow-roll inflation where we study the couplings of Higgs-type and $Z$-type bosons to the inflaton in an effective field theory (EFT) framework.
	In section \ref{eft} we describe parallel considerations in the  more agnostic EFT approach mentioned above. 	The two levels of effective descriptions are then used in Sections \ref{hNG} and \ref{sectiononZ} (supplemented by technical Appendices \ref{appscalarmode}-\ref{appvectorng}) to derive some of the detailed forms of NG due to Higgs-type and $Z$-type exchanges respectively. We conclude in Section \ref{conclu}.
	
	\paragraph{Hubble Units} In this paper, the Hubble scale during inflation is denoted by $H$. To reduce clutter, from now on we will set $H\equiv1$ in most of the numbered equations, with a few exceptions where we explicitly write it for the sake of clarity. Factors of $H$ can be restored via dimensional analysis. However, we will refer explicitly to $H$ in the text throughout, again for ease of reading, and in the unnumbered equations within the text. 
	
	\section{Preliminaries}\label{prelim}
	
	\subsection{The in-in Formalism for Cosmological Correlators }
	Primordial NG induced by inflaton fluctuations are calculated as ``in-in''  expectation values 
	\cite{Weinberg:2005vy} of certain gauge-invariant (products of) operators at a fixed instant of time towards the end of inflation, denoted by $t_f$. The expectation needs a  specification of the quantum state. The notion of ``vacuum'' is ill-defined because spacetime expansion gives a time-dependent Hamiltonian, $\mathsf{H}(t)$. However, for very short distance modes/physics at some very early time $t_i$, the expansion is negligible and we can consider the state to be the Minkowski vacuum, $|\Omega \rangle$. As such modes redshift to larger wavelengths at $t_f$, the state at $t_f$ can then be taken to be given by $U(t_f, t_i) |\Omega \rangle$, where 
	\begin{equation}
	U(t_f, t_i) = T e^{-i\int\limits_{t_i}^{t_f} dt \mathsf{H}(t) }.
	\end{equation}
	In order to capture arbitrarily large wavelengths at $t_f$ in this manner, we formally take $t_i \rightarrow -\infty$.
	(For free fields, the state defined in this way at finite times, is the Bunch-Davies ``vacuum''.)
	Then the desired late-time expectation value of a gauge invariant operator $Q$ is given in the Schroedinger picture by, $\langle \Omega |U(t_f, t_i = - \infty)^\dagger Q U(t_f, t_i = - \infty) |\Omega \rangle$. 
	
	Now the calculation of the expectation value becomes standard. First, we go over to the interaction picture, and second we employ the standard trick of continuing the early evolution slightly into  complex time to 
	project the free vacuum $\vert 0 \rangle$   onto the interacting vacuum  $\vert \Omega\rangle$.  Thus we arrive at the in-in master formula,
	\begin{equation}\label{ininmasterformula}
	\langle \Omega |U(t_f, t_i)^\dagger Q U(t_f, t_i) |\Omega \rangle = \langle 0\vert \bar{T} e^{+i\int\limits_{-\infty(1+i\epsilon)}^{t_f} dt_2 \mathsf{H}_I^{\text{int}}(t_2) } Q_I(t_f) T e^{-i\int\limits_{-\infty(1-i\epsilon)}^{t_f} dt_1 \mathsf{H}_I^{\text{int}}(t_1) }\vert 0\rangle.
	\end{equation}
	In the above, the subscript $I$ denotes that the corresponding operator is to be evaluated in the interaction picture. Finally, $\mathsf{H}^{\text{int}}(t)$ is the interaction part of the Hamiltonian of the fluctuations, i.e. $\mathsf{H}=\mathsf{H}_0+\mathsf{H}^{\text{int}}$ with $\mathsf{H}_0$ being quadratic in fluctuations. We note that the anti-time ordered product also appears in \eqref{ininmasterformula}. The perturbative expansion of cosmological correlators of the above general type is facilitated as usual by expanding in products of Wick contractions, given by in-in propagators. This leads to a diagrammatic form, illustrated in Fig. \ref{fig:single-field}. 
	
	\subsection{Useful Gauges for General Coordinate Invariance}
	Metric and inflaton fluctuations are not gauge invariant under diffeomorphisms. Hence we now review two useful gauges and a gauge invariant quantity characterizing the scalar perturbations during inflation. Our discussion will be brief and for more details the reader is referred to \cite{Maldacena:2002vr,Weinberg:1102255}. For simplicity, we will specialize here to single-field slow-roll inflation, but the considerations are more general.
	
	The metric of dS space is given by,
	\begin{equation}
	ds^2=-dt^2+ a^2(t)d\vec{x}^2,
	\end{equation}
	with $a(t)=e^{Ht}$ being the scale factor in terms of Hubble scale $H$.
	To discuss the gauge choices, it is useful to decompose the spatial metric $h_{ij}dx^idx^j$ in presence of inflationary backreaction as follows \cite{Weinberg:1102255},
	\begin{equation}\label{spatialmetric}
	h_{ij}=a^2(t)\left((1+A) \delta_{ij}+\frac{\partial^2 B}{\partial x^i \partial x^j}+\partial_j C_i +\partial_i C_j+\gamma_{ij}\right),
	\end{equation}
	where, $A,B,C_i,\gamma_{ij}$ are two scalars, a divergenceless vector, and a transverse traceless tensor perturbation respectively. The inflaton field can also be decomposed into a classical part $\phi_0(t)$ and a quantum fluctuation $\xi(t,\vec{x})$, 
	\begin{equation}
	\phi(t,\vec{x})=\phi_0(t)+\xi(t,\vec{x}).
	\end{equation}
	Using the transformation rules of the metric and scalar field, it can be shown that the quantity \cite{Weinberg:2008hq},
	\begin{equation}\label{R}
	\mathcal{R} \equiv \frac{A}{2}-\frac{1}{\dot{\phi}_0}\xi,
	\end{equation}
	is gauge invariant. This is the quantity that is conserved on superhorizon scales for single field inflation  \cite{Salopek:1990jq,Wands:2000dp,Weinberg:2003sw,Senatore:2012ya,Assassi:2012et}. Although $\mathcal{R}$ seems to depend on more than one scalar fluctuation, there is only one physical scalar fluctuation which is captured by it. 
	This is because among the five scalar fluctuations in the metric plus inflaton system, two are non-dynamical constraints and two more can be gauged away by appropriate diffeomorphisms, leaving only one fluctuation. To make this manifest, we can do gauge transformations which set either $A$ or $\xi$ to zero in \eqref{R} to go to spatially flat and comoving gauge respectively. The first of these will be most useful for simplifying in-in calculations involving Hubble-scale massive particles external to the inflation dynamics, while the second one is useful for constraining the squeezed limit of the NG due to inflationary dynamics itself.
	
	\paragraph{Spatially flat gauge \cite{Maldacena:2002vr}}
	In this gauge the spatial metric \eqref{spatialmetric} becomes
	\begin{equation}\label{spatiallyflat}
	h_{ij} = a^2(t)\left(\delta_{ij}+\gamma_{ij}\right).
	\end{equation}
	Gauge invariant answers can be obtained by writing $\xi$ in terms of $\mathcal{R}$ using \eqref{R}, which becomes in this gauge,
	\begin{equation}
	\mathcal{R}=-\frac{1}{\dot{\phi}_0}\xi.
	\end{equation}

	\paragraph{Comoving gauge \cite{Maldacena:2002vr}}
	In this gauge the spatial metric \eqref{spatialmetric} looks like 
	\begin{equation}
	h_{ij} = a^2(t)\left((1+A)\delta_{ij}+\gamma_{ij}\right),
	\end{equation}
	with quantum inflaton field $\xi=0$. This means the gauge invariant quantity ${\mathcal R}$ evaluated in the new gauge becomes,
	\begin{equation}\label{conversion}
	{\mathcal R} = \frac{A}{2},
	\end{equation}
	which lets us rewrite the spatial metric \eqref{spatialmetric} as
	\begin{equation}\label{comovinggauge}
	h_{ij} = a^2(t)\left((1+2\mathcal{R})\delta_{ij}+\gamma_{ij}\right),
	\end{equation}
	with $\mathcal{R}$ being conserved after horizon exit (in single-field inflation). 
	\subsection{Observables}\label{observables}
	Having discussed the gauge choices, we now move on to discussing the observables. The power spectrum for the density perturbations is given by,
	\begin{equation}
	P_{k}\equiv\langle\mathcal{R}(\vec{k})\mathcal{R}(-\vec{k})\rangle^\prime,
	\end{equation}
	where the $\prime$ denotes the notation that momentum conserving delta functions are taken away i.e. 
	\begin{equation}
	\langle\mathcal{R}(\vec{k}_1)\cdots\mathcal{R}(\vec{k}_n)\rangle = (2\pi)^3\delta^3(\vec{k}_1+\cdots+\vec{k}_n)\langle\mathcal{R}(\vec{k}_1)\cdots\mathcal{R}(\vec{k}_n)\rangle^\prime
	\end{equation} 
	The power spectrum can be evaluated to be
	\begin{equation}\label{power}
	P_k=\frac{1}{\dot{\phi}_0^2}\frac{1}{2k^3},
	\end{equation}
	where the R.H.S. is to be evaluated at the moment of horizon exit $k=aH$ for a given $k$-mode. Since different $k$-modes exit the horizon at different times and $\frac{H^4}{\dot{\phi}_0^2}$ has a slow time dependence, the combination $k^3P_k$ is not exactly $k$-independent, and we can write
	\begin{equation}
	k^3P_k\propto \left(\frac{k}{k_*}\right)^{n_s-1}
	\end{equation}
	where $1-n_s$ is the tilt of the power spectrum and $k_*$ is a ``pivot'' scale. From Planck data \cite{Ade:2015lrj} we get, $n_s\approx 0.96$ and $\frac{H^4}{\dot{\phi}_0^2}=8.7\times 10^{-8}$ at $k_*=0.05 \text{ Mpc}^{-1}$. In position space, the power spectrum takes the form, 
	\begin{equation}\label{powerpos}
	\langle\mathcal{R}(\vec{x}_1)\mathcal{R}(\vec{x}_2)\rangle \sim \frac{1}{|x_1-x_2|^{n_s-1}}
	\end{equation}
	
	To calculate the bispectrum we will be interested in evaluating $\langle \mathcal{R}({\vec{k}_1})\mathcal{R}({\vec{k}_3})\mathcal{R}({\vec{k}_3})\rangle$. By translational invariance the three momenta form a triangle, and by rotational invariance we are only interested in the shape and size of the triangle, not in the orientation of the triangle. Furthermore since we also have approximate scale invariance, we do not care about the overall size of the triangle, so effectively the momentum dependence of bispectrum is governed only by the ratios $\frac{k_3}{k_1}$ and $\frac{k_2}{k_1}$. We denote the bispectrum by the function $B(k_1,k_2,k_3)$,
	\begin{equation}\label{Bfunction}
	B(k_1,k_2,k_3)=	\langle \mathcal{R}({\vec{k}_1})\mathcal{R}({\vec{k}_3})\mathcal{R}({\vec{k}_3})\rangle^\prime.
	\end{equation}
	It is convenient to define a dimensionless version of this, 
	\begin{equation}\label{Ffunction}
	F(k_1,k_2,k_3)=\frac{B(k_1,k_2,k_3)}{P_{k_1}P_{k_3}}.
	\end{equation}
	The crude estimate of cosmic variance \eqref{cosmicvar} translates to $\delta F \sim 10^{-4} - 10^{-3}$. It is often conventional in the literature to typify the size of NG by the  value of $F$ at the equilateral point, 
	\begin{equation}\label{fnl}
	f_{\text{NL}} \equiv \frac{5}{18} F(k,k,k).
	\end{equation}

	Since we are mostly interested here in the squeezed limit for future signals, 
	$k_3 \ll k_1, k_2$, we will explicitly compute $F$ in that limit, referring to $f_{\text{NL}}$  only in the context of current NG limits (see subsection \ref{minimalgoldstone}).
	In terms of the quantum inflaton field $\xi$, the function $F$ can be rewritten as, 
	\begin{equation}
	F(k_1,k_2,k_3)=-\dot{\phi}_0\frac{\langle\xi({\vec{k}_1})\xi({\vec{k}_2})\xi({\vec{k}_3})\rangle'}{\langle\xi({\vec{k}_1})\xi({-\vec{k}_1})\rangle' \langle\xi({\vec{k}_3})\xi({-\vec{k}_3})\rangle'}\vert_{k_3\ll k_1,k_2},
	\end{equation}
	and where the R.H.S. is evaluated at the point of horizon exit for each mode.

	\section{Squeezed Limit of Cosmological Correlators}\label{squeezed}
	
	\subsection{NG from Single Field Inflation in the Squeezed Limit}\label{singlefield}
	
	In single field inflation, NG in the squeezed limit is proportional to the tilt of the inflaton power spectrum \cite{Maldacena:2002vr,Creminelli:2004yq,Creminelli:2011rh}, i.e.
	\begin{equation}\label{fnl_singlefield}
	F^{\text{Single Field}}(k_1, k_2, k_3)\vert_{k_3 \ll k_1,k_2}  = (1-n_s)+\mathcal{O}\left(\frac{k_3}{k_1}\right)^2.
	\end{equation}
	Let us go to comoving gauge \eqref{comovinggauge} to demonstrate this. We are interested in computing   $\langle\mathcal{R}_h(\vec{k}_1)\mathcal{R}_h(\vec{k}_2)\mathcal{R}_s(\vec{k}_3)\rangle^\prime$, where the subscript $h(s)$ means the associated momentum is hard(soft). We define position space coordinates $\vec{x}_i$ to be conjugate to momentum $\vec{k}_i$. In the limit $k_3\ll k_1,k_2$ we are interested in an ``Operator Product Expansion (OPE)'' regime, $\vert \vec{x}_1-\vec{x}_2\vert \ll \vert \vec{x}_1-\vec{x}_3\vert $. 
	\begin{figure}[h]
		\centering
		\includegraphics[width=0.5\linewidth]{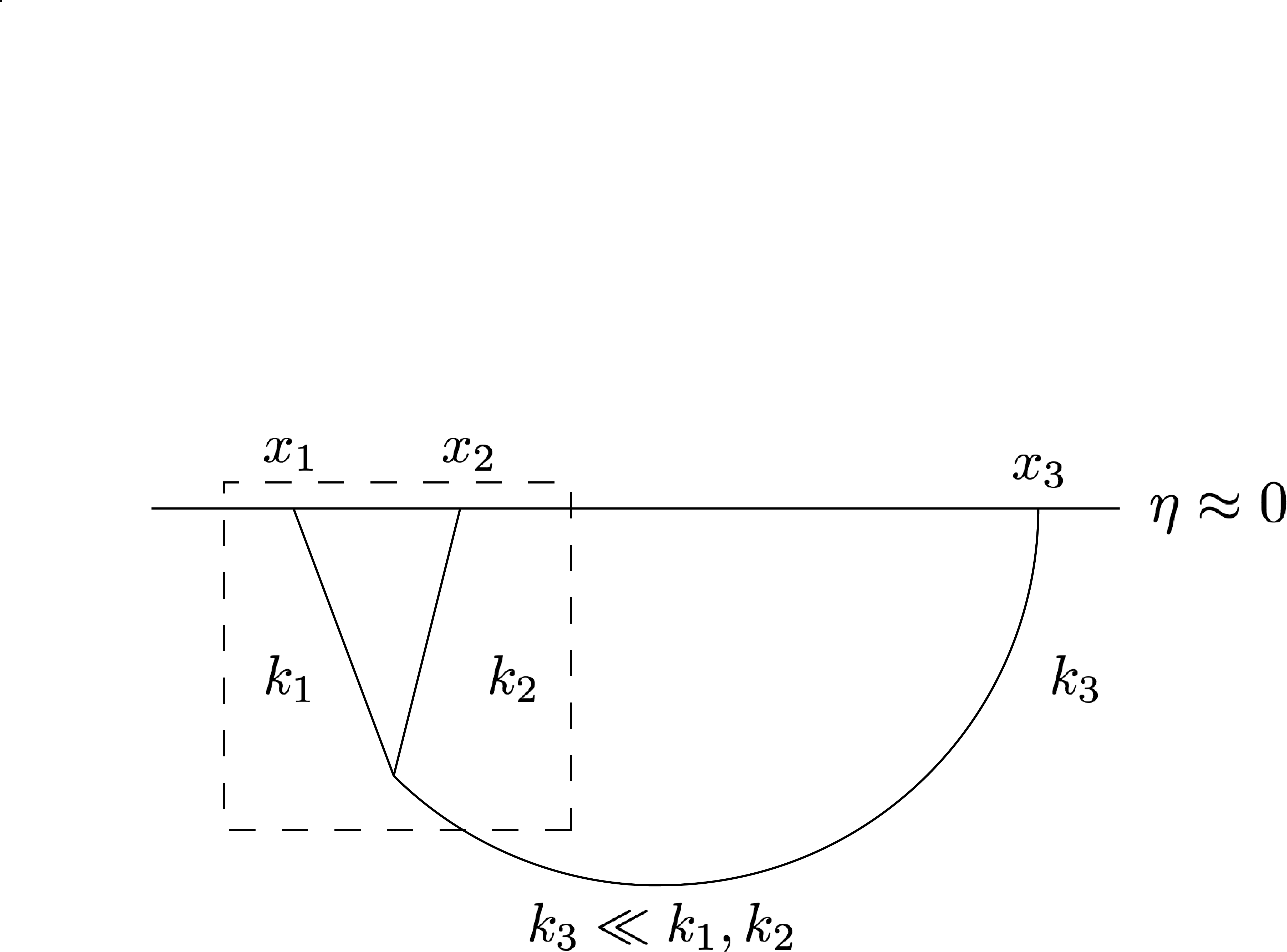}
		\caption{NG in single-field inflation}
		\label{fig:single-field}
	\end{figure}
	Consider just the leading tree-level structure of the associated diagram in Fig. \ref{fig:single-field}, and first focus on just the boxed subdiagram. We see that for this subdiagram the soft line is just a slowly-varying
	background field in which we are computing a hard 2-point correlator. 
	Thus,
	\begin{equation}
	\langle\mathcal{R}_h(x_1)\mathcal{R}_h(x_2)\mathcal{R}_s(x_3)\rangle \approx \langle \langle\mathcal{R}_h(x_1)\mathcal{R}_h(x_2)\rangle_{\mathcal{R}_s(\frac{x_1+x_2}{2})}\mathcal{R}_s(x_3)  \rangle.
	\end{equation}
	The effect of the soft mode $\mathcal{R}_s$ is just to do the transform $\vec{x}\rightarrow (1+\mathcal{R}_s)\vec{x}$ of \eqref{comovinggauge} within the leading 2-point function of \eqref{powerpos}:
	\begin{equation}
	\langle\mathcal{R}_h(x_1)\mathcal{R}_h(x_2)\rangle_{\mathcal{R}_s}\sim \frac{1}{\left(|x_1-x_2|(1+\mathcal{R}_s)\right)^{n_s-1}} \approx \frac{1}{\left(|x_1-x_2|\right)^{n_s-1}}(1-n_s)\mathcal{R}_s\left(\frac{x_1+x_2}{2}\right).
	\end{equation}
	To get the middle expression, we have taken $\mathcal{R}_s$ to be approximately constant over distances of order $|\vec{x}_1 - \vec{x}_2|$, a good approximation since $k_3\rightarrow 0$. The last expression follows by expanding in (small) ${{\cal R}_s}$, evaluated at the midpoint $(\vec{x_1} + \vec{x_2})/2$. 
	We have also dropped a $\mathcal{R}_s$-independent piece since that goes away when we consider the three point function. 
	
	Thus the three point function becomes,
	\begin{multline}
	\langle\mathcal{R}_h(x_1)\mathcal{R}_h(x_2)\mathcal{R}_s(x_3)\rangle\approx \langle \langle\mathcal{R}_h(x_1)\mathcal{R}_h(x_2)\rangle_{\mathcal{R}_s(\frac{x_1+x_2}{2})}\mathcal{R}_s(x_3)  \rangle \\ \approx (1-n_s)\frac{1}{\left(|x_1-x_2|\right)^{n_s-1}} \frac{1}{\left(|x_1-x_3|\right)^{n_s-1}}.
	\end{multline}
	Fourier transforming to momentum space,
	\begin{multline}
	\langle\mathcal{R}_h(\vec{k}_1)\mathcal{R}_h(\vec{k}_2)\mathcal{R}_s(\vec{k}_3)\rangle^\prime \sim (1-n_s) \frac{1}{k_1^{4-n_s}}\frac{1}{k_3^{4-n_s}} \\ \sim (1-n_s) \langle\mathcal{R}_h(\vec{k}_1)\mathcal{R}_h(-\vec{k}_1)\rangle^\prime \langle\mathcal{R}_s(\vec{k}_3)\mathcal{R}_s(-\vec{k}_3)\rangle^\prime,
	\end{multline}
	leading to \eqref{fnl_singlefield}. Subleading corrections proportional to $\left(\frac{k_3}{k_1}\right)$ are absent 
	by rotational invariance, so the leading corrections are order 
	$\left(\frac{k_3}{k_1}\right)^2$.
	
	The importance of the above expression lies in the fact that in the squeezed limit any value of $F^{\text{Single Field}}$ bigger than ${\cal O}(1-n_s)$ will signal the presence of new physics beyond single-field inflationary dynamics. In particular, next we comment on what can happen to the squeezed limit if we have multiple light ($m\ll H$) fields ( ``multifield inflation'') or $m\sim H$ fields (``quasi single field inflation'' \cite{Chen:2009zp})  during inflation.	
	
	\subsection{NG from Multifield Inflation in the Squeezed Limit}
	
	If we have light fields  with $m\ll H$, other than the inflaton, then during inflation those fields can lead to larger NG in the squeezed limit than (\ref{fnl_singlefield}), see \cite{Byrnes:2010em} and references therein. This can be understood again via similar  in-in diagrammatics to Fig. \ref{fig:single-field} . In this case it is again true that we have to evaluate the hard two point function in the background of \textit{some} soft mode, and correlate the result with a $\mathcal{R}$ soft mode. However, since $\mathcal{R}_s$ is no longer the only soft mode in the theory,
	\begin{equation}
	\langle\mathcal{R}_h(x_1)\mathcal{R}_h(x_2)\rangle_{\text{soft mode}} \neq \langle\mathcal{R}_h(x_1)\mathcal{R}_h(x_2)\rangle_{\mathcal{R}_s}.
	\end{equation}
	Thus the derivation in the previous subsection does \textit{not} go through. Consequently $F^{\text{Multi Field}}$ in the squeezed limit is no longer constrained to be order $(1-n_s)$, but rather it becomes model dependent.
	
	\subsection{NG from Hubble-scale Masses in the Squeezed Limit}
	
	The situation changes quite a lot if we have particles with $m\sim H$. Such particles can modify the bispectrum in a way that in the squeezed limit $F$ contains a \textit{non-analytic} part,
	\begin{equation}\label{fnl_qsfi}
	F^{\text{nonanalytic}}  \propto f(\mu)\left(\frac{k_3}{k_1}\right)^{\frac{3}{2}+i\mu}+f(-\mu)\left(\frac{k_3}{k_1}\right)^{\frac{3}{2}-i\mu},
	\end{equation} 
	where, $\mu=\sqrt{\frac{m^2}{H^2}-\frac{9}{4}}$ and $f(\mu)$ is a calculable function of the mass of the particle, which is of the order 1 when $\mu\sim 1$ but is ``Boltzmann suppressed'' $\sim e^{-\pi\mu}$ for large $\mu$. We have a proportionality sign in \eqref{fnl_qsfi} because there are model dependent prefactors which can take either large or small values, thus from \eqref{fnl_qsfi} itself we can not get a complete estimate of NG. We will spell out the model dependent prefactors later. 
	
	The crucial aspect of \eqref{fnl_qsfi} is that $F$ now contains a \textit{non-analytic} dependence on $\left(\frac{k_3}{k_1}\right)$ along with other analytic terms. Importantly this non-analytic behavior can \textit{not} be captured by any single or multifield inflation models where all the masses are much smaller than $H$.
	
	This dependence also encodes the information about the mass of the Hubble scale particle, via the exponent $\mu$, \cite{Chen:2009zp,Baumann:2011nk,Chen:2012ge,Assassi:2012zq,Noumi:2012vr,Arkani-Hamed:2015bza}. If the massive particle has a nonzero spin(s), then $F^{non-analytic}$ has an additional factor dependent on Legendre polynomials, $P_s(\cos\theta)$, where $\hat{k}_1\cdot\hat{k}_3=\cos\theta $. If we can measure this angular dependence precisely enough then we can get the information about spin as well \cite{Arkani-Hamed:2015bza,Lee:2016vti}.
	Furthermore, such angular dependence is absent in purely single-field and some of the multifield inflation models. This can in principle help us in distinguishing the ``signal'' of $m\sim H$ particles from the ``background'' of $m\ll H$ particles.
	
	We see as we go to the region, $m\ll H$, the leading behavior reverts to being analytic $\approx\left(\frac{k_3}{k_1}\right)^{3/2-3/2}$, and indistinguishable from purely single-field or multi-field inflation. 
	This means it is observationally challenging to reach the region $m \ll H$ and still distinguish and measure $m$ accurately. Also, at the other extreme, for $m\gg H$ cosmological production is strongly Boltzmann suppressed, so observation will again be difficult. Therefore we are led to a window around $H$ for doing spectroscopy of masses and spins.

\begin{figure}[h]
		\centering
		\includegraphics[width=1\linewidth]{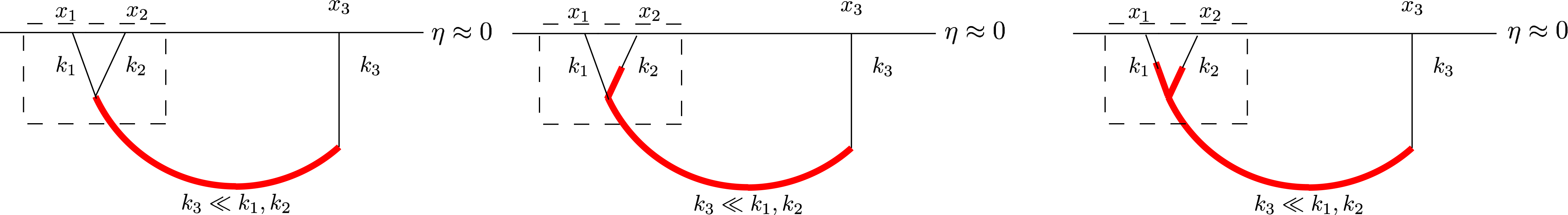}
		\caption[Non-Gaussianity due to Massive Particles]{From left to right: (a) Single Exchange Diagram, (b) Double Exchange Diagram, (c) Triple Exchange Diagram. Note that all these diagrams rely on mixing between the inflaton fluctuation and massive scalar in the (implicit) non-trivial background of rolling $\phi_0(t)$.}
		\label{fig:three-diagrams}
	\end{figure}

	
	Let us briefly explain the form of \eqref{fnl_qsfi}, first concentrating on just the soft $k_3$-dependence.  
	In presence of new particles with $m\sim H$ there are additional contributions to the bispectrum beyond those in Fig. \ref{fig:single-field}. At tree level we can have three diagrammatic forms, as shown in Fig. \ref{fig:three-diagrams}. These are called single, double and triple exchange diagram based on the number of massive propagators \cite{Lee:2016vti}.
	\begin{figure}[h]
		\centering
		\includegraphics[width=0.8\linewidth]{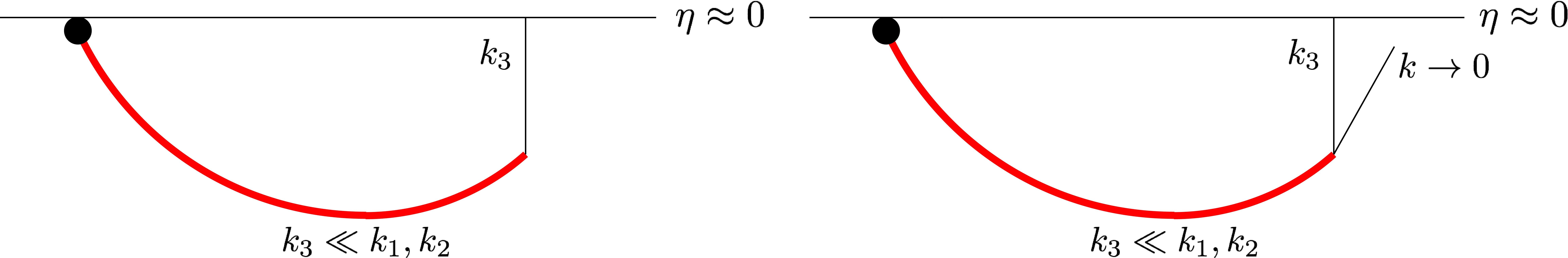}
		\caption{From left to right: (a) ``OPE'' approximation of three point function in squeezed limit as a two point function. The $\phi_0$ background causing mixing is not explicitly shown, as in Fig. \ref{fig:three-diagrams}.  (b) The same ``OPE'' approximation expressed an inflaton-$h$  three point function with one inflaton leg set to zero momentum to now explicitly represent the background $\phi_0$.}
		\label{fig:ope}
	\end{figure}
	In the squeezed limit, we are once again interested in calculating $\langle\mathcal{R}_h(\vec{k}_1)\mathcal{R}_h(\vec{k}_2)\mathcal{R}_s(\vec{k}_3)\rangle^\prime$ i.e. correlation of two hard modes with a soft mode. 
	In position space, this again corresponds to an ``OPE'' limit, $\vert x_{12}\vert \ll \vert x_{13}\vert $, where the hard subdiagram is given by an effective local vertex, depicted in Fig. \ref{fig:ope} by the round black blob. The strength of this effective vertex is then given by the hard two-point function in the background of the massive but $k_3$-soft field, which is predominantly $k_3$-independent. Tracking only the $k_3$-dependence is then given by the two-point correlator shown in Fig. \ref{fig:ope}.

	The leading $k_3$ dependence can be worked out by the scaling properties of the fields involved, which 
	can be read off from their classical late time asymptotics. For a general scalar field, 
	\begin{equation}
	\chi(\eta,\vec{x})\vert_{\eta\rightarrow 0}=(-\eta)^{\Delta_1} \mathcal{O}_1(\vec{x}) + (-\eta)^{\Delta_2} \mathcal{O}_2(\vec{x}),
	\end{equation}
	where, $\Delta_{1,2}=\frac{3}{2}\pm i \sqrt{\frac{m^2}{H^2}-\frac{9}{4}}$.
	This means, $\mathcal{O}_{1(2)}(\vec{x}) $ can be thought of as an operator with scale dimension $\Delta_{1(2)}$. So we will denote $\mathcal{O}_{1(2)}\equiv \mathcal{O}_{\Delta_{1(2)}}$.
	
	As $\eta\rightarrow 0$, dS isometry generators in 4D acts as generators of the conformal group in 3D space. 
	However, the leading effect of the inflationary background is to break this conformal invariance of late-time correlators, but only weakly for slow roll. Using the simple scaling symmetry subgroup of the 3D conformal invariance, we find \footnote{There can be subleading slow-roll $\sim (1 -n_s)$ corrections to the exponent which are neglected here.} \footnote{Note that 4D dimensionful parameters, such as the Planck scale, do not break this 3D conformal or scale invariance.} 
	\begin{equation}
	\langle \mathcal{O}_\Delta(\vec{x}_2) \mathcal{R}_{s}(\vec{x}_3)\rangle_{\text{inf}} \propto |x_{23}|^{-\Delta}.
	\end{equation}
	However, it is well known that 2-point correlators of differing scale dimension vanish if conformal invariance is exact, therefore the implicit proportionality ``constant''  is suppressed by slow-roll parameters here. 
	Fourier transforming and writing $\Delta=\frac{3}{2}\pm i\mu$, we see $F$ should have the factor $k_3 ^{\frac{3}{2}\pm i\mu}$. We can now put back the $k_1$ dependence, which again by the above scale invariance can only  enter into the expression for $F$ as shown in \eqref{fnl_qsfi}.
	
	\section{Gauge-Higgs Theory and Cosmological Collider Physics}\label{gaugeccp}
	
	\subsection{The Central Plot and its Connections to the Literature}\label{relationtolit}
	
	Having commented on NG and the squeezed limit in general, we focus on what kind of signature a gauge theory coupled to the inflaton will have on NG. In particular we study signatures of Higgs scalars and gauge bosons. Non-trivial spin of heavy particles  in the context of slow-roll inflation was first considered in 
	Ref. \cite{Arkani-Hamed:2015bza}, primarily for even spin.
	In Ref. 
	\cite{Lee:2016vti} both even and odd spin were considered. In both \cite{Arkani-Hamed:2015bza} and \cite{Lee:2016vti} no assumptions were made on 
	the origins of the heavy masses.  
	Here, 
	we will impose the stringent constraints following from assuming that the heavy masses arise via the Higgs mechanism of weakly-coupled gauge field theory, in particular for spins 0 and 1. 
	In particular, the relevant non-linear terms coupling the gauge-Higgs sector and the inflaton will be more suppressed 
	by requiring gauge invariance than would be the   case for massive fields unconnected to a Higgs mechanism.
	
	For example, consider the interaction of a pair of massive spin-1 particles, $Z_\mu$,  with a pair of inflaton fields.  Without considering a gauge theoretic origin for  $Z$, a  low dimension interaction respecting (approximate) inflaton shift symmetry has the form, 
	\begin{equation}
	\frac{1}{\Lambda^2}(\partial\phi)^2 Z^\mu Z_\mu,  
	\end{equation}
	where $\Lambda$ is of order the cutoff of EFT. However, if  $Z_\mu$ is a Higgsed gauge boson, the analogous interaction must arise from 
	\begin{equation}
	\frac{1}{\Lambda^4}(\partial\phi)^2|D\mathcal{H}|^2.
	\end{equation}
	Crucially the interaction between $Z_\mu$ and inflaton has to happen in the presence of the Higgs field $\mathcal{H}$ since we are assuming inflaton to be a gauge singlet. Assuming the gauge theory is spontaneously broken, we see that the gauge theory interaction has a suppression of the order $m_Z^2/\Lambda^2$ compared to the non-gauge theoretic case.\footnote{The non-gauge theoretic case can be viewed as the limit of the gauge case  where $m_Z \sim \Lambda$. For example, a QCD $\rho$ meson or a spin-one superstring excitation cannot be housed in point-particle EFT, except in the marginal sense where the effective cutoff is $\Lambda \sim m_Z$, where the constraints of gauge invariance disappear.} 
	This argument can be generalized to all the gauge boson interaction terms that we consider below. This also makes a general point that it can be harder to see NG due to gauge sector particles compared to non-gauge theoretic states. This is especially true for spin-1 particles as we saw above. 
	
	In \cite{Chen:2016uwp,Chen:2016hrz} the signature of  gauge theory was considered, focusing on unbroken electroweak symmetry during the inflationary phase as well as the scenario of Higgs inflation (in which the inflaton is identified with the physical Higgs field). In a general gauge theory with unbroken gauge symmetry, the gauge bosons will be massless up to (small) loop corrections \cite{Burgess:2009bs,Chen:2016nrs}. Non-trivial spectroscopy must then proceed via gauge-charged matter, which can only appear in loops by charge conservation and the singlet nature of the inflaton. Such loops are difficult (but depending on specific models, may not be impossible \footnote{For example one can imagine working in an effective theory of inflation with its cutoff $\Lambda\gtrsim H$, however if we have a cutoff very close to Hubble then the gauge theory spectrum is no longer separated from the states coming from some UV completion of the field theory, and measurements of NG cannot be translated robustly into
		information about the gauge theory alone. Such a scenario, of course, is still interesting, but we do not focus on that in this paper.}) to observe for several reasons. First, at one  loop \cite{Arkani-Hamed:2015bza}, $F^{\text{loop}}\sim \tilde{f}(\mu)\left(\frac{k_3}{k_1}\right)^{3+2i\mu}+\tilde{g}(\mu)\left(\frac{k_3}{k_1}\right)^{3-2i\mu}+\cdots$, so the fall-off is faster compared to \eqref{fnl_qsfi} as one goes to smaller $k_3$. Second, for heavier masses the Boltzmann suppression goes as $e^{-2\pi\mu}$ because there is now a pair of massive particles involved. Thirdly, we will obviously have the loop factor ($\sim \frac{1}{16\pi^2}$) suppression.
	
	This gives us motivation to look for bigger tree level effects which will be present if gauge symmetry is broken spontaneously during inflation. In \cite{Chen:2016uwp,Chen:2016hrz} such a scenario was mentioned although the primary focus was on Higgs-inflation-like scenarios in which the Higgs VEV is very large compared to $H$ and consequently the massive gauge bosons are too heavy to be seen via NG due to Boltzmann suppression. The situation is much better if one keeps the gauge theory and inflaton sectors distinct,  with gauge symmetry spontaneously broken and Higgs VEV not too much larger than $H$. This is the case we focus on, and we will see that such scenarios can give rise to observable NG for both spin-0 and spin-1 particles.
	
	Since the Hubble scale during inflation can be very high ($H\lesssim 5\times10^{13}\text{GeV}$), inflation and the study of NG provides an exciting arena to hunt for new particles. In this regard two distinct possibilities arise. We discuss them next.

	\subsection{High Energy Physics at the Hubble Scale}
	
	We could imagine a scenario in which there exists some new spontaneously broken gauge theory at $H$. 
	Then some of the gauge-charged matter and gauge-fields may become singlets under the residual unbroken gauge symmetry. Bosons of this type, spin-0 and spin-1, can therefore have Hubble scale masses, couple to the inflaton, and 
	leave their signatures on NG at {\it tree-level}. For simplicity here, we focus on 
	spontaneously broken $U(1)$ gauge theory with no residual gauge symmetry, but is straightforward to  
	generalize to the nonabelian case. For example, we can imagine a scalar in the fundamental representation of $SU(N)$ breaking the symmetry to $SU(N-1)$. Then the gauge boson associated with the broken diagonal generator plays the role of the massive $U(1)$ gauge boson that we consider now. 
	
	Let us focus on the case of single-field slow-roll inflation. We write an effective theory with cutoff $\Lambda$. Since we are interested in effects of gauge theory on NG, we will write down higher derivative interaction terms between the gauge sector and inflaton. But we will not be explicit about higher derivative terms containing gauge sector fields alone or the inflaton alone, although we will ensure that such terms are within EFT control. 
	
	The lagrangian containing the inflaton $\phi$ (with an assumed shift symmetry), the Higgs ($\mathcal{H}$)  and gauge bosons
	(not necessarily the SM Higgs and gauge bosons) has the form
	
	\begin{align}\label{lag}
	\mathcal{L}=&\frac{1}{2}M_{\text{pl}}^2 R+\mathcal{L}_{\text{Gauge Theory}}  -\frac{1}{2}(\partial\phi)^2 -V(\phi) + \mathcal{L}_{\text{int}}^{\text{inf}} + \mathcal{L}_{\text{int}}^{\text{inf-gauge}},
	\end{align}
	where $\mathcal{L}_{\text{Gauge Theory}}$ contains all the terms (including higher derivative terms) containing gauge theory fields alone. $V(\phi)$ is a generic slow roll potential. $\mathcal{L}_{\text{int}}^{\text{inf}}$ contains higher derivative terms containing inflaton alone. For our purpose the interesting interaction terms between gauge theory and the inflaton are contained in
	$\mathcal{L}_{\text{int}}^{\text{inf-gauge}}$, which we write below assuming an UV cutoff $\sim \Lambda$ and a set of dimensionless EFT coefficients $c_i$,
	
	\begin{multline}\label{lintinf}
	\mathcal{L}_{\text{int}}^{\text{inf-gauge}} = \frac{c_1}{\Lambda}\partial_\mu\phi(\mathcal{H}^\dagger D^\mu\mathcal{H})+ \frac{c_2}{\Lambda^2}(\partial\phi)^2 \mathcal{H}^\dagger \mathcal{H} +  \frac{c_3}{\Lambda^4}(\partial\phi)^2|D\mathcal{H}|^2 +
	\frac{c_4}{\Lambda^4}(\partial\phi)^2Z_{\mu\nu}^2 \\
	+\frac{c_5}{\Lambda^5}(\partial\phi)^2\partial_\mu\phi (\mathcal{H}^\dagger D^\mu \mathcal{H})+
	\cdots
	\end{multline}
	In $\mathcal{L}_{\text{int}}^{\text{inf}}$, the first term gives a quadratic mixing between Higgs and $Z^0$. It also couples Higgs, $Z$ and the inflaton. But it does not contain any quadratic mixing between the inflaton and $Z$; and also none between the inflaton and Higgs. But we do see, from Fig. \ref{fig:three-diagrams}, that we need one or more quadratic mixings between the inflaton and the massive particle of interest. Such quadratic mixing does arise from the second and the fifth term, which give quadratic mixing of the inflaton with Higgs and $Z$ respectively. The third term gives, among other interactions, the interaction between an inflaton and a pair of Zs. We have not written operators coming from the expansion in $\left(\frac{\mathcal{H}^\dagger \mathcal{H}}{\Lambda^2}\right)$ since these will be subdominant to the terms we have already considered.
	
	To unpack \eqref{lintinf} we can go to the unitary gauge for $U(1)$ gauge theory and write down some of the relevant terms,
	
	\begin{multline}\label{lintinfprime}
	\mathcal{L}_{\text{int}}^{\text{inf-gauge}}=\rho_{1,Z} Z^0 h + \rho_{1,h}h\dot{h}+ \frac{\rho_{1,Z}}{2v}Z^0 h^2 + \frac{\rho_{1,Z}}{\dot{\phi}_0}\partial_\mu\xi Z^\mu h  \\
	+\alpha \mathcal{H}^\dagger \mathcal{H} - \rho_2 \dot{\xi}h +\frac{\rho_2 v}{4\dot{\phi}_0}(\partial\xi)^2- \frac{\rho_2}{2v}\dot{\xi}h^2 + \frac{\rho_2}{2 \dot{\phi}_0}(\partial\xi)^2h \\
	-\frac{c_3\dot{\phi}_0}{\Lambda^4}\dot{\xi}\left((\partial h)^2+m^2 Z^\mu Z_\mu\right) \\
	-\frac{2c_4\dot{\phi}_0}{\Lambda^4}\dot{\xi}Z_{\mu\nu}^2\\
	+\rho_{5,Z} \dot{\xi}Z^0(1+\frac{2h}{v})+\frac{\rho_{5,Z}}{\dot{\phi}_0}\dot{\xi}\partial_\mu\xi Z^\mu-\frac{\rho_{5,Z}}{2 \dot{\phi}_0}(\partial\xi)^2 Z^0+\rho_{5,h} \dot{\xi}\dot{h}(\frac{h}{v}+1)-\frac{\rho_{5,h} }{\dot{\phi}_0}\dot{\xi}\partial^\mu\xi \partial_\mu h -\frac{\rho_{5,h}}{2 \dot{\phi}_0}(\partial\xi)^2\dot{h}
	\end{multline}
	where we have expanded the Higgs field in unitary gauge $\mathcal{H} = \left(0 \hspace{1em}\frac{(h+v)}{\sqrt{2}}\right)^T$ and the  inflaton field $\phi=\phi_0+\xi$. The inflationary background gives a correction to the Higgs quadratic term via the parameter $\alpha = -\frac{c_2 \dot{\phi}_0^2}{\Lambda^2}$. We also have several quadratic mixing parameters, $\rho_i$, 
	\begin{multline}\label{quamixing}
	\rho_{1,Z} = -\frac{\text{Im}(c_{1}) \dot{\phi}_0 m_Z}{\Lambda};\quad \rho_{1,h} = -\frac{\text{Re}(c_{1}) \dot{\phi}_0 }{2\Lambda};
	\quad \rho_2 = \frac{2 c_2 \dot{\phi}_0 v}{\Lambda^2} ;\\
	\rho_{5,Z} = \frac{\text{Im}(c_{5}) \dot{\phi}_0^2 m_Z v}{\Lambda^5};\quad\rho_{5,h} = \frac{\text{Re}(c_{5}) \dot{\phi}_0^2 v}{\Lambda^5}.
	\end{multline}
	
	\subsection{Heavy-lifting of Gauge-Higgs Theory}\label{uplift}
	
	Until now, we have been discussing theories with Higgs physics intrinsically of order $H$. 
	Now although a future detection of $m\sim H$ particles via NG will be very interesting in its own right, 
	given that $H$ may well be orders of magnitude beyond the energies of foreseeable particle colliders, we would not have valuable complementary access to this physics in the lab. But as discussed in the introduction, the alternative is the ``heavy-lifted'' scenario,   in which $m \sim H$ during inflation and again yields observable NG, and yet $m \ll H$ in the current post-inflationary era and therefore conceivably is accessible to collider and other ``low-energy'' probes. 
	
	Given a gauge theory at low energy, we can consider adding a non-minimal coupling of the Higgs to gravity, $c R \mathcal{H}^\dagger \mathcal{H}$ to the lagrangian \eqref{lag}, where we will consider $c$ of order one. This gives a Higgs effective potential of the form, 
	\begin{equation}
	V_{\text{eff}}(\mathcal{H}) = \lambda_h |\mathcal{H}|^4 -\mu_h^2 |\mathcal{H}|^2 - c R \mathcal{H}^\dagger \mathcal{H}   ,
	\end{equation}
	While the curvature is negligible in the current era, during inflation we have $R \approx 12 H^2$, so that for $c > 0$, the symmetry breaking scale setting gauge-Higgs physical masses is naturally of order $H$. We can also see how this ``heavy-lifting'' mechanism appears in Einstein frame in which the inflaton and Higgs potential get modified to
	\begin{equation}
	(V(\phi)+V(\mathcal{H}))\rightarrow (V(\phi)+V(\mathcal{H})) /\Omega^4 \approx V(\phi)(1-\frac{4 c \mathcal{H}^\dagger \mathcal{H}}{M_{\text{pl}}^2}) + V(\mathcal{H}),
	\end{equation}
	where $\Omega^2 = 1+\frac{2 c \mathcal{H}^\dagger \mathcal{H}}{M_{\text{pl}}^2}$ is the Weyl scaling factor used to get to Einstein frame and we have kept the leading correction in $c \mathcal{H}^\dagger \mathcal{H}/M_{\text{pl}}^2$\footnote{There may in addition be direct Higgs-inflaton couplings even before the Weyl-rescaling to Einstein frame, in which case the Einstein frame couplings may be modified from that above. However, even this modification would have to share similar features, namely that 
		during inflation the Higgs mass parameter is effectively raised to the $H^2$-scale and 
		in the current post-inflationary era the Higgs mass parameter is much smaller in order to fit the current electroweak data. Therefore, we will not pursue this more general modified lagrangian, for simplicity}. For NG, the discussion in the previous subsection then carries over from this point.
	
	As we elaborated in the introduction, one interesting fact about the heavy-lifting mechanism is that it is testable. This requires a knowledge of the couplings of the gauge theory sector in the current era, where they may be accessible at collider energies, and a measurement of $H$ during inflation, as for example via the  primordial tensor power spectrum. We can then use the renormalization group to run those couplings up to $H$, and thereby predict the mass ratios of spin-0 and spin-1 $h$ and $Z$ type particles (bosons charged under the full gauge symmetry which are singlets of the unbroken gauge symmetry) as they were in the inflationary epoch when they contributed to NG. Here the richer the set of $h$ and $Z$ type particles, and hence the larger the set of mass ratios, the less precision we would need to measure each ratio in NG in order to be convinced that we are seeing the \textit{same} gauge theory in both regimes.
	
	\section{NG in Single Field Slow Roll Inflation}\label{slowroll}

	We saw in the previous section that the leading interaction between inflaton and gauge theory is captured by \eqref{lintinf} and \eqref{lintinfprime}. These can be used to estimate the magnitudes of NG induced by $h$ and $Z$. However, the parameters appearing in those two lagrangians have to satisfy several consistency requirements. We first discuss such restrictions and then proceed with the estimation of NG. Our discussion in this section will be in the context of slow roll inflation.
	
	\subsection{Cutoff and Coupling Strengths of Effective Theory }
	
	We start with the restriction on $\Lambda$, which we saw in the previous section sets the most optimistic suppression scale for higher-dimensional interactions relevant to NG. We imagine that $\Lambda$ roughly represents the mass scale of heavy particles that have been integrated out to give the effective non-renormalizable couplings we need between the gauge sector and inflaton. We can therefore think of them as $\Lambda$-mass ``mediators'' of the requisite effective interactions. But in general, if such mediators couple substantively to both the inflaton and to the gauge sector, they will also mediate inflaton (non-renormalizable) self-interactions, roughly powers of $\left(\frac{(\partial\phi)^2}{\Lambda^4}\right)$. In order for the effective expansion in these powers to be controlled, we should require $\Lambda$ to exceed the inflationary kinetic energy  \cite{Creminelli:2003iq},
	\begin{equation}
	\Lambda > \sqrt{\dot{\phi}_0}.
	\end{equation}
	In our ensuing discussion of single-field inflation, we will take this bound to hold. We will assume an approximate inflaton shift symmetry during inflation, allowing the $\Lambda^4$ to be only as big as the slowly-rolling kinetic energy rather than a larger scale.
	
	The potential energy of the inflaton field $V(\phi)$ gives rise to an even higher energy scale $V^{\frac{1}{4}}$, which is bigger than $\sqrt{\dot{\phi}_0}$. Approximate shift symmetry during inflation keeps this scale from spoiling the EFT expansion in higher-dimension operators, but after inflation this symmetry may be significantly broken and the higher scale can then affect dynamics significantly. In particular, EFT with  $\Lambda < V^{\frac{1}{4}}$ can break down at reheating, signaling that the $\Lambda$-scale mediators can be reheated and subsequently decay. However, the NG produced and described by the controlled effective theory during inflation are already locked in on superhorizon scales and are insensitive to the subsequent post-inflationary breakdown of the EFT. 
	
	Furthermore, in theories involving large ``vacuum'' expectation values, non-renormalizable operators in the UV theory can become super-renormalizable (or marginal) in the IR, once some fields are set to their expectation values. There is then the danger of such effective super-renormalizable couplings becoming strong in the IR, and outside perturbative control, or becoming effective mass terms which are too large phenomenologically and have to be fine-tuned to be smaller. This general concern is realized in the present context, because of the large classical expectation given by $\dot{\phi}_0(t) \gg H^2$, as well as large $\langle {\cal H} \rangle > H$ within some of the interesting parameter space. We find that these issues are avoided for sufficiently small $c_i$ in \eqref{lintinf} with, 
	\begin{equation}
	c_i \sim {\cal O}(H/\sqrt{\dot{\phi}_0}),
	\end{equation}
	which we take to hold from now on.
	We go into more detail on such restrictions in the next subsection.
	
	To concretely illustrate the above considerations, consider the following set up. We imagine a theory, with a cutoff $\Lambda^\prime \gtrsim V^{\frac{1}{4}}> \sqrt{\dot{\phi}_0}$, in which the inflaton and Higgs do not interact directly. Thus a term like $\frac{1}{\Lambda^{\prime 2}}(\partial\phi)^2\mathcal{H}^\dagger\mathcal{H}$ is absent in the lagrangian. However, we assume the presence of a ``mediator'' gauge-singlet particle $\sigma$ with mass $m_\sigma \sim \sqrt{\dot{\phi}_0}$, which talks to both the inflaton and Higgs separately via the terms,
	\begin{equation}
	\frac{1}{\Lambda^\prime}(\partial\phi)^2\sigma + \mu_\sigma \sigma \mathcal{H}^\dagger \mathcal{H}.
	\end{equation}
	Then below $m_\sigma$, we can integrate $\sigma$ out to write an effective coupling between the inflaton and Higgs,
	\begin{equation}
	\frac{\mu_\sigma}{\Lambda^\prime m_\sigma^2}(\partial\phi)^2\mathcal{H}^\dagger \mathcal{H} \equiv \frac{c_2}{\Lambda^2}(\partial\phi)^2 \mathcal{H}^\dagger \mathcal{H}.
	\end{equation}
	Now in the previous paragraphs we have stated that the choice of $\Lambda \gtrsim \sqrt{\dot{\phi}_0} $ and $c_i\sim {\cal O}(H/\sqrt{\dot{\phi}_0})$ will lead to a controlled effective theory expansion. These parameters are reproduced naturally if we take,
	$\mu_\sigma \sim H; \Lambda^\prime \sim V^{\frac{1}{4}}\sim \epsilon^{-1/4}\sqrt{\dot{\phi}_0}; m_\sigma \sim \sqrt{\dot{\phi}_0}$ in the theory containing the mediator. 
	In the above we have take $\epsilon \lesssim 10^{-2}$ consistent with current bounds \cite{Ade:2015lrj}.
	Note that we also induce the $(\partial \phi)^4$ operator, but with strength $\epsilon^{1/2}/\dot{\phi}_0^2$, so that the inflaton derivative expansion is controlled within the effective theory with $\sigma$ integrated out.

	This shows as a proof-of-principle how $\Lambda$ can represent the mass scale of heavy particles which are integrated out in the inflation-era effective theory, consistent with the even higher mass scale $V^{1/4}$ driving the accelerated expansion. It is possible that reheating later accesses this higher mass scale and produces $\sigma$ particles, but these rapidly decay and do not affect the NG signals derived in the effective theory with $\sigma$ integrated out.



	\subsection{Visibility of a Higgs Scalar}\label{higgsslowroll}
	
	Let us start with \eqref{lintinf} to discuss in detail the coupling between the inflaton and a Higgs scalar, $h$. As we briefly mentioned in the previous section, the first term of \eqref{lintinf} does not give rise to any quadratic mixing between the inflaton and $h$. This can seen by going to the unitary gauge for Higgs and using the equation of motion for the inflaton. So we move on to the second term of \eqref{lintinf},
	\begin{equation}\label{higgscoupling}
	\frac{c_2}{\Lambda^2}(\partial\phi)^2\mathcal{H}^\dagger \mathcal{H}.
	\end{equation}
	To be more precise, we have separated the strength of the interaction into a  dimensionless coupling $c_2$, and the physical cutoff $\Lambda$. 
	One can think of $c_2$ as being typical of the dimensionless strength of couplings between the gauge-Higgs sector and the inflaton sector. 
	Quantum loops are taken to be cutoff at 
	$\Lambda$. 
	
	The coupling \eqref{higgscoupling} gives rise to several terms, 
	\begin{multline}\label{higgsexpansion}
	\alpha \mathcal{H}^\dagger \mathcal{H}-\frac{\rho_2 v}{2} \dot{\xi}- \frac{\rho_2^2}{8\alpha}(\partial\xi)^2- \rho_2 \dot{\xi}h \\ +\frac{\alpha}{\dot{\phi}_0}\dot{\xi}h^2 + \frac{\rho_2}{2\dot{\phi}_0}(\partial\xi)^2h \\
	-\frac{\alpha}{2\dot{\phi}_0^2}(\partial\xi)^2h^2,
	\end{multline}
	where,
	\begin{equation}
	\rho_2 = \frac{2 c_2 v \dot{\phi}_0 }{\Lambda^2}; \hspace{2em} \alpha =- \frac{c_2 \dot{\phi}_0^2}{\Lambda^2}.
	\end{equation} 
	In \eqref{higgsexpansion} we have put in several terms that we dropped previously in \eqref{lintinfprime} for brevity. Also looking at  \eqref{lintinfprime} we see that we have dropped terms involving $\rho_5$ which are subleading compared to terms involving $\rho_2$. Coming back to the first line of \eqref{higgsexpansion}, we see that we have a quadratic mixing (denoted by $\rho_2$) between the inflaton and  $h$. We also have a term contributing to classical Higgs potential given by the parameter $\alpha$.
	Writing the scalar potential as
	\begin{equation}
	V(\mathcal{H}) = -\mu_h^2\mathcal{H}^\dagger \mathcal{H} + \lambda_h (\mathcal{H}^\dagger,\mathcal{H})^4 
	\end{equation}
	we see $\Delta\mu_h^2 = -\alpha$, so $\alpha$ should be thought of as a tuning parameter, which we ideally do not want to be much bigger than $H^2$ in order to avoid fine-tuning. Now we are in a position to summarize the different restrictions on the parameters.
	
	\paragraph{Classical Restrictions}
	We have several restrictions on the parameters $\rho_2$ and $\alpha$,
	\begin{itemize}
		\item We have a tadpole for $\dot{\xi}$, but as long as we have $\rho_2 v \ll \dot{\phi}_0$ it does not give dangerously large kinetic energy into the effective theory.
		\item To have perturbative control, we should require $\rho_2 < H$.
		\item To not have large modification to inflaton kinetic term we should have $\frac{\rho_2^2}{4\alpha}\ll1$.
		\item Also to have a controlled effective theory we should have $v\lesssim \Lambda$.
	\end{itemize}
	
	\paragraph{Quantum Corrections}
	\begin{itemize}
		\item From the quartic interaction between Higgs and inflaton we have, 
		\begin{equation}
		\Delta m_h^2 \sim \frac{1}{16\pi^2}c_2\Lambda^2 < m_{h}^2.
		\end{equation}
		\item From the same quartic interaction we have, 
		\begin{equation}
		\Delta\lambda_h\sim \frac{1}{16\pi^2}c_2^2	< \lambda_h.
		\end{equation}	
	\end{itemize}
	The inequalities above just impose the constraint of quantum stability, or absence of 
	loop-level fine tuning.
	
	\paragraph{Estimates for NG}
	
	A scalar particle can give rise to a nontrivial squeezed limit via three possible diagrams at tree level, as shown in Fig. \ref{fig:three-diagrams}. We can estimate the parametric strength of the associated NG quickly for each of the three diagrams using \eqref{higgsexpansion}:
	\begin{equation}\label{fnlhiggs}
	F^{\text{single}}_{h}\sim \rho_2^2; \hspace{2em} F^{\text{double}}_{h}\sim \rho_2^2 \alpha; \hspace{2em} F^{\text{triple}}_{h}\sim \rho_2^3 \lambda_h v \dot{\phi}_0 \sim \rho_2^2\alpha,
	\end{equation}
	where the right-hand sides are further modulated by  
	functions of kinematic shape as sketched in \eqref{fnl_qsfi}, and detailed later in Section \ref{hNG}. We have used the subscript $h$ to denote that we are estimating NG due to $h$. Importantly, when we do not have any classical tuning of the Higgs mass i.e. $\alpha \lesssim H^2$, all the diagrams give a similar contribution.
	
	\paragraph{Specific Parameter Choice}
	To have a feeling for all the above constraints and estimates we now focus on a benchmark parameter choice: $c_2=\frac{H}{\sqrt{\dot{\phi}_0}},\lambda_h=\frac{H^2}{2\dot{\phi}_0},\Lambda=3 \sqrt{\dot{\phi}_0}$, that we use later in Section \ref{hNG}. Such a choice gives depending on $m_h$, $F \sim {\cal O}(0.1)$, which should be observable. One can also check with this choice that all the above constraints are satisfied with  no fine-tuning of parameters. 
		
	
	\subsection{Visibility of a Massive Gauge Boson}
	
	Let us start with lagrangian \eqref{lintinf} again to get a coupling between the  inflaton and a massive gauge boson $Z$. To organize the couplings, it is useful to look for the essential quadratic mixing between the  inflaton and the timelike/longitudinal component of $Z$ first. 
	
	\paragraph{Quadratic Mixing} There is no such mixing when we consider terms up to dimension four. At dimension five, we get the term $\frac{c_1}{\Lambda}\partial_\mu\phi \mathcal{H}^\dagger D_\mu \mathcal{H}$ from \eqref{lintinf}. This term gives a coupling between $h$, the inflaton and $Z$. But after using the equation of motion for $Z$, we do not get the desired quadratic mixing. However, this term does give a mixing between $h$ and $Z^0$, where the superscript $0$ refers to the time component, not the charge of the $Z$ (which is always taken to be neutral for our purposes, as discussed earlier),
	\begin{equation}
	\frac{c_1}{\Lambda}\partial_\mu\phi \mathcal{H}^\dagger D^\mu \mathcal{H} \supset \rho_{1,Z} h Z^0,
	\end{equation}
	with, $\rho_{1,Z}=-\frac{ \text{Im}(c_{1}) \dot{\phi}_0 m_Z}{\Lambda}$. To look for quadratic mixing between the inflaton and $Z$ we have to go to yet higher order terms. The leading operators come at dimension nine due to shift symmetry of the inflaton couplings, 
	\begin{equation}
	\frac{c_5}{\Lambda^5}(\partial\phi)^2 \partial_\mu\phi \mathcal{H}^\dagger
	D^\mu \mathcal{H}.
	\end{equation} 
	This term gives a quadratic mixing both between $h$ and $Z^0$ and also between the inflaton and $Z^0$. The former is subleading compared to what we already have from the dimension five operator. So focusing on the latter, we have
	\begin{equation}
	\rho_{5,Z}\dot{\xi}Z^0,
	\end{equation}
	where, $\rho_{5,Z}=\frac{c_{5,I}\dot{\phi}_0^2m_Zv}{\Lambda^5}\sim\rho_{1,Z}\frac{ v \dot{\phi}_0}{\Lambda^4}$. In the last relation we have taken the EFT coefficients to be $\sim \frac{H}{\sqrt{\dot{\phi}}_0}$.
	
	\paragraph{Cubic Interactions} For brevity we will not write down all possible terms after expanding the lagrangian in unitary gauge. Rather we will focus on the terms that contribute to diagrams in Fig. \ref{fig:three-diagrams}. Focusing on cubic interactions between just the inflaton and $Z$ we do not get any contribution from the dimension five term, $\frac{c_1}{\Lambda}\partial_\mu\phi \mathcal{H}^\dagger D_\mu \mathcal{H}$. The leading operators then comes in at dimension eight,
	\begin{equation}
	\frac{c_3}{\Lambda^4}(\partial\phi)^2 |D\mathcal{H}|^2 \supset  -\frac{c_3m_Z^2 \dot{\phi}_0}{\Lambda^4}\dot{\xi}Z_\mu^2+\cdots ,
	\end{equation}
	and
	\begin{equation}
	\frac{c_4}{\Lambda^4}(\partial\phi)^2 Z_{\mu\nu}^2 \supset  -\frac{2c_4 \dot{\phi}_0}{\Lambda^4}\dot{\xi}Z_{\mu\nu}^2+\cdots .
	\end{equation}
	We also have another possible cubic interaction coming from the same dimension nine operator we considered above,
	\begin{equation}
	\frac{c_5}{\Lambda^5}(\partial\phi)^2 \partial_\mu\phi \mathcal{H}^\dagger D_\mu \mathcal{H} \supset -\frac{\rho_{5,Z}}{2\dot{\phi}_0}(\partial\xi)^2 Z^0 +\frac{\rho_{5,Z}}{\dot{\phi}_0}\dot{\xi}\partial_\mu\xi Z^\mu + \cdots .
	\end{equation} 
	Lastly we also have another dimension nine operator,
	\begin{equation}
	\frac{c_6}{\Lambda^5}\partial_\mu\phi \mathcal{H}^\dagger D^\mu \mathcal{H} |D\mathcal{H}|^2  \supset \frac{\rho_{5,Z} m_Z^2}{4 \dot{\phi}_0} Z^0 Z^\mu Z_\mu + \cdots,
	\end{equation}
	where, in the last relation we have taken $c_6\sim c_5$ just for simplicity.
	
	To summarize we collect the essential terms in the Lagrangian, 
	\begin{multline}
	\rho_{1,Z} h Z^0 + \rho_{5,Z}\dot{\xi}Z^0 -\frac{c_3m_Z^2 \dot{\phi}_0}{\Lambda^4}\dot{\xi}Z_\mu^2 -\frac{2c_4 \dot{\phi}_0}{\Lambda^4}\dot{\xi}Z_{\mu\nu}^2 -\frac{\rho_{5,Z}}{2\dot{\phi}_0}(\partial\xi)^2 Z^0 +\frac{\rho_{5,Z}}{\dot{\phi}_0}\dot{\xi}\partial_\mu\xi Z^\mu \\+ \frac{\rho_{5,Z} m_Z^2}{4 \dot{\phi}_0} Z^0 Z^\mu Z_\mu + \cdots,
	\end{multline}
	with, $\rho_{5,Z}=\frac{c_{5,I}\dot{\phi}_0^2m_Zv}{\Lambda^5}\sim\rho_{1,Z}\frac{ v \dot{\phi}_0}{\Lambda^4}$.
	
	\paragraph{Estimates of NG} Just like the case of scalars, we give the estimates for the three diagrams in Fig. \ref{fig:three-diagrams} (assuming $c_{i}\sim \frac{H}{\sqrt{\dot{\phi}_0}}$):
	\begin{equation}
	F^{\text{single}}_{Z}\sim \left(\frac{\rho_{1,Z} v \dot{\phi}_0}{\Lambda^4}\right)^2; \hspace{0.5em} F^{\text{double}}_{Z}\sim F^{\text{single}}_{Z} \times \frac{\rho_{1,Z} \dot{\phi}_0}{\Lambda^3}; \hspace{0.5em} F^{\text{triple}}_{Z}\sim\left(F^{\text{single}}_{Z}\right)^{3/2}\times \rho_{1,Z}\frac{ v \dot{\phi}_0}{\Lambda^4}
	\end{equation}
	We note for perturbativity, $\rho_{1,Z}<H^2$. Since $\Lambda> \sqrt{\dot{\phi}_0}$ and $v < \Lambda$, the single exchange diagram is expected to dominate over the other two. But even the single exchange diagram itself is too small to be observable. This is because it involves two suppressions, firstly due to the factor of $\rho_{1,Z}^2$. Secondly, there is a suppression, $\left(\frac{v \dot{\phi}_0 H}{\Lambda^4}\right)^2$ where $v, \sqrt{\dot{\phi}_0}< \Lambda$.
	Even assuming $\rho_{1,Z}\lesssim H^2$ and $v,\sqrt{\dot{\phi}_0}\lesssim \Lambda$ we have,  $F^{\text{single}}_{Z}\lesssim \frac{H^2}{\dot{\phi}_0}\lesssim 10^{-3}$, likely unobservably small. However, we will see in the next subsection that there is a loop-hole in this pessimistic conclusion.

	\subsection{Gauge Theory with a Heavy Higgs Scalar}\label{heavyhiggsz}
	
	In the above analysis we have seen that the couplings between just the inflaton and $Z$ are too small to give any observable NG. However there can be bigger effects for the $Z$ if the Higgs scalar $h$ becomes somewhat heavier than the Hubble scale, so that we can integrate it out. As an example, the inflaton can mix with the $Z$ via a virtual $h$ exchange to have a quadratic mixing of the form,
	$\frac{\rho_{1,Z}\rho_2}{m_h^2}\dot{\xi}Z^0$. 
	Similarly there can be  a cubic interaction between the inflaton and $Z$ of the form, $\frac{\rho_2 m_Z^2}{v m_h^2}\dot{\xi}Z_\mu^2$, and a similar term involving $Z$ field strength, which we do not write down for brevity. To summarize, we have the following interactions below the mass of $h$,
	\begin{equation}\label{heavyhiggs}
	\frac{\rho_{1,Z}\rho_2}{m_h^2}\dot{\xi}Z^0 + \frac{\rho_{1,Z}\rho_2}{m_h^2 \dot{\phi}_0}\dot{\xi}\partial_\mu\xi Z^\mu+\frac{\rho_2 m_Z^2}{v m_h^2}\dot{\xi}Z_\mu^2 +\cdots .
	\end{equation}
	\paragraph{Estimates of NG} Taking $\rho_{1,Z},\rho_2 \lesssim 1$ in Hubble units, we get the following estimates for NG,
	\begin{equation}
	F^{\text{single}}_{Z}\lesssim \frac{1}{m_h^4};\hspace{2em} F^{\text{double}}_{Z}\lesssim \frac{\dot{\phi}_0}{v m_h^6}.
	\end{equation}
	The triple exchange diagram, however, comes out to be smaller than the single exchange diagram. In this case it can be estimated as, $F^{\text{triple}}_{Z}\sim \frac{\dot{\phi}_0H^4}{m_h^6}\times\frac{H^3}{vm_h^2}$.
	
	Coming to the single and double exchange, as a benchmark choice of parameters which we use in Section \ref{sectiononZ}, we take $m_h= 3H$; $v= \sqrt{\dot{\phi}_0}; \Lambda = 3\sqrt{\dot{\phi}_0}$; $\text{Im} c_1= 6 H/\sqrt{\dot{\phi}_0}$; and $c_2 = \frac{9}{2} H/\sqrt{\dot{\phi}_0}$.
	Then we have, $F^{\text{single}}_{Z}\sim {\cal O}(0.01)$ and $F^{\text{double}}_{Z}\sim {\cal O}(0.1)$ depending on $m_h$.
	This scenario can thus lead to a very weak but probably observable NG due to $Z$. The ``price'' we pay is that the Higgs scalar $h$ which mediates the $Z$-inflaton interaction, with $m_h = 3 H$, will itself be too Boltzmann suppressed to observe in NG. Of course this does not preclude having observable $h$-like NG from other lighter Higgs scalars in multi-scalar Higgs theories. For the above parameter choice there is no classical or quantum tuning.

	
	~
	
	In conclusion, we see that in single-field slow-roll inflation, we can 
	get small but observable NG. 
	
	\section{NG in the Effective Goldstone Description of Inflationary Dynamics}\label{eft}
	
	The suppression of NG in single-field slow-roll inflation is due to the fact that the effective cutoff $\Lambda$ is constrained to be at least as large as the inflaton kinetic energy  scale, $\sqrt{\dot{\phi}_0} \gg H$, in order to perturbatively control the derivative expansion of the inflaton. However, it is possible that the dynamics takes some other form than standard single-field slow-roll inflation, and the cutoff of effective field theory, $\Lambda$, may then be lower, 
	yielding stronger NG. We now turn to a more ``agnostic'' approach to the inflationary dynamics so as to explore this possibility.  An elegant and powerful approach at this level is provided by the  Effective Goldstone description \cite{Cheung:2007st}. It is based on the central requirement that 
	the hot big bang has to emerge from the inflationary phase. In a relativistic theory this means that there must be a physical local ``clock'' field during inflation which dictates when inflation ends at each point in space. Such a clock field chooses a physical time coordinate during inflation and breaks the time diffeomorphism of dS spontaneously. The inflaton can then be thought of as the Goldstone boson associated with this spontaneous breaking. In this way, successful classical inflation is treated as a ``black-box'' input, 
	providing a background process in which the gauge-Higgs dynamics coupled to quantum inflaton fluctuations play out.
	
	We first review the construction of the EFT  of this Goldstone field in the absence of the Gauge-Higgs sector. Then we extend this construction to couple the Goldstone field to a Gauge-Higgs sector and estimate the magnitude of NG due to $h$ and $Z$ particles.
	
	\subsection{Minimal Goldstone Inflationary Dynamics}\label{minimalgoldstone}
	\subsubsection{Leading Terms in the Effective Theory and Power Spectrum}
	We start by writing in the effective lagrangian all the terms that are consistent with the unbroken 3D spatial diffeomorphisms on a fixed time slice. Such terms include 4D scalars and also any 3D diffeomorphism invariants made out of the following variables:
	\begin{equation*}
	t, g^{00}, g^{0\mu}V_\mu, K_{\mu\nu},
	\end{equation*}
	where, $V_\mu$ is any vector and $K_{\mu\nu}$ is the extrinsic curvature of the time slice. The above set of terms are allowed because they behave as scalars under spatial diffeomorphisms. This time slicing can be thought of as analogous to the unitary gauge of a spontaneously broken gauge theory in which the Goldstone boson is absent because it is ``eaten up'' by the gauge field i.e. the metric on the time slice. To restore the Goldstone boson, we do a transformation along the broken generator, which in the present context is a time translation,
	\begin{equation}
	t\rightarrow t+\pi(x),
	\end{equation} 
	and promote $\pi(x)$ to the quantum field denoting the Goldstone boson i.e the inflaton. Under such a time translation we record the transformation rules of the various terms mentioned above,
	\begin{equation}
	\begin{aligned}
	b(t)&\rightarrow b(t+\pi(x))=b(t)+\dot{b(t)}\pi(x)+\cdots \\
	g^{00}&\rightarrow \frac{\partial(t+\pi)}{\partial x^\mu}\frac{\partial(t+\pi)}{\partial x^\nu}g^{\mu\nu}\\
	g^{0\mu}V_\mu &\rightarrow \frac{\partial(t+\pi)}{\partial x^\nu}g^{\nu\mu}V_\mu.
	\end{aligned}
	\end{equation}
	
	The appearance of $\pi(x)$ in the specific combination $t+\pi(x)$ implies that we can restore 4D diffeomorphism by letting, $\pi(x)\rightarrow\pi(x)-\xi(t,\vec{x})$ under a time translation $t\rightarrow t+\xi(t,\vec{x})$ so that the combination $t+\pi(x)$ behaves as a scalar. In the above we have not written the transformation of the extrinsic curvature terms, because as we will show below, we will be interested in a regime where one can ignore terms involving extrinsic curvature. Let us now use the above strategy for restoring 4D diffeomorphism invariance to get the effective lagrangian for $\pi(x)$.
	
	We will expand around a background quasi-dS metric, 
	\begin{equation}
	ds^2=-dt^2+e^{2 H(t)}d\vec{x}^2.
	\end{equation}
	The leading effective lagrangian for small fluctuations around this metric and up to two-derivative order (neglecting extrinsic curvature as noted above) is then given by 
	\begin{equation}
	S =\int d^4x\sqrt{-g}\left( \frac{1}{2}M_\text{pl}^2 R - b(t) (g^{00} + 1) -\Lambda(t)\right).
	\end{equation} 
	The associated Einstein equations then look like
	\begin{equation}
	\begin{aligned}
	H^2 &=\frac{1}{3 M_\text{pl}^2}(\Lambda(t)+2b(t))\\
	\dot{H}+H^2 &=\frac{1}{3 M_\text{pl}^2}(\Lambda(t)-b(t)).
	\end{aligned}
	\end{equation}
	
	The above two equations fix the time-dependent couplings, $\Lambda(t)$ and $b(t)$, which when substituted back gives,
	\begin{equation}\label{Slin}
	S =\int d^4x\sqrt{-g}\left(\frac{1}{2}M_\text{pl}^2 R+M_\text{pl}^2\dot{H}g^{00}-(3M_\text{pl}^2H^2+M_\text{pl}^2\dot{H})\right).
	\end{equation}
	
	We can restore the Goldstone field in the above action by doing the time translation $t\rightarrow t+\pi$ under which,
	\begin{equation}
	g^{00}\rightarrow \frac{\partial(t+\pi)}{\partial x^\mu}\frac{\partial(t+\pi)}{\partial x^\nu}g^{\mu\nu} = (1+\dot{\pi})^2 g^{00}+2(1+\dot{\pi})\partial_i\pi g^{0i}+(\partial_i\pi)(\partial_j\pi)g^{ij}.
	\end{equation}
	The above transformation contains mixing of metric perturbations with the inflaton $\pi(x)$, but as we will justify soon, we can neglect such mixings. In that approximation, the transformation of $\delta g^{00}\equiv g^{00}+1$ and $g^{0\mu}V_\mu$ simplifies,
	\begin{equation}
	\begin{aligned}
	\delta g^{00} &\rightarrow -2\dot{\pi}+(\partial\pi)^2\\\label{metrictrans}
	g^{0\mu}V_\mu &\rightarrow  -(1+\dot{\pi})V_0+a^{-2}\partial_i\pi V_i.
	\end{aligned}
	\end{equation}
	
	Using  this, and working to leading order in $\epsilon \ll 1$, 
	the action (\ref{Slin}) reduces to\footnote{The term linear in $\dot{\pi}$ cancels with a similar term coming from the expansion $H(t+\pi)$ after an integration by parts. We have also dropped a few subleading terms coming from the expansion $M_{\text{pl}}^2\dot{H}(t+\pi)$ and $M_{\text{pl}}^2H^2(t+\pi)$}
	\begin{equation}\label{freegoldstone}
	S=\int d^4x\sqrt{-g}\left(\frac{1}{2}M_\text{pl}^2 R+M_\text{pl}^2\dot{H}(\partial\pi)^2-3M_\text{pl}^2H^2\right).
	\end{equation}
	This yields a quadratic action for the Goldstone boson. 
	
	To canonically normalize the Goldstone boson, we can define the field $\pi_c(x)$,  
	\begin{equation}\label{canonicalnorm}
	\pi_c(x)= \sqrt{2}M_\text{pl}(-\dot{H})^{\frac{1}{2}}\pi \equiv f_\pi^2 \pi,
	\end{equation} 
	where $f_\pi$ itself is approximately constant in time as 
	$\epsilon, \eta \ll 1$. 
	This definition of $f_\pi$ generalizes the ``decay constant'' of chiral lagrangians for Goldstone bosons of internal symmetries. In single-field inflation it is simply given by  $f_\pi^4=\dot{\phi}_0^2=2\epsilon H^2 M_{\text{pl}}^2=-2\dot{H}M_{\text{pl}}^2$, but here we are not assuming single-field inflationary dynamics, and indeed our effective lagrangian makes no explicit reference to $\phi$. 
	The last term in the Goldstone action \eqref{freegoldstone} act as the dominant energy density during inflation,
	\begin{equation}
	\rho=3M_\text{pl}^2H^2,
	\end{equation}
	which is the familiar relation. 
	
	Before calculating physical quantities we have to relate $\pi(x)$ to the gauge invariant quantity $\mathcal{R}$. This can be done by first noticing that in the absence of $\pi(x)$, i.e. in the unitary gauge, the spatial metric is the same as \eqref{comovinggauge},
	\begin{equation}\label{met1}
	h_{ij} = a^2(t)\left((1+2\mathcal{R})\delta_{ij}+\gamma_{ij}\right).
	\end{equation}
	To introduce $\pi(x)$ we again do the transformation $t\rightarrow t+\pi$. To relate $\pi$ to $\mathcal{R}$ we demand that in presence of $\pi$ the spatial metric should not contain any 3D-scalar metric fluctuations, so that it should be given by
	\begin{equation}\label{met2}
	h_{ij} = a^2(t)\left(\delta_{ij}+\gamma_{ij}\right).
	\end{equation}
	This gives the leading order relation,
	\begin{equation}\label{rwithpi}
	\mathcal{R}=-\pi.
	\end{equation}
	
	Using \eqref{freegoldstone}, \eqref{canonicalnorm} and \eqref{rwithpi} we can calculate the inflaton power spectrum,  
	\begin{align}
	\langle\pi_c (\vec{k})\pi_c(-\vec{k})\rangle^\prime=\frac{1}{2k^3}\qquad
	\Rightarrow \langle\mathcal{R}(\vec{k})\mathcal{R}(-\vec{k})\rangle^\prime=\frac{1}{2f_\pi^4 k^3},
	\end{align}
	which matches the single-field slow-roll calculation \eqref{power} \footnote{This is assuming that subsequent terms in the EFT do not contribute significantly to the quadratic lagrangian for the inflaton. If that is not the case, then the power spectrum depends on a combination of the scales $f_\pi$ and $c_s$, the speed of propagation for the inflaton fluctuation}, but now more agnostically with regard to the inflationary dynamics.
	
	Before moving on to inflaton interaction terms, we pause to justify why we have ignored terms involving extrinsic curvature and mixing of the inflaton with metric perturbations.
	The transformation of $K_{\mu\nu}$ can be obtained by writing it down in terms of the induced metric $h_{\mu\nu}$ on the time slice, and the normal vector $n^\mu$ to the surface \cite{Wald},
	\begin{equation}
	K_{\mu\nu}=\frac{1}{2}\big( n^\sigma\nabla_\sigma h_{\mu\nu}+h_{\mu\sigma}\nabla_\nu n^\sigma + h_{\sigma\nu}\nabla_\mu n^\sigma\big).
	\end{equation} 
	We note that being a type of curvature, $K_{\mu\nu}$ always has an extra derivative acting on the metric component. This means compared to terms like $\delta g^{00}$, scalars like $(\delta K_{\mu\nu})^2$ or $(\delta K^\mu_\mu)^2$ will have an extra $\frac{E^2}{\Lambda^2}$ suppression\footnote{$K^\mu_\mu$ is suppressed only by $\frac{E}{\Lambda}$, however a term involving $K^\mu_\mu$ can be reduced to a term containing $g^{00}$, and thus gives no new information \cite{Cheung:2007st}} where $\Lambda$ is the cutoff of the effective theory presumed to suppress these higher-derivative terms. Since $E\sim H$ and we will be considering situations with $\Lambda \gtrsim 10 H$, we can ignore contributions coming from extrinsic curvature.
	
	We now turn to the mixing of the inflaton with metric fluctuations. 
	From the transformation of $g^{00}$ we get a term of the form, $h^{00}\dot{\pi}$ along with $\dot{\pi}^2$. However from Einstein eqs. \cite{Maldacena:2002vr},  $h^{00}\sim\sqrt{\epsilon}\frac{H}{M_{\text{pl}}}$, which means the mixing term is suppressed compared to $\dot{\pi}^2$ by a factor of $\epsilon$. Since we are interested in the regime, $E\sim H$ and $\epsilon\ll 1$,  we can drop such mixing terms, which simplifies the transformation laws as advertised earlier,
	\begin{equation}\label{deltag00}
	\begin{aligned}
	\delta g^{00}\rightarrow & -2\dot{\pi}+(\partial\pi)^2 \\
	g^{0\mu}V_\mu\rightarrow & +(1+\dot{\pi})V^0+\partial_i\pi V^i.
	\end{aligned}
	\end{equation}

	\subsubsection{Higher Order Terms}
	Let us now move on to discuss higher order corrections to the quadratic-in-$\pi$ lagrangian discussed above. Ignoring terms involving extrinsic curvature, we have terms of the form $M_n^4(\delta g^{00})^n$ where $M_n$'s are some mass scales. In particular we have for $n=2$,
	\begin{equation}
	M_2^4(\delta g^{00})^2 \supset 4M_2^4\dot{\pi}^2 + \cdots
	\end{equation}
	This term modifies the kinetic term for the inflaton. However such modifications are small when $M_2^4\lesssim f_\pi^4$ which we will take to be the case.\footnote{It can happen that $M_2^4\gg f_\pi^4$ in which limit the inflaton fluctuations propagate with a speed $c_s\ll 1$. While we will restrict to cases with $c_s\approx 1$, our analysis can be easily extended to include $c_s\ll 1$.}
	Then we can simplify the transformation of $\delta g^{00}$ even further by noting that $(\partial\pi)^2\sim \left(\frac{\partial\pi_c}{f_\pi^2}\right)^2\sim \frac{H^4}{f_\pi^4}\ll \dot{\pi}\sim\frac{\dot{\pi}_c}{f_\pi^2}\sim \frac{H^2}{f_\pi^2}$, where we have used \eqref{canonicalnorm} and the fact that $\pi_c\sim H$. Thus we can write,
	\begin{equation}
	\delta g^{00}\rightarrow  -\frac{2\dot{\pi}_c}{f_\pi^2}.
	\end{equation}
	
	Since $\delta g^{00}$ is dimensionless, a power counting rule in the EFT is not manifest. However this can be fixed by defining the dimension 2 operator $\delta g^{00}_{c}$ which transforms as,
	\begin{equation}\label{g00can}
	\delta g^{00}_{c} \equiv -\frac{1}{2}f_\pi^2 \delta g^{00} \rightarrow  \dot{\pi}_c.
	\end{equation}
	We will take a power-counting rule that higher-dimensional operators in terms of $\pi_c$ are suppressed by powers of $\Lambda$, with order one coefficients. Let us illustrate this power counting rule by the example of the dimension six operator arising from $M_3^4(\delta g^{00})^3$.
	By our power counting rule we expect this term to go as
	\begin{equation}
	M_3^4(\delta g^{00})^3\sim\frac{\bar{d}_1}{\Lambda^2}(\delta g^{00}_{c})^3 \rightarrow \frac{\bar{d}_1}{\Lambda^2}\dot{\pi}_c^3,
	\end{equation}
	where $\bar{d}_1$ is an $\mathcal{O}(1)$ EFT coefficient.
	At higher orders we have an expansion like
	\begin{equation}
	\bar{d}_2\frac{(\delta g^{00}_{c})^4}{\Lambda^4} + \bar{d}_3\frac{(\delta g^{00}_{c})^5}{\Lambda^6} +
	\cdots .
	\end{equation}
	
	Importantly, non-observation of NG in Planck data puts a bound on the cutoff $\Lambda$ of the EFT.  For example, the dimension-6 operator $\frac{\bar{d}_1}{\Lambda^2}\dot{\pi}_c^3$ that we discussed above induces an inflaton three-point function of the form,
	\begin{eqnarray}\label{pidotcubed}
	F^{\dot{\pi}^3}(k_1,k_2,k_3) = -2\bar{d}_1 \frac{ f_\pi^2}{\Lambda^2}\left(\frac{k_1^3 k_3^3}{k_1k_2k_3 (k_1+k_2+k_3)^3}\right),
	\end{eqnarray}
	from which we can calculate $f_{\text{NL}}^{\dot{\pi}^3}=-\frac{5\bar{d}_1}{243}\frac{f_\pi^2}{\Lambda^2}$ (as defined in  \eqref{fnl}). From the Planck bound \cite{Ade:2015ava} $f_{\text{NL}}^{\text{equil}}=-4\pm 43$ we get the mild constraint $\Lambda > H$, where we have assumed $\bar{d}_1\sim 1$.\footnote{Of course the Planck analysis did not use exactly the shape of NG in \eqref{pidotcubed}. However, the equilateral template \cite{Creminelli:2005hu} they did use is ``close'' enough to \eqref{pidotcubed}, as measured by the standard
		``cosine'' parameter \cite{Babich:2004gb}.}
	

	\subsection{Incorporating Gauge-Higgs Theory into the Goldstone Effective Description}
	
	We now couple the EFT to a gauge-Higgs theory and focus on inflaton-$h$ and inflaton-$Z$ couplings in turns, and discuss the estimates of NG.

	\subsubsection{Visibility of a Higgs Scalar}
	When we include the Gauge-Higgs theory in the EFT we encounter a new dimension-3 operator that we can write down on the fixed time slice,
	\begin{equation}
	\lambda_1\mathcal{H}^\dagger D^0 \mathcal{H}.
	\end{equation}
	After introducing $\pi(x)$ this gives rise to
	\begin{equation}
	\mathcal{H}^\dagger D^0 \mathcal{H} \rightarrow \mathcal{H}^\dagger D^0 \mathcal{H} + \frac{1}{f_\pi^2}\partial_\mu\pi_c \mathcal{H}^\dagger D^\mu \mathcal{H}.
	\end{equation}
	The first term on the RHS gives, apart from some tadpoles which are safe in the sense discussed in subsection \ref{higgsslowroll}, a modification to the Higgs quadratic term $h^2$, a quadratic mixing between $h$ and $Z^0$ and a cubic interaction between $h$ and $Z^0$. The second term also gives a cubic interaction between inflaton, $h$ and $Z^0$. However, it does not couple inflaton to $h$ alone, as can be seen by using the equation of motion for the inflaton.
	
	Thus we consider next the marginal (in terms of the canonical inflaton field as in \eqref{g00can}) operator $\lambda_2 \delta g^{00}_c \mathcal{H}^\dagger \mathcal{H}$, which upon introducing $\pi(x)$ gives
	\begin{equation}
	\lambda_2 \delta g^{00}_c \mathcal{H}^\dagger \mathcal{H} \rightarrow \frac{1}{2}\lambda_2 \dot{\pi}_c (v^2+2hv+h^2).
	\end{equation}
	For $v\sim H$ and $\lambda_2<1$, the inflaton tadpole above is safe again in the same sense as discussed in subsection \ref{higgsslowroll}. 
	
	At subsequent orders we have,
	\begin{equation}
	\begin{aligned}
	\frac{d_1}{\Lambda}\delta g^{00}_c \mathcal{H}^\dagger D^0 \mathcal{H} &\rightarrow -\frac{\text{Re}(d_{1})v}{2\Lambda}\dot{\pi}_c\dot{h}-\frac{\text{Re}(d_{1})}{2\Lambda}\dot{\pi}_ch\dot{h}+\frac{\text{Re}(d_{1})v}{2\Lambda f_\pi^2} \dot{\pi}_c\partial_\mu\pi_c\partial^\mu h+\cdots,\\
	\frac{d_2}{\Lambda^2}(\delta g^{00}_c)^2\mathcal{H}^\dagger \mathcal{H} &\rightarrow \frac{d_2 v^2}{2\Lambda^2}\dot{\pi}_c^2 + \frac{d_2 v}{\Lambda^2}\dot{\pi}_c^2h+\cdots,\\
	\frac{d_3}{\Lambda^2}\delta g^{00}_c|D\mathcal{H}|^2 &\rightarrow \frac{d_3}{2\Lambda^2}\dot{\pi}_c(\partial h)^2+\cdots.
	\end{aligned}
	\end{equation}
	
	In the following, just for technical simplicity, we will use a set of benchmark values such that the inflaton-$h$ quadratic mixing is predominantly given by $\lambda_2$ instead of $\text{Re}(d_{1})$. Then, the leading operators for Higgs-inflaton interactions have the form, 
	\begin{equation}\label{higgscouplingeft}
	\lambda_2 v \dot{\pi}_c h + \frac{1}{2}\lambda_2\dot{\pi}_ch^2 + \frac{d_2 v}{\Lambda^2} \dot{\pi}_c^2 h+\cdots.
	\end{equation}
	
	\paragraph{Estimates of NG}
	As before we can get quick estimates for NG given by the three diagrams shown in Fig. \ref{fig:three-diagrams} (assuming $d_2\sim 1$):
	\begin{equation}\label{fnlhiggs2}
	F^{\text{single}}_{h}\sim \frac{\lambda_2 v^2 f_\pi^2}{\Lambda^2}; \hspace{2em} F^{\text{double}}_{h}\sim \lambda_2^3 v^2 f_\pi^2; \hspace{2em} F^{\text{triple}}_{h}\sim \lambda_h v \lambda_2^3 v^3 f_\pi^2 \sim \lambda_2^3 v^2 f_\pi^2.
	\end{equation}
	We see for a sample choice of parameters, $\lambda_2\lesssim 1$, $\Lambda \lesssim 10H$ and $v\sim H$, we can easily achieve a promising $F_h\sim \mathcal{O}(1)$. Furthermore, with the above choices loop corrections are small.
	
	\subsubsection{Visibility of a Massive Gauge Boson}
	For $Z$ we do not have any relevant or marginal pure inflaton-$Z$ interaction. Inflaton-$Z$ interactions coming from the term $\mathcal{H}^\dagger D^0\mathcal{H}$ after restoring $\pi(x)$ vanish by equations of motion. So the leading inflaton-$Z$ coupling is given by
	\begin{equation}
	\frac{d_1}{\Lambda}\delta g^{00}_c \mathcal{H}^\dagger D^0 \mathcal{H} \rightarrow -\frac{\text{Im}(d_{1})m_Z v}{2\Lambda}\dot{\pi}_c  Z^0- \frac{\text{Im}(d_{1})m_Z v}{2\Lambda f_\pi^2}\dot{\pi}_c\partial_\mu\pi_c  Z^\mu\cdots,
	\end{equation}
	which gives a quadratic mixing between inflaton and $Z$. At dimension 6 we have the operators,
	\begin{equation}
	\begin{aligned}
	\frac{d_3}{\Lambda^2}\delta g^{00}_c |D\mathcal{H}|^2 &\rightarrow \frac{d_3m_Z^2}{2\Lambda^2} \dot{\pi}_c Z_{\mu}^2+\cdots,\\
	\frac{d_4}{\Lambda^2}\delta g^{00}_c Z_{\mu\nu}^2 &\rightarrow \frac{d_4}{\Lambda^2} \dot{\pi}_c Z_{\mu\nu}^2+\cdots,
	\end{aligned}
	\end{equation}
	where $Z_{\mu\nu}$ is the $Z$ field strength.
	
	We can summarize the inflaton-$Z$ interaction as
	\begin{equation}\label{inf-Z-goldstone}
	-\frac{\text{Im}(d_{1})m_Z v}{2\Lambda}\dot{\pi}_c  Z^0- \frac{\text{Im}(d_{1})m_Z v}{2\Lambda f_\pi^2}\dot{\pi}_c\partial_\mu\pi_c  Z^\mu
	+ \frac{d_3m_Z^2}{2\Lambda^2} \dot{\pi}_c Z_{\mu}^2+\frac{d_4}{\Lambda^2} \dot{\pi}_c Z_{\mu\nu}^2+\cdots.
	\end{equation}

	\paragraph{Estimates of NG}
	The estimates for the single and the double exchange diagram in Fig. \ref{fig:three-diagrams} are
	\begin{equation}
	F^{\text{single}}_{Z}\sim  \frac{v^2}{\Lambda^2}; \hspace{2em}
	F^{\text{double}}_{Z}\sim  \frac{v^2f_\pi^2}{\Lambda^4}.
	\end{equation}
	We see for the choice $\Lambda\sim 10 H$, $v \sim H$, the double exchange contribution dominates over the single exchange and can give $F_Z\sim {\cal O}(0.1)$. For $\Lambda\sim 5 H$, $v \sim H$, $F_Z\sim {\cal O}(1)$.
		As with the case of single field slow roll, the triple exchange diagram comes out to be smaller than the single exchange diagram. It can be estimated as, $F^{\text{triple}}_{Z}\sim \frac{v^3}{\Lambda^3}\frac{f_\pi^2}{\Lambda^3}$.
	
	To summarize, we have demonstrated a controlled EFT with $\Lambda\sim 5 - 10H$ can give rise to observable NG due to $h$ and $Z$ particles. Also, this scenario does not suffer from large destabilizing quantum corrections. In the context of slow roll inflation we saw that to see NG due to $Z$ we had to consider a considerably heavier associated physical Higgs $h$, which is itself too Boltzmann-suppressed to see in NG. However in the context of the more general Goldstone description this is not necessary. That is, in the Goldstone description it is possible to see NG for both a $Z$ and its associated $h$, while in single-field inflation the associated $h$ would be too Boltzmann-suppressed to be visible. This therefore allows us to more thoroughly verify the heavy-lifting mechanism for a greater part of the gauge-Higgs spectrum.
	
	\section{Detailed Form of NG Mediated by $h$}\label{hNG}
	
	In the previous sections we have given only crude estimates for NG due to $h$ and $Z$. In this section, we derive the detailed expressions for $F(k_1,k_2,k_3)$ \eqref{Ffunction}. We will consider the case of single-field slow-roll inflation as well as the more general  effective Goldstone description of inflation.
	
	We begin by first considering the general Goldstone description of inflation. 
	Then, the Higgs-inflaton couplings from the previous section are (taking EFT coefficient $d_2= 1$)
	\begin{equation}\label{LHiggs}
	\lambda_2 v \dot{\pi}_c h + \frac{1}{2}\lambda_2\dot{\pi}_ch^2 + \frac{ v \dot{\pi}_c^2 h }{\Lambda^2} ,
	\end{equation}
	which gives rise to single, double and triple exchange diagrams as shown in Fig. \ref{fig:three-diagrams}. A similar single exchange diagram and an identical triple exchange diagram have been calculated in \cite{Arkani-Hamed:2015bza} and \cite{Chen:2009zp} respectively. Thus here we focus on calculating the double exchange diagram using the mixed propagator formalism developed in \cite{Chen:2017ryl}. We also modify the existing calculation of the single exchange diagram for our particular case. 
	
	In the squeezed limit, $F(k_1,k_2,k_3)$ is only a function of $\frac{k_3}{k_1}$ and has the form 
	\begin{equation}\label{Fquasisingle}
	F = f(\mu)\left(\frac{k_3}{k_1}\right)^{\frac{3}{2}+i\mu} + f(\mu)^*\left(\frac{k_3}{k_1}\right)^{\frac{3}{2}-i\mu}.
	\end{equation}
	Now we give the detailed expressions for $f(\mu)$ for different diagrams, leaving the details of the calculation for appendix \ref{appscalarng}.
	
	\subsection{Single Exchange Diagram}\label{higgscalc}
	
	As derived in appendix \ref{appscalarng} in \eqref{singleheft},
	\begin{multline}
	F^{\text{single}}_{h} = - \frac{1}{8}\times\lambda_2\left(\frac{v f_\pi}{\Lambda}\right)^2\times\\
	\left(\Gamma\left(\frac{1}{2}+i\mu\right)^2 \Gamma\left(-2i\mu\right)\left(\frac{1}{2}+i\mu\right)\left(\frac{3}{2}+i\mu\right)(1+i\sinh(\pi\mu))\left(\frac{k_3}{k_1}\right)^{\frac{3}{2}+i\mu}+(\mu\rightarrow-\mu)\right).
	\end{multline}
	The strength of the NG can be characterized by recasting the above equation as \eqref{Fquasisingle} and evaluating the quantity $\frac{5}{18}|f(\mu)|$ to conform with \eqref{fnl}. We denote the resulting strength by $|f^{\text{single}}_{h}|$ and it is sampled below for various masses for the benchmark values, $\lambda_2=0.2;\lambda_h=0.5;\Lambda=8H$ :
	\begin{center}
		\begin{tabular}{|r|l|}
			\hline
			mass & $|f^{\text{single}}_{h}|$ \\
			\hline \hline
			1.6 H & 1.453 \\
			1.9 H & 0.420 \\
			2.2 H & 0.183 \\
			\hline
		\end{tabular}
	\end{center}
	Of course, for $m \gg H$, the NG become Boltzmann suppressed and unobservable.

\begin{figure}[h]
		\centering
		\includegraphics[width=0.6\linewidth]{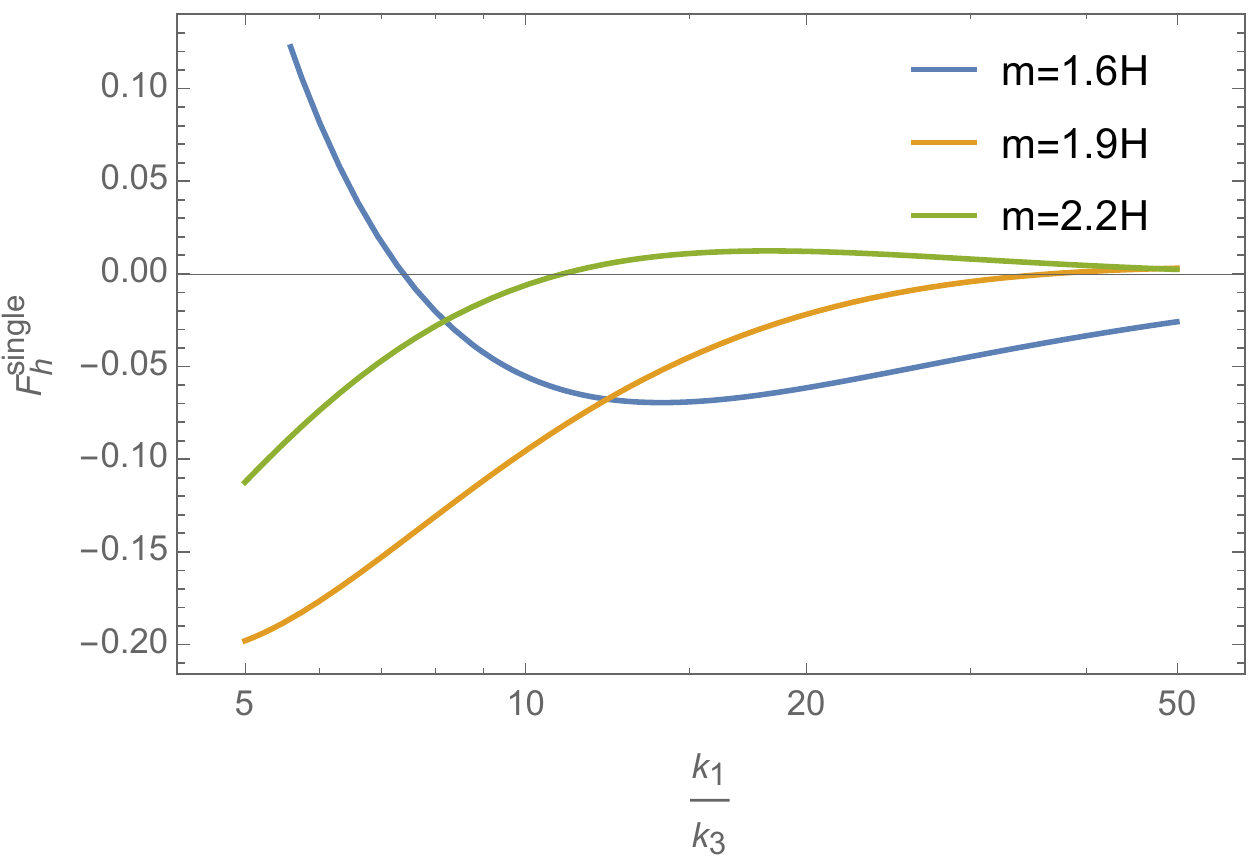}
		\caption{Dimensionless three-point function $F_h^{\text{single}}$ \eqref{Ffunction} for different masses in Goldstone Effective description \eqref{singleheft} with $\lambda_2=0.2;\lambda_h=0.5;\Lambda=8H$.}
		\label{fig:singleexchangescalarfinalgoldstone1}
	\end{figure}

\begin{figure}[h]
		\centering
		\includegraphics[width=0.6\linewidth]{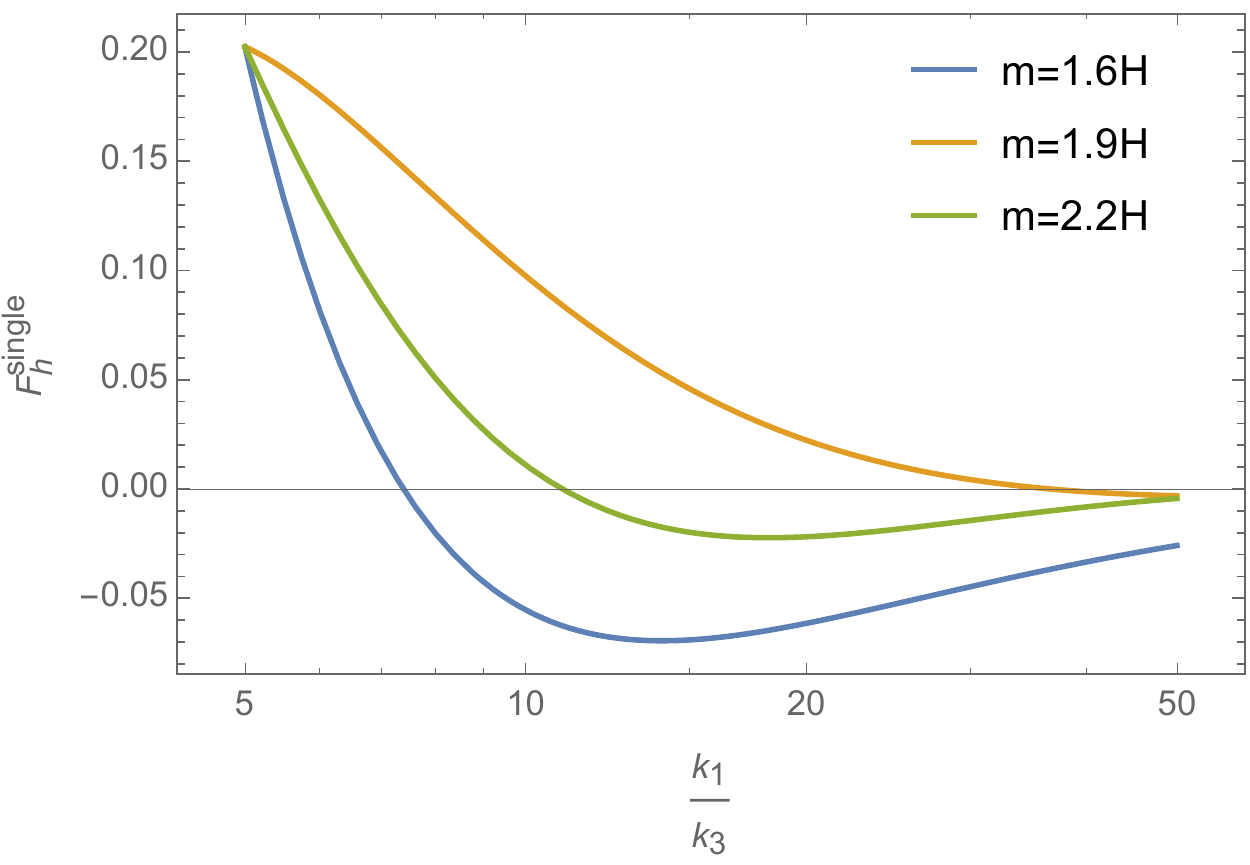}
		\caption{Shape sensitivity of $F_h^{\text{single}}$ to $m_h$. We have chosen three plausible sets of parameters for which $F_h^{\text{single}}$ agree at the fiducial ratio $\frac{k_1}{k_3}=5$. This illustrates our ability to discriminate among different masses.}
		\label{fig:singleexchangescalarfinalgoldstone2}
	\end{figure}

	We see that we can generically have $|f^{\text{single}}_{h}| \sim 0.1-1$, which can be accessible. Coming to the shape of NG, as we have mentioned before, $\frac{k_3}{k_1}$ dependence of $F_h$ encodes the mass information of $h$, and to verify the heavy-lifting mechanism it is crucial to determine the mass with reasonable precision. In \cite{Meerburg:2016zdz} such an analysis was done in the context of 21-cm cosmology. From their estimates, we see for $|f^{\text{single}}_{h}| >0.1$ we should be able to determine the mass at 10 percent level or better. 
	We illustrate our results in Figs. \ref{fig:singleexchangescalarfinalgoldstone1} and \ref{fig:singleexchangescalarfinalgoldstone2}. 
	

\begin{figure}[h]
		\centering
		\includegraphics[width=0.6\linewidth]{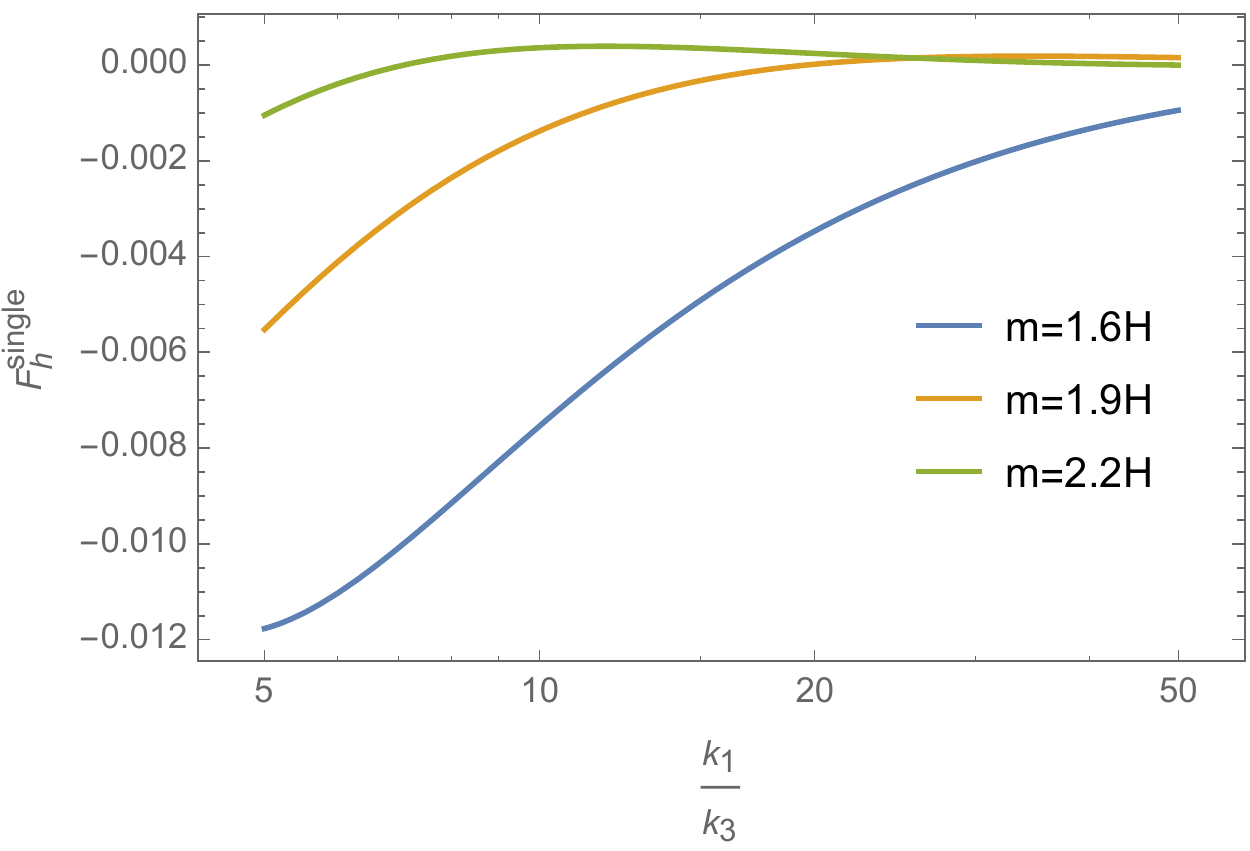}
		\caption{Dimensionless three-point function $F_h^{\text{single}}$ \eqref{Ffunction} for different masses in Single-field Slow-roll description \eqref{singlehsr} with $c_2=\frac{H}{\sqrt{\dot{\phi}_0}},\lambda_h=\frac{H^2}{2\dot{\phi}_0},\Lambda=3 \sqrt{\dot{\phi}_0}$.}
		\label{fig:singleexchangescalarfinalslow1}
	\end{figure}

\begin{figure}[h]
		\centering
		\includegraphics[width=0.6\linewidth]{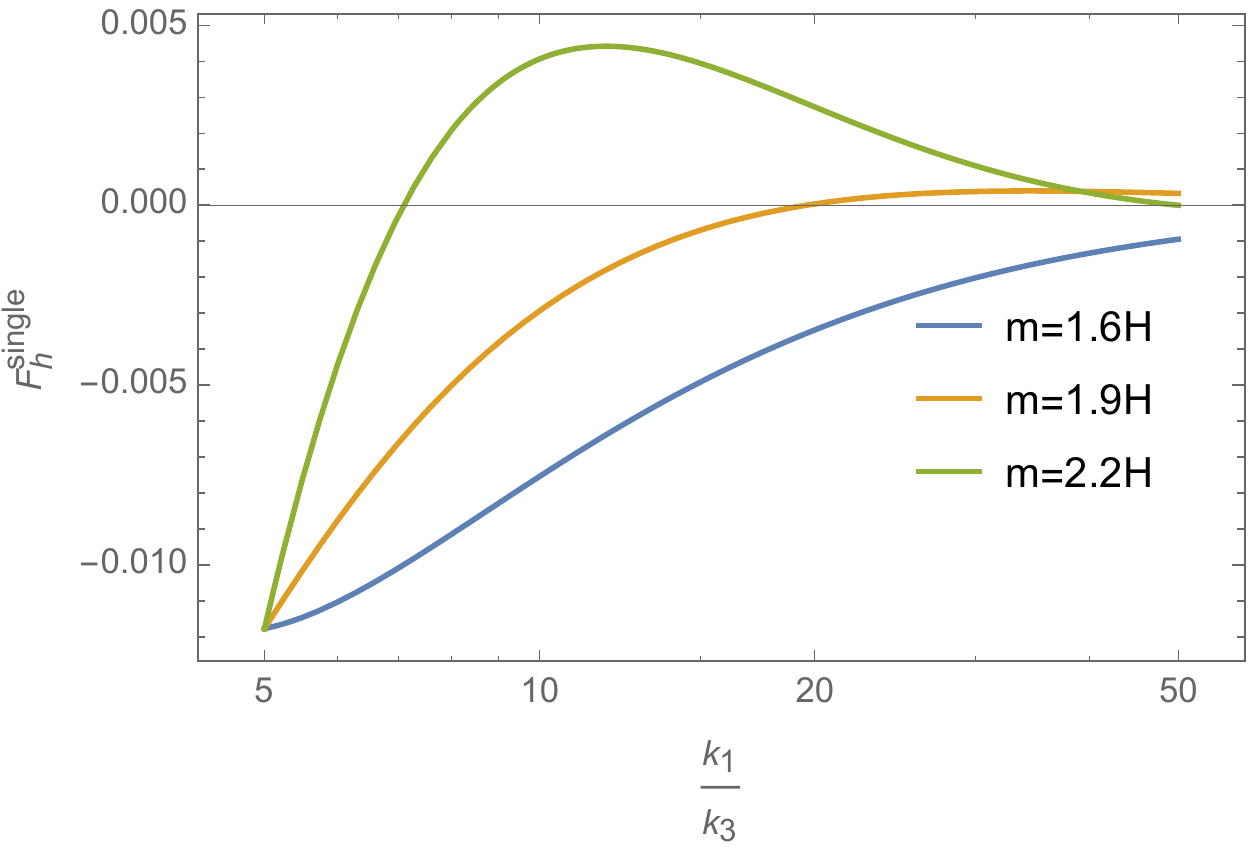}
		\caption{Shape sensitivity of $F_h^{\text{single}}$ to $m_h$. We have chosen three plausible sets of parameters for which $F_h^{\text{single}}$ agree at the fiducial ratio $\frac{k_1}{k_3}=5$. This illustrates our ability to discriminate among different masses.}
		\label{fig:singleexchangescalarfinalslow2}
	\end{figure}

	In the special case of single-field slow-roll inflation, the lagrangian reads from lagrangian \eqref{higgsexpansion},
	\begin{equation}
	- \rho_2 \dot{\xi}h +\frac{\alpha}{\dot{\phi}_0}\dot{\xi}h^2 + \frac{\rho_2}{2\dot{\phi}_0}(\partial\xi)^2h 
	\end{equation} 
	From the above by similar methods we find from \eqref{singlehsr},
	\begin{multline}
	F^{\text{single}}_{h} = - \frac{1}{4}\times\rho_2^2\\
	\left(\Gamma\left(\frac{1}{2}+i\mu\right)^2 \Gamma\left(-2i\mu\right)\left(\frac{3}{2}+i\mu\right)\left(\frac{5}{2}+i\mu\right)(1+i\sinh(\pi\mu))\left(\frac{k_3}{k_1}\right)^{\frac{3}{2}+i\mu}+(\mu\rightarrow-\mu)\right).
	\end{multline}
	where, $\rho_2=\frac{2c_2v\dot{\phi}_0}{\Lambda^2}$.
	Now, we can again evaluate $|f^{\text{single}}_{h}|$ for some benchmark values, $c_2=\frac{H}{\sqrt{\dot{\phi}_0}},\lambda_h=\frac{H^2}{2\dot{\phi}_0},\Lambda=3 \sqrt{\dot{\phi}_0}$:
	\begin{center}
		\begin{tabular}{|r|l|}
			\hline
			mass & $|f^{\text{single}}_{h}|$ \\
			\hline \hline
			1.6 H & 0.047 \\
			1.9 H & 0.008 \\
			2.2 H & 0.003 \\
			\hline
		\end{tabular}
	\end{center}
	The above parameter choice implies classical Higgs mass tuning at the 25 percent level, and there are no large quantum corrections.
	For the function $F^{\text{single}}_{h}$ we illustrate our results in Figs. \ref{fig:singleexchangescalarfinalslow1} and \ref{fig:singleexchangescalarfinalslow2}.
	
	\subsection{Double Exchange Diagram}
\begin{figure}[h]
		\centering
		\includegraphics[width=0.6\linewidth]{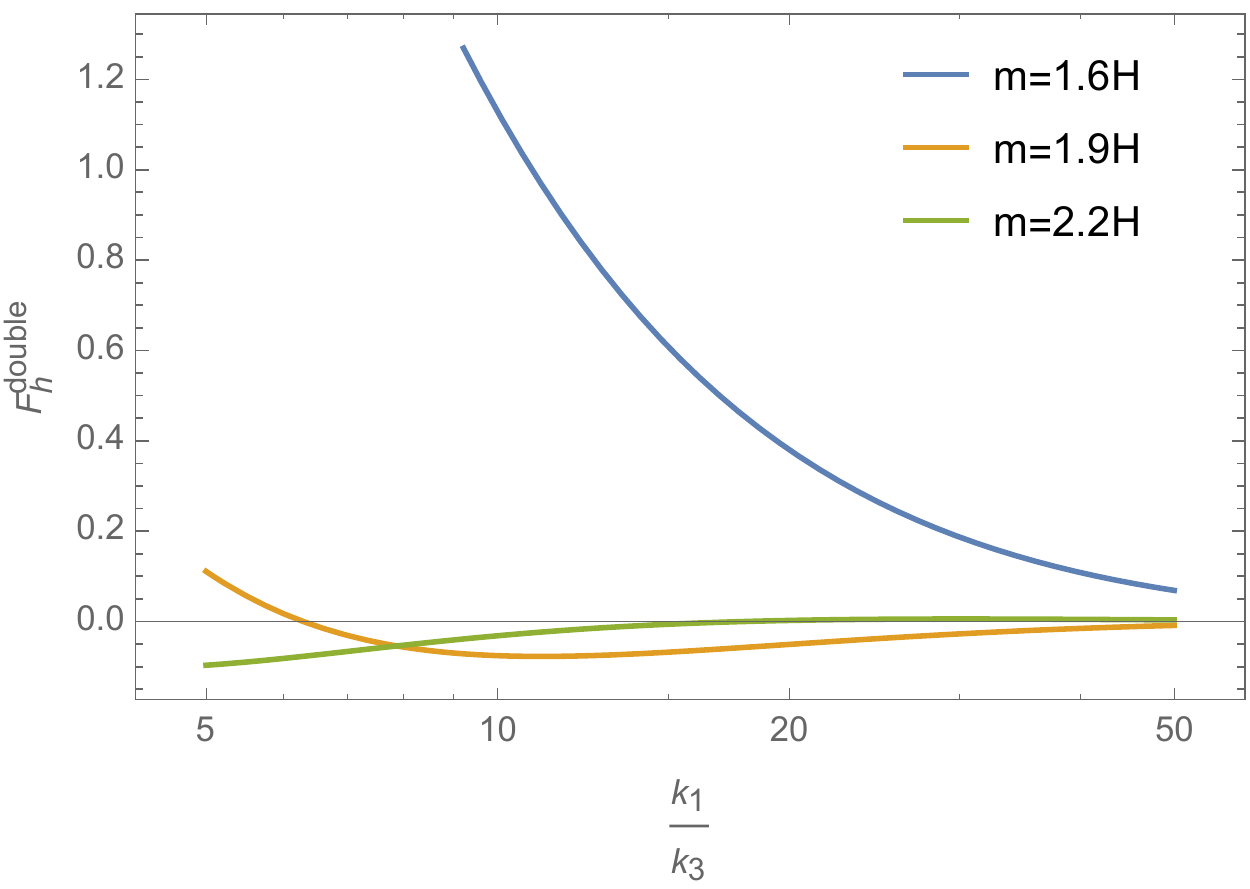}
		\caption{Dimensionless three-point function $F_h^{\text{double}}$ \eqref{Ffunction} for different masses in Goldstone Effective description \eqref{doubleheft} with $\lambda_2=0.2;\lambda_h=0.5$.}
		\label{fig:doubleexchangescalarfinalgoldstone1}
	\end{figure}

\begin{figure}[h]
		\centering
		\includegraphics[width=0.6\linewidth]{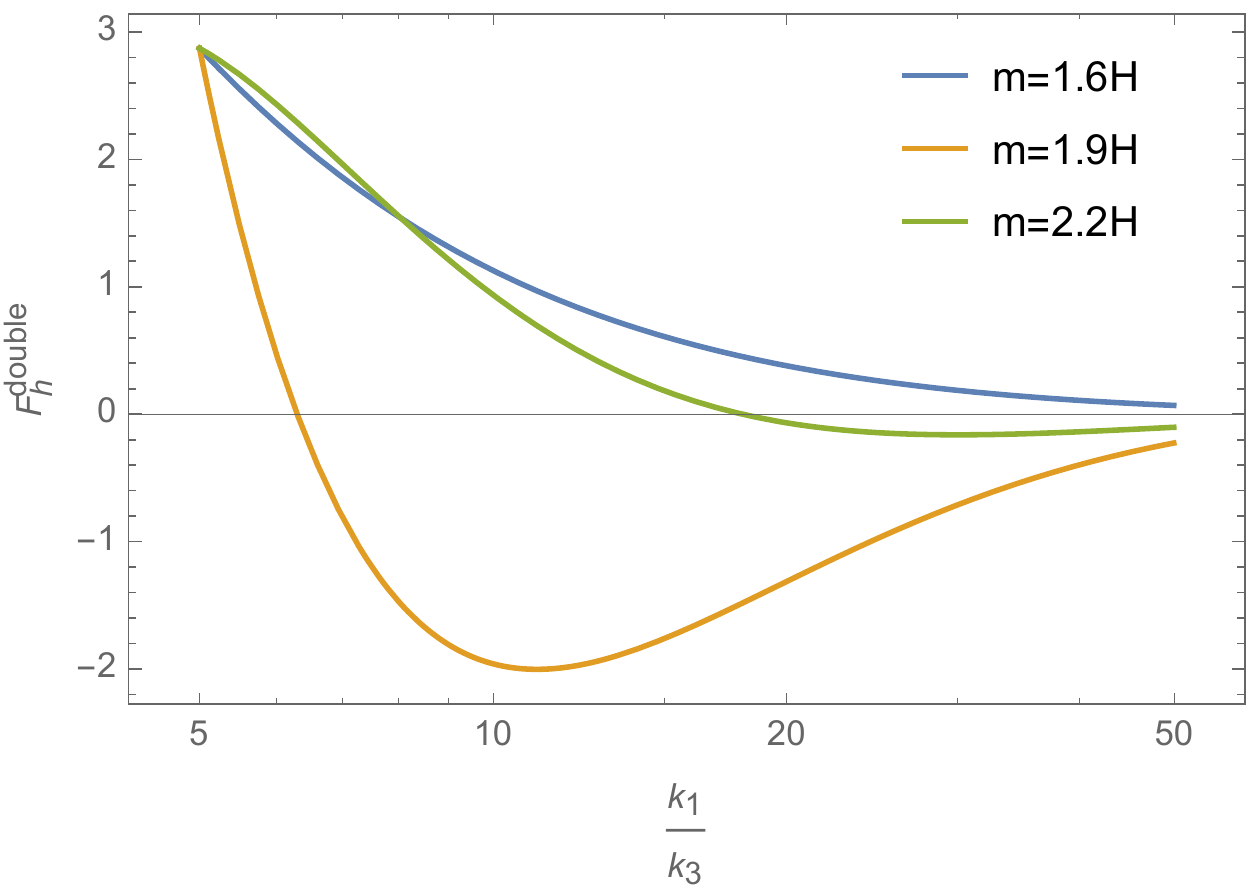}
		\caption{Shape sensitivity of $F_h^{\text{double}}$ to $m_h$. We have chosen three plausible sets of parameters for which $F_h^{\text{double}}$ agree at the fiducial ratio $\frac{k_1}{k_3}=5$. This illustrates our ability to discriminate among different masses.}
		\label{fig:doubleexchangescalarfinalgoldstone2}
	\end{figure}

	As derived in appendix \ref{appscalarng} in \eqref{doubleheft}, 
	\begin{equation}
	F^{\text{double}}_{h} =  \lambda_2(\lambda_2 v f_\pi)^2 \frac{i\pi^2}{32}\left(A(\mu)s(\mu)-A^*(-\mu)s^*(-\mu)\right)\left(\frac{k_3}{k_1}\right)^{3/2}\left(\frac{k_3}{2k_1}\right)^{i\mu}+(\mu\rightarrow-\mu)
	\end{equation}
	where, $A(\mu)$ and $s(\mu)$ are mass dependent coefficients: 
	$A(\mu)=-2\sqrt{2/\pi}\text{sech}(\pi\mu)\Gamma(-i\mu)\sin(\frac{\pi}{4}+\frac{i\pi\mu}{2})$; and $s(\mu)$ can be represented by the integral,
	$s(\mu)=\int_{0}^{\infty}\frac{dx}{x^2}e^{-ix}J_{+}(x)x^{3/2+i\mu}$
	where, $J_{+}(x)$ is a somewhat complicated function given in \eqref{Jplus}. We exemplify the strength of NG below for the benchmark values, $\lambda_2=0.2;\lambda_h=0.5$:
	\begin{center}
		\begin{tabular}{|r|l|}
			\hline
			mass & $|f^{\text{double}}_{h}|$ \\
			\hline \hline
			1.6 H & 4.972 \\
			1.9 H & 0.647 \\
			2.2 H & 0.171 \\
			\hline
		\end{tabular}
	\end{center}
	We illustrate the momentum dependence of $F_h^{\text{double}}$ in 
	Fig. \ref{fig:doubleexchangescalarfinalgoldstone1} and \ref{fig:doubleexchangescalarfinalgoldstone2}. 
	In the special case of single-field slow-roll inflation, using lagrangian \eqref{higgsexpansion}, $F_h^{\text{double}}$ takes an identical form to \eqref{doubleheft} except the coupling constants are now different \eqref{doublehsr},
	\begin{equation}
	F^{\text{double}}_{h} =  \alpha\rho_2^2 \frac{i\pi^2}{16}\left(A(\mu)s(\mu)-A^*(-\mu)s^*(-\mu)\right)\left(\frac{k_3}{k_1}\right)^{3/2}\left(\frac{k_3}{2k_1}\right)^{i\mu}+(\mu\rightarrow-\mu)
	\end{equation}
	The strength of the NG then, for the same set of benchmark values, $c_2=\frac{H}{\sqrt{\dot{\phi}_0}},\lambda_h=\frac{H^2}{2\dot{\phi}_0},\Lambda=3 \sqrt{\dot{\phi}_0}$:
	\begin{center}
		\begin{tabular}{|r|l|}
			\hline
			mass & $|f^{\text{double}}_{h}|$ \\
			\hline \hline
			1.6 H & 0.117 \\
			1.9 H & 0.015 \\
			2.2 H & 0.003 \\
			\hline
		\end{tabular}
	\end{center}
	The shape dependence is identical to Figs. \ref{fig:doubleexchangescalarfinalgoldstone1} and \ref{fig:doubleexchangescalarfinalgoldstone2}, so not shown explicitly.

	\subsection{Triple Exchange Diagram}
\begin{figure}[h]
		\centering
		\includegraphics[width=0.6\linewidth]{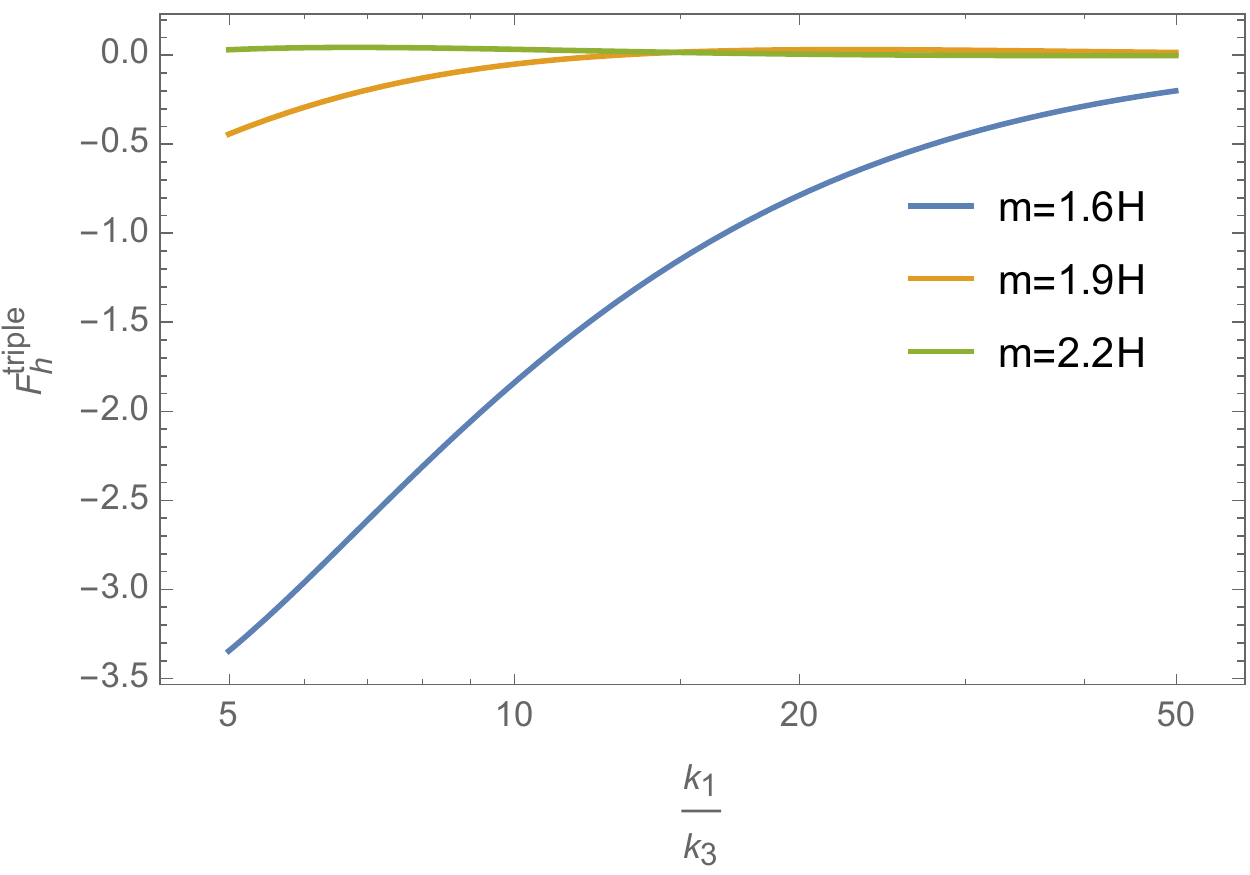}
		\caption{Dimensionless three-point function $F_h^{\text{triple}}$ \eqref{Ffunction} for different masses in Goldstone Effective description \eqref{tripleheft} with $\lambda_2=0.2;\lambda_h=0.5$.}
		\label{fig:tripleexchangescalarfinalgoldstone1}
	\end{figure}

\begin{figure}[h]
		\centering
		\includegraphics[width=0.6\linewidth]{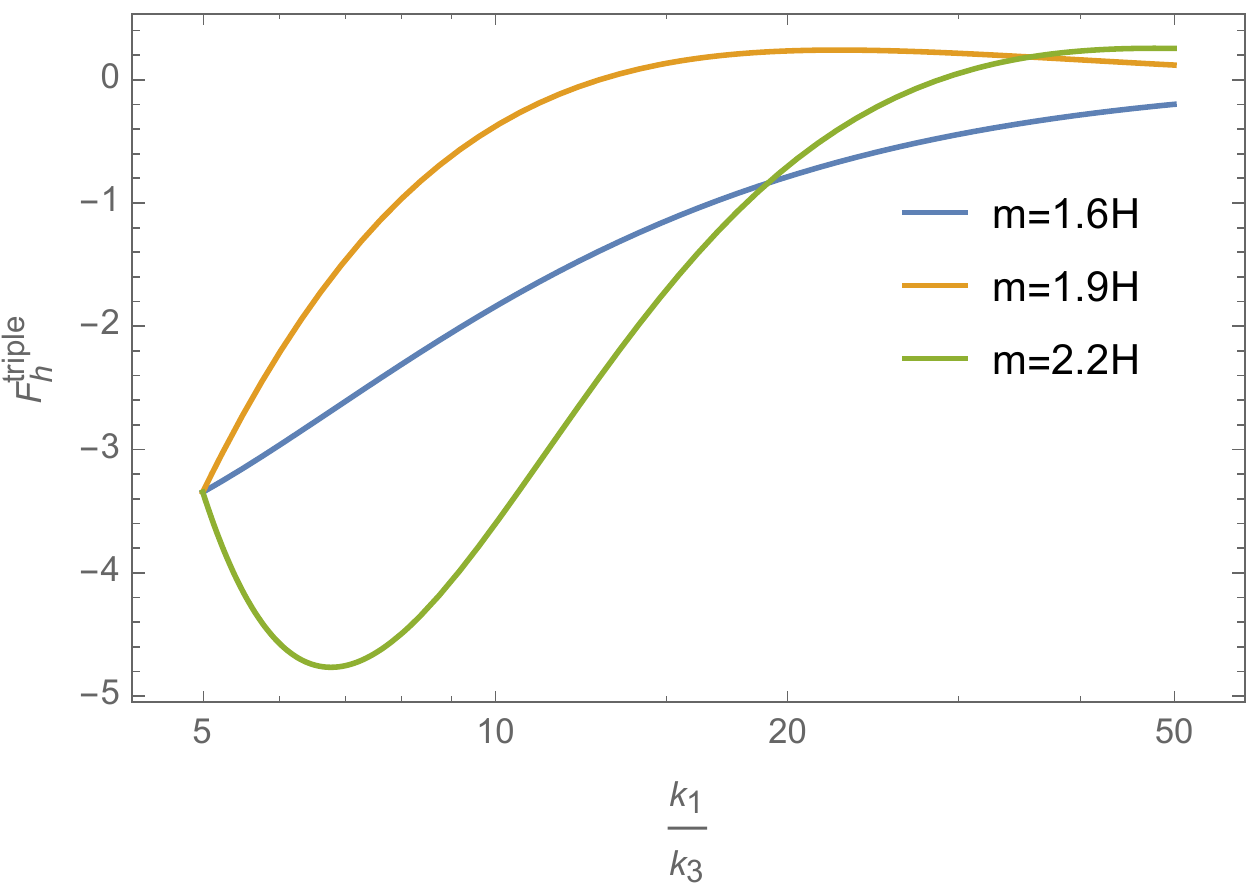}
		\caption{Shape sensitivity of $F_h^{\text{triple}}$ to $m_h$. We have chosen three plausible sets of parameters for which $F_h^{\text{triple}}$ agree at the fiducial ratio $\frac{k_1}{k_3}=5$. This illustrates our ability to discriminate among different masses.}
		\label{fig:tripleexchangescalarfinalgoldstone2}
	\end{figure}

	The triple exchange diagram has been calculated in \cite{Chen:2017ryl}, but we include it here for completeness and comparison to the other diagrams. As derived in appendix \ref{appscalarng} in \eqref{tripleheft},
	\begin{equation}
	F^{\text{triple}}_{h}=\frac{\pi^3\lambda_2^3v^3f_\pi^2\lambda_h v}{128}(+i)\left(A(\mu)t(\mu)-A^*(-\mu)t^*(-\mu)\right)\left(\frac{k_3}{k_1}\right)^{\frac{3}{2}}\left(\frac{k_3}{2k_1}\right)^{i\mu}+(\mu\rightarrow-\mu)
	\end{equation}
	where, $A(\mu)$ is the same coefficient as introduced above and $t(\mu)=\int_{0}^{\infty}\frac{dx}{x^4}J_{+}(x)^2x^{\frac{3}{2}+i\mu}$. We exemplify the strength of NG below for the benchmark values, $\lambda_2=0.2$ and $\lambda_h=0.5$:
	\begin{center}
		\begin{tabular}{|r|l|}
			\hline
			mass & $|f^{\text{triple}}_{h}|$ \\
			\hline \hline
			1.6 H & 10.1 \\
			1.9 H & 0.772 \\
			2.2 H & 0.148 \\
			\hline
		\end{tabular}
	\end{center}
	We illustrate the momentum dependence of $F_h^{\text{triple}}$ in 
	Fig. \ref{fig:tripleexchangescalarfinalgoldstone1} and \ref{fig:tripleexchangescalarfinalgoldstone2}. 
	In the special case of single-field slow-roll inflation, using lagrangian \eqref{higgsexpansion}, $F_h^{\text{triple}}$ takes an identical form except the coupling constants are now different \eqref{triplehsr},
	\begin{equation}
	F^{\text{triple}}_{h}=\frac{\pi^3\rho_2^3\dot{\phi}_0\lambda_h v}{128}(+i)\left(A(\mu)t(\mu)-A^*(-\mu)t^*(-\mu)\right)\left(\frac{k_3}{k_1}\right)^{\frac{3}{2}}\left(\frac{k_3}{2k_1}\right)^{i\mu}+(\mu\rightarrow-\mu)
	\end{equation}
	The strength of the NG then, for the same set of benchmark values,
	$c_2=\frac{H}{\sqrt{\dot{\phi}_0}},\lambda_h=\frac{H^2}{2\dot{\phi}_0},\Lambda=3 \sqrt{\dot{\phi}_0}$:
	\begin{center}
		\begin{tabular}{|r|l|}
			\hline
			mass & $|f^{\text{triple}}_{h}|$ \\
			\hline \hline
			1.6 H & 0.239 \\
			1.9 H & 0.018 \\
			2.2 H & 0.003 \\
			\hline
		\end{tabular}
	\end{center}
	The shape dependence is identical to Figs. \ref{fig:tripleexchangescalarfinalgoldstone1} and \ref{fig:tripleexchangescalarfinalgoldstone2}, so not shown explicitly.
	\section{Detailed Form of NG Mediated by $Z$}\label{sectiononZ}

	To discuss the form of $F$ for NG mediated by $Z$, we again first focus on the Goldstone effective description as before, and specialize to the single-field slow-roll description following that. Since the triple exchange diagram is too small to make any observable contribution we will restrict ourselves to single and double exchange diagrams.
	
	The Goldstone effective lagrangian needed for this case is given by \eqref{inf-Z-goldstone} which we rewrite,
	\begin{equation}\label{inf-Z-goldstone2}
	-\frac{\text{Im}(d_{1})m_Z v}{2\Lambda}\dot{\pi}_c  Z^0- \frac{\text{Im}(d_{1})m_Z v}{2\Lambda f_\pi^2}\dot{\pi}_c\partial_\mu\pi_c  Z^\mu
	+ \frac{d_3m_Z^2}{2\Lambda^2} \dot{\pi}_c Z_{\mu}^2+\frac{d_4}{\Lambda^2} \dot{\pi}_c Z_{\mu\nu}^2+\cdots
	\end{equation}
	In this case in the squeezed limit, $F(k_1,k_2,k_3)$ is a function of $\frac{k_3}{k_1}$ and also the angle between $\hat{k}_3$ and $\hat{k}_1$,
	\begin{equation}
	F = \left(f(\mu)\left(\frac{k_3}{k_1}\right)^{\frac{5}{2}+i\mu} + f(\mu)^*\left(\frac{k_3}{k_1}\right)^{\frac{5}{2}-i\mu}\right)\sin^2\theta
	\end{equation}
	where, $\theta=\hat{k}_3\cdot\hat{k}_1$. We also see that $F$ falls faster with $\frac{k_3}{k_1}$. The angle dependence, in principle, gives an important handle to determine the spin-1 nature of $Z$. Recently in \cite{MoradinezhadDizgah:2017szk} it was analyzed to what extent future galaxy surveys can constrain mass and spin. A forecast using 21-cm cosmology would also be important and possibly more constraining. 
	
	Now we give the expressions for $f(\mu)$ for
	the single exchange diagram, leaving the details for the appendix \ref{appvectorng}. The computation of double exchange diagram will not be performed in this paper, however using the mixed propagator formalism \cite{Chen:2017ryl} it can be done. Here, we will only give some reasonable estimates.
	
	\subsection{Single Exchange Diagram}
	
	As derived in Appendix \ref{appvectorng} in \eqref{singlezeft},
	\begin{gather}
	F^{\text{single}}_{Z}=\left(\frac{v}{2\Lambda}\right)^2\frac{1}{16\pi}\sin^2\theta\Gamma(\frac{3}{2}+i\mu)\Gamma(\frac{3}{2}-i\mu)\cosh(\pi\mu) \times\nonumber\\
	\left((7-5i\mu+16\mu^2+4i\mu^3)\Gamma(\frac{3}{2}+i\mu)^2\Gamma(-2-2i\mu)(1-i\sinh(\pi\mu))\left(\frac{k_3}{k_1}\right)^{\frac{5}{2}+i\mu}+(\mu\rightarrow-\mu)\right),
	\end{gather}
	where, $\theta=\hat{k}_3\cdot\hat{k}_1$. We illustrate the strength of NG, for the parameter choices, $v=3H;\Lambda=8H$:
	\begin{center}
		\begin{tabular}{|r|l|}
			\hline
			mass & $|f^{\text{single}}_{Z}|$ \\
			\hline \hline
			0.4 H & 0.003 \\
			0.8 H & 0.001 \\
			\hline
		\end{tabular}
	\end{center}
	We see the strengths are quite weak, hence 21-cm cosmology is critical if we are to see NG due to the single exchange diagram. Note that even an imprecise measurement should be readily distinguishable from scalar-mediated NG and NG purely due to the inflationary dynamics (analytic in $\frac{k_3}{k_1}$), due to the non-trivial angular dependence. 
	
	We now discuss single-field slow-roll inflation. The relevant lagrangian for a non-negligible $Z$-mediated signal arises when the associated Higgs scalar $h$ is heavy enough that its on-shell propagation is Boltzmann suppressed, but can be integrated out to yield new $Z$ vertices, as in 
	\eqref{heavyhiggs}. It 
	has an identical structure to the Goldstone lagrangian \eqref{inf-Z-goldstone2} above, as shown in \ref{appvectorng}. Hence $F$ can be obtained just by the replacement, 
	\begin{equation}
	\frac{v m_Z}{2\Lambda}\rightarrow \frac{\rho_{1,Z}\rho_2}{m_h^2}.
	\end{equation}
	We see for $\rho_{1,Z}=1,\rho_2=1,m_h=3H$ we have roughly the same strength of NG as the effective Goldstone theory. However, we get parametrically bigger NG in both effective theories from the double exchange diagram in Fig. \ref{fig:three-diagrams}, which we now discuss.

	\subsection{Double Exchange Diagram}
	
	As we mentioned above, in this paper we will give only an estimate of the double exchange diagram. 
	As we have explained in Sec. \ref{squeezed}, in the  squeezed limit diagrams factorize into contributions from hard and soft processes. This means in Fig. \ref{fig:three-diagrams} (b), the $Z$ propagator having hard momenta $k_2$ is expected to be a function of $\mathcal{O}(1)$ (in Hubble units). In that approximation the diagram then has the same topology as the single exchange diagram. However, as can be seen from the lagrangian \eqref{inf-Z-goldstone2}, the parametric strength of the diagram goes like
	\begin{equation}
	\left(\frac{v}{2\Lambda}\right)^2\times \frac{f_\pi^2}{\Lambda^2},
	\end{equation}
	which has the enhancement by $\frac{f_\pi^2}{\Lambda^2}$. 
	Thus, while we saw that the single-exchange contribution was at best marginally detectable in the future, the double-exchange contribution should be much more promising in magnitude for $\Lambda \sim 5 - 10H$, $v \sim 2-3H$, with $f_Z \sim 0.1 - 1$.
	We leave a precise calculation of this for later work, to hopefully confirm this expectation. 
	
	Moving to the case of single-field slow-roll inflation, from  \eqref{heavyhiggs} arising from integrating out the associated heavy $h$, we see that the double-exchange diagram is parametrically enhanced over single-exchange by a factor of $\frac{H\dot{\phi}_0}{v m_h^2}$, so that $f_Z \sim 0.01$ for $v\sim \sqrt{\dot{\phi}_0}; m_h\sim 3H$. This should yield a weak but detectable signal. 
	
	\section{Concluding Remarks and Future Directions}\label{conclu}
	
	Cosmological Collider Physics builds on the distinctive non-analytic momentum dependence of primordial NG mediated by particles with  masses $m \sim H$, in contrast to the analytic dependence of NG due purely to the inflationary dynamics, driven by fields  with $m \ll H$.  In this paper, we focused on 
	the question of whether gauge-theories with such ultra-high $\sim H$ mass scales could be detected by this means, since such theories are obviously very highly motivated.
	If the gauge symmetry is unbroken  during inflation, gauge-charged states can only affect primordial NG via very small loop-level effects, difficult to observe. 
	However, we showed that when the gauge-symmetry is (partially) Higgsed, the Higgs-type spin-0 and $Z$-type spin-1 bosons can contribute at {\it tree level} to potentially observable NG. 
	The simplest effective vertices one can write connecting the gauge-Higgs states to the inflaton so as to mediate NG are non-renormalizable, suppressed at least by powers of the cutoff of the inflationary EFT, $\Lambda$, representing the threshold of even heavier physics that has been integrated out.
		 The largest NG will then come by considering the lowest consistent $\Lambda$. 
	We studied these NG within two effective descriptions of the inflationary dynamics: a) generic slow-roll inflation models, and b) the effective Goldstone  description of inflaton quantum fluctuations. In slow-roll, the minimal cutoff $\Lambda$ is given by the scale of kinetic energy of the rolling inflaton field, $\sqrt{\dot{\phi}_0} \sim 60 H$. The effective Goldstone description is more agnostic about inflationary dynamics, 
	treating this as a given classical background process, in which case
	$\Lambda$ can  be as low as a few $H$. 
	Of course, the detailed strengths of NG, $F$, that we get in the two cases are model-parameter dependent, but we can briefly summarize using the results in Sections \ref{hNG} and \ref{sectiononZ}:
	\begin{center}
		\begin{tabular}{|c|c|c|c|}
			\hline
			 & Goldstone EFT  & Goldstone EFT  & Slow-roll Models  \\
			 $F$ & with $\Lambda\sim 5H$& with $\Lambda\sim 10H$ & with $\Lambda\sim 60H$ \\
			\hline 
			$h$ & $1-10$ & $0.1-1$ & $0.01-0.1$  \\
			$Z$ & $0.1-1$ & $0.01-0.1$ & $0.001-0.01$ \\
			\hline
		\end{tabular}
	\end{center}
	The dimensionless bispectrum $F$ (see \eqref{Bfunction},\eqref{Ffunction}) given above is the maximum value taken in the squeezed regime. Based on the above table, several remarks are in order. While the above choices for EFT cutoffs lead to an observable strength of NG, we cannot make the cutoffs much bigger, since the NG falls rapidly as a function of squeezing and 
		the observable precision is limited by cosmic variance, $\delta F \sim 10^{-4} - 10^{-3}$,  \eqref{cosmicvar}.  The scale of Higgsing, $v$, is also relevant to our theoretical control. 
	Higgsing obviously relaxes the tight constraints of gauge invariance, allowing tree-level NG. But there are non-trivial constraints of the gauge structure following from having to expand observables in powers of
	$v/\Lambda$. In the UV limit $v \sim \Lambda$, the constraints of gauge-invariance disappear altogether.
	To stay in theoretic control, we have chosen $\frac{v}{\Lambda} \lesssim \frac{1}{3}$ in our studies. 
	
	We have used effective non-renormalizable vertices for this paper, but it is obviously of great interest and importance to seek a more UV-complete level of theoretical description to have greater confidence in the opportunity to detect gauge theory states in NG. We see that the strength of NG is bigger when it is mediated by $h$'s compared to mediation by $Z$'s. Furthermore, if cosmological collider physics turns out to be in a purely gauge-theoretic domain, then we would not see any states with spin $>1$, and their associated angular dependences. 
	Spin $ > 2$ mediated NG would signal a breakdown of point-particle field theories, perhaps signaling the onset of string theoretic structure. On the other hand, observing spins $0,1$ only, with stronger spin-0 signals, would give strong evidence for the structure studied above. While the (NM)SM gives only one $h$ and one $Z$, extensions of it (for example, even just some colored scalars) or whole new gauge sectors are capable of giving multiple $h/Z$-type states to observe. 
	
	We have argued that a strong possibility for $m_{\rm gauge-theory} \sim H$ is that they arise via a ``heavy-lifting'' mechanism from much lower-scale gauge theories in the current era. If these gauge theories are already seen at lower-scale terrestrial experiments, then the 
	renormalization group allows us to predict expected mass ratios in NG. In principle, such corroboration 
	would provide spectacular evidence for the large range of validity of such gauge theories, and the absence of intervening (coupled) states. However, we cannot hope to get a very precise measurement of such mass ratios, given cosmic variance. But if we are ever in the position to predict even a few such ratios, modestly precise measurements in NG would still be compelling. Alternatively, of course, we may discover wholly unexpected gauge-structure within the NG, at least dimly seen.

	There are multiple future directions which remain to pursue. There is obviously the need for an explicit calculation of the double-exchange diagram involving $Z$-type particles which would provide a check for our estimates. Cosmological correlations derived from inflationary expansion are famously nearly spatially scale-invariant. But in large regimes of slow-roll inflation or in the Goldstone description, the correlators are actually nearly spatially {\it conformally} invariant, that is they are close to the isometries of dS spacetime. 
	In this paper, we have assumed this regime of inflationary dynamics. But it is possible to relax this assumption of approximate conformal invariance, and just keep approximate scale invariance, for example allowing a small speed of inflaton fluctuations, $c_s\ll 1$, which can give rise to larger NG \cite{Alishahiha:2004eh,Chen:2006nt,Cheung:2007st}, even allowing us to probe loop effects of charged states. This remains to be explored. 
	There is also the generic question of how efficiently we can use NG templates to look for simultaneous presence of spin-0 and spin-1 particles, with a ``background'' of inflationary NG as well as 
	late-time effects.  Recent preliminary studies in these  directions appear in \cite{Meerburg:2016zdz,MoradinezhadDizgah:2017szk} which suggest that some of the stronger signals we describe above would be visible with reasonable precision.

	We can view the heavy-lifting mechanism as leveraging un-naturalness, by noting that the low-dimension Higgs mass term of elementary Higgs fields is very ``unstable'' to curvature-related corrections. In that sense, confirming heavy-lifting of an unnatural gauge-Higgs theory, such as the (NM)SM, would be a strong sign that naturalness is massively violated  in Nature. Of course, the validity of naturalness
	is one of the burning debates and concerns within fundamental physics. But it is also possible that terrestrial experiments show us a natural theory, such as a supersymmetric gauge theory. One can then consider the possibility of heavy-lifting of such a natural theory. Depending on the nature of supersymmetry-breaking it is possible that the lifted gauge theory exhibits a different pattern of supersymmetry soft breaking and associated Higgsing than the unlifted theory in the current era.  We leave an investigation of supersymmetric gauge-Higgs theory for the future. 
	
	We have seen that invaluable information on the gauge-theoretic structure of the laws of nature can be imprinted on cosmological NG, but we have also seen that these signals are extremely weak given cosmic variance. To have any chance of seeing and deciphering such exciting physics will require pushing experimental precision and understanding of systematic uncertainties to the their limits. Heavy-lifting indeed!

	\section*{Acknowledgments}
	The authors would like to thank Nima Arkani-Hamed, Julian Munoz, Marc Kamionkowski, David E. Kaplan for helpful discussions. This research was supported in part by the NSF under Grant No. PHY-1620074 and by the Maryland Center for Fundamental Physics (MCFP).
	\appendix
	
	\section{Scalar Fields in dS Space}\label{appscalarmode}
	The metric for the Poincare patch of dS spacetime in Hubble units can be written as
	\begin{equation}
	ds^2 = \frac{-d\eta^2+d\vec{x}^2}{\eta^2}.
	\end{equation}
	
	\subsection{Massive Fields}
	
	We want to get the mode functions for a quantum field in dS. This can be obtained by first solving the classical equation of motion and then by canonically quantizing the theory. Let us start by writing the equation of motion,
	\begin{gather}
	\partial_\mu(\sqrt{-g}g^{\mu\nu}\partial_\nu\phi)=\sqrt{-g}m^2\phi \\ \label{eomforscalar}
	\Rightarrow\partial_\eta^2\phi-\frac{2}{\eta}\partial_\eta\phi-\partial_i^2\phi+\frac{m^2}{\eta^2}\phi=0.
	\end{gather}
	This can be solved in terms of Hankel (or equivalently, Bessel) functions. After Fourier transforming to $\vec{k}$-space, we can write a general classical solution as, 
	\begin{align}\label{generalsoln}
	\phi(\eta,\vec{k})=c_1(-\eta)^{\frac{3}{2}}H_{i\mu}^{(1)}(-k\eta)+c_2(-\eta)^{\frac{3}{2}}H_{i\mu}^{(2)}(-k\eta),
	\end{align}
	where, $\mu=\sqrt{\frac{m^2}{H^2}-\frac{9}{4}}$. 
	
	As usual, to canonically quantize the theory, we elevate the coefficients $c_1, c_2$ to linear combinations of creation and destruction operators, $a_{\vec{k}}^\dagger, a_{\vec{k}}$, on the Bunch-Davies vacuum. The quantum field thereby has the form,
	\begin{equation}
	\phi(\eta,\vec{k})=f_k(\eta)a_{\vec{k}}^\dagger + \bar{f}_k(\eta)a_{-\vec{k}},
	\end{equation}
	where the mode functions, $f_k(\eta)$ and $\bar{f}_k(\eta)$ (or equivalently, the linear combinations referred to above), are determined as follows. We first find the conjugate momentum $\pi=\frac{\partial\mathcal{L}}{\partial_\eta\phi}$ and demand, $[\pi(\eta,\vec{x}),\phi(\eta,\vec{y})]=i\delta^3(\vec{x}-\vec{y})$ and $[a_{\vec{k}},a^\dagger_{\vec{k}^\prime}]=(2\pi)^3\delta^3(\vec{k}-\vec{k}^\prime)$. This gives the Wronskian condition on the mode functions at $\eta\rightarrow-\infty$,
	\begin{equation}\label{Wronskian}
	\bar{f}_k(\eta)f^\prime_k(\eta)-\bar{f}^\prime_k(\eta) f_k(\eta) = i\eta^2.
	\end{equation}
	To impose the Bunch Davies vacuum we demand $f_k(\eta)\propto e^{ik\eta}$, and using the Wronskian condition \eqref{Wronskian} we can also fix the normalization of $f_k(\eta)$ up to a phase. In summary, we demand
	\begin{equation}
	\lim_{\eta\to -\infty}f_k(\eta)=(-\eta)\sqrt{\frac{1}{2k}}e^{ik\eta}.
	\end{equation}
	This can be satisfied by choosing
	\begin{align}\label{mode1}
	f_k(\eta)= (+ie^{-i\pi/4})\frac{\sqrt{\pi}}{2}e^{\pi\mu/2}(-\eta)^{\frac{3}{2}}H_{i\mu}^{(2)}(-k\eta),
	\end{align}
	and
	\begin{align}\label{mode2}
	\bar{f}_k(\eta)= (-ie^{i\pi/4})\frac{\sqrt{\pi}}{2}e^{-\pi\mu/2}(-\eta)^{\frac{3}{2}}H_{i\mu}^{(1)}(-k\eta).
	\end{align}
	Here, we have introduced some phase factors which are just conventions, which drop out when
	we calculate propagators. 
	
	We note that in \eqref{ininmasterformula} we have both time and anti-time ordered expressions appearing. Since a propagator involves two mode functions, we can have a total of four types of propagators depending on the mode functions coming from either the time or anti-time ordered part. We denote time(anti-time) ordering by a $+(-)$ sign. As an example, a propagator $G_{+-}(k,\eta,\eta^\prime)$ means the mode function with argument $\eta(\eta^\prime)$ is coming from time(anti-time) ordering. Similarly, $G_{++}(k,\eta,\eta^\prime)$ means both the mode functions are coming from time ordering. Thus we can write
	\begin{equation}
	\begin{aligned}
	G_{++}(k,\eta,\eta^\prime)&=\bar{f}_k(\eta^\prime)f_k(\eta)\theta(\eta^\prime-\eta)+\bar{f}_k(\eta)f_k(\eta^\prime)\theta(\eta-\eta^\prime) \\
	G_{+-}(k,\eta,\eta^\prime)&=\bar{f}_k(\eta^\prime)f_k(\eta) \\
	G_{-+}(k,\eta,\eta^\prime)&=\bar{f}_k(\eta)f_k(\eta^\prime)\\ 
	G_{--}(k,\eta,\eta^\prime)&=\bar{f}_k(\eta^\prime)f_k(\eta)\theta(\eta-\eta^\prime)+\bar{f}_k(\eta)f_k(\eta^\prime)\theta(\eta^\prime-\eta).
	\end{aligned}
	\label{++}
	\end{equation}
	Among these four, $G_{+-},G_{++}$ are conjugates of $G_{-+},G_{--}$ respectively, and so we only have two independent propagators.

	\subsection{Inflaton Mode Functions}

	Mode functions for massless fields, in particular the inflaton, follow by using $\mu=3 i/2$ in \eqref{mode1} and \eqref{mode2}, which gives
	
	\begin{equation}\label{infmodes}
	\begin{aligned}
	f_k(\eta)&=\frac{(1-ik\eta)e^{ik\eta}}{\sqrt{2k^3}}\\
	\bar{f}_k(\eta)&=\frac{(1+ik\eta)e^{-ik\eta}}{\sqrt{2k^3}}.
	\end{aligned}
	\end{equation}

	\subsection{Some Useful Relations for Diagrammatic Calculations}

	For later use we also note a few relations involving Hankel and hypergeometric functions that arise upon evaluating the Feynman diagrams for the NG correlators of interest. We can write the following integral involving Hankel functions in terms of a hypergeometric function (valid for real $\mu$ and $\nu\equiv i\mu < \frac{1}{2}$),
	\begin{multline}\label{h2}
	e^{\pi\mu/2}\int_{0}^{\infty}dx x^n e^{-ipx}H_{i\mu}^{(2)}(x)=\\ \frac{(-i/2)^n}{\sqrt{2}\Gamma(n+3/2)}\sqrt{2/\pi}\Gamma(n+1+i\mu)\Gamma(n+1-i\mu){}_2F_1(n+1+i\mu,n+1-i\mu,n+3/2,(1-p)/2)
	\end{multline}
	\begin{multline}\label{h1}
	e^{-\pi\mu/2}\int_{0}^{\infty}dx x^n e^{+ipx}H_{i\mu}^{(1)}(x)=\\ \frac{(+i/2)^n}{\sqrt{2}\Gamma(n+3/2)}\sqrt{2/\pi}\Gamma(n+1-i\mu)\Gamma(n+1+i\mu){}_2F_1(n+1-i\mu,n+1+i\mu,n+3/2,(1-p)/2).
	\end{multline}
	It is useful later to approximate these expressions for large $p$
	by using limiting forms of Hankel and hypergeometric functions,
	\begin{equation}
	e^{\pi \mu/2}H_{i\mu}^{(2)}(z)=\frac{i}{\pi}\left(\Gamma(i\mu)(z/2)^{-i\mu}e^{\pi\mu/2}+\Gamma(-i\mu)(z/2)^{+i\mu}e^{-\pi\mu/2}\right)
	\end{equation}
	
	\begin{equation}
	e^{-\pi \mu/2}H_{i\mu}^{(1)}(z)=-\frac{i}{\pi}\left(\Gamma(-i\mu)(z/2)^{+i\mu}e^{\pi\mu/2}+\Gamma(+i\mu)(z/2)^{-i\mu}e^{-\pi\mu/2}\right).
	\end{equation}
	We also need large negative argument expansion of hypergeometric function,
	\begin{equation}\label{hypergeolimit}
	{}_2F_1(a,b,c;z)=\frac{\Gamma(b-a)\Gamma(c)}{\Gamma(b)\Gamma(c-a)}(-z)^{-a}+\frac{\Gamma(c)\Gamma(a-b)}{\Gamma(a)\Gamma(c-b)}(-z)^{-b}.
	\end{equation}

	\section{NG due to $h$ Exchange}\label{appscalarng}
	
	\subsection{Calculation of Single Exchange Diagram}
	
	We will use the in-in formula \eqref{ininmasterformula} to calculate NG due to the single exchange diagram which is depicted in Fig. \ref{fig:three-diagrams} (a). 
	We begin by reviewing this calculation in the context of single-field slow-roll inflation, as originally performed in \cite{Arkani-Hamed:2015bza}. 
	The relevant terms in the lagrangian \eqref{higgsexpansion} for such a diagram are
	\begin{equation}\label{higgssingle}
	\mathcal{L}\supset -\rho_2 \dot{\xi}h + \frac{\rho_2}{2\dot{\phi}_0}(\partial\xi)^2 h + \cdots .
	\end{equation} 
	Note in \eqref{ininmasterformula} we have both time and anti-time ordering. Thus each vertex can contribute either from time or anti-time ordering. So an in-in diagram with $n$ vertices gives rise to $2^n$ subdiagrams. These subdiagrams differ in the type of propagators used for the massive particle and inflatons. For example if both the vertices are coming from time ordering, we should use $G_{++}$ for the massive propagator as defined in \eqref{++}. We call the subdiagram containing $G_{++}$ to be $I_{++}$. Thus for the single exchange diagram we have four subdiagrams which we denote by $I_{++},I_{+-},I_{-+},I_{--}$ depending on which kind of massive propagator has been used. However, to compute the entire three point function due to single exchange diagram, we have to consider only two subdiagrams, since the other two are related by complex conjugation. For example, we will calculate only $I_{--}$ and $I_{+-}$ which are related to $I_{++}$ and $I_{-+}$ respectively by complex conjugation. We sum all four contributions to get the final answer. To clarify the above comments, we write the expressions for four subdiagrams schematically,
	
	\begin{gather}
	I_{\pm \pm}\propto (\pm i)(\pm i)\int \frac{d\eta}{\eta^4}\int \frac{d\eta^\prime}{\eta^{\prime 4}}g_{\pm}(k_3;\eta^\prime)\tilde{g}_{\pm}(k_1,k_2;\eta)G_{\pm\pm}(k_3;\eta,\eta^\prime).
	\end{gather}
	The prefactors $\mp i$ arise depending on whether we use $e^{-i \int H dt}$ for time ordering or $e^{+i \int H dt}$ for anti-time ordering. $g_{\pm},\tilde{g}_{\pm}$ are inflaton bulk-boundary propagators (which we define below); and $G_{\pm\pm}$ are bulk-bulk propagators \eqref{++} for $h$.
	
	\subsubsection{Calculation of $I_{+-}$}
	
	Let us start with $I_{+-}$ diagram. This diagram factorizes into a product of two integrals with one coming from time ordering and another from anti time ordering. 
	\paragraph{Anti-time Ordered Contribution}
	We first calculate inflaton contribution using inflaton mode function \eqref{infmodes},
	\begin{equation}
	g_{-}(k;\eta)\equiv\langle\dot{\xi}(\eta,\vec{k})\xi(\eta_0\rightarrow 0,-\vec{k})\rangle = -\frac{\eta^2k^2}{2k^3}e^{-ik\eta},
	\end{equation}
	to write the anti time ordered contribution as 
	\begin{equation}
	\text{Anti-time Ordered Contribution}=
	(+i)\int_{-\infty}^{0}\frac{d\eta^\prime}{\eta^{\prime 4}}\left(-\frac{\eta^{\prime 2}k_3^2}{2k_3^3}e^{-ik_3\eta^\prime}\right)\bar{f}_{k_3}(\eta^\prime).
	\end{equation}
	Using the mode functions \eqref{mode2} and relation \eqref{h1} we get
	\begin{equation}
	\text{Anti-time Ordered Contribution} = -\frac{1}{2\sqrt{2}}\frac{1}{k_3^{\frac{3}{2}}}\Gamma\left(\frac{1}{2}+i\mu\right)\Gamma\left(\frac{1}{2}-i\mu\right).
	\end{equation}
	\paragraph{Time Ordered Contribution}
	Let us first calculate the inflaton contribution again. Now we have to do a little more work since based on \eqref{higgssingle} we see that we have to find the contraction which can be schematically written as $\langle\xi\xi \vert (\partial\xi)^2\rangle$. Writing $\tilde{g}_{+}(k_1,k_2,\eta)$ as,
	\begin{equation}
	\tilde{g}_{+}(k_1,k_2,\eta) \equiv \langle \xi(\eta_0,\vec{k}_1) \xi(\eta_0,\vec{k}_2) (\partial_\mu\xi(\eta,-\vec{k}_1)\partial^\mu\xi(\eta,-\vec{k}_2))\rangle,
	\end{equation}
	we get, using inflaton mode function \eqref{infmodes},
	\begin{multline}
	\tilde{g}_{+}(k_1,k_2,\eta) = -\left(-\frac{1}{2k_1^3}\right)\left(-\frac{1}{2k_2^3}\right)e^{ik_{12}\eta}k_1^2k_2^2\eta^4 \\
	+\eta^2(-i k_{1i})(-i k_{2i})\frac{(1-ik_1\eta)(1-ik_2\eta)}{2k_1^3 2k_2^3}e^{ik_{12}\eta},
	\end{multline}
	where we have defined, $k_{12}=k_1+k_2$. We can simplify this by removing some $\eta$-dependent factors by writing the above as an operator $\mathcal{D}$ acting on $e^{i k_{12}\eta}$, where
	\begin{equation}
	\mathcal{D} \equiv k_1^2 k_2^2\partial_{k_{12}}^2 + (-\vec{k}_1\cdot\vec{k}_2)(1-k_{12}\partial_{k_{12}}+k_1 k_2 \partial_{k_{12}}^2).
	\end{equation}
	Then,
	\begin{equation}
	\tilde{g}_{+}(k_1,k_2,\eta)=\frac{\eta^2}{4 k_1^3 k_2^3} \mathcal{D} e^{i k_{12}\eta}.
	\end{equation}
	
	Thus the time ordered contribution looks like
	\begin{equation}
	\text{Time Ordered Contribution}=(-i)\frac{1}{4k_1^3 k_2^3}\mathcal{D}\int_{-\infty}^{0}\frac{d\eta}{\eta^2}e^{ik_{12}\eta}f_{k_3}(\eta).
	\end{equation}
	The integral can be evaluated using \eqref{h2} and \eqref{mode1} to get a hypergeometric function. Since we will be interested in squeezed limit, $k_3\ll k_1,k_2$, we can expand the answer using \eqref{hypergeolimit}. We then get
	\begin{multline}
	\text{Time Ordered Contribution}\\=\frac{1}{4\sqrt{2}k_1^3k_2^3\sqrt{k_3}}\mathcal{D}\left(\frac{\Gamma(-2i\mu)\Gamma(1/2+i\mu)}{\Gamma(1/2-i\mu)}\left(\frac{p}{2}\right)^{-1/2-i\mu}+\frac{\Gamma(2i\mu)\Gamma(1/2-i\mu)}{\Gamma(1/2+i\mu)}\left(\frac{p}{2}\right)^{-1/2+i\mu}\right).
	\end{multline}
	The action of $\mathcal{D}$ simplifies in the squeezed limit, 
	\begin{equation}
	\mathcal{D}k_{12}^\alpha = \frac{1}{8}(\alpha-1)(\alpha-2)k_{12}^{2+\alpha},
	\end{equation}
	using which,
	\begin{multline}
	\text{Time Ordered Contribution} = \frac{1}{4\sqrt{2}k_1^3k_2^3\sqrt{k_3}} \\
	\times \frac{1}{8}(3/2+i\mu)(5/2+i\mu)\times
	\frac{\Gamma(-2i\mu)\Gamma(1/2+i\mu)}{\Gamma(1/2-i\mu)}k_{12}^2\left(\frac{k_{12}}{2k_3}\right)^{-1/2-i\mu} +(\mu\rightarrow -\mu).
	\end{multline} 
	
	Now we are ready to put together both the contributions:
	\begin{equation}
	I_{+-}=\frac{\rho_2^2}{\dot{\phi}_0}\frac{1}{64 k_1^2 k_2^2 k_3^2}\Gamma(1/2+i\mu)^2 \Gamma(-2i\mu)(3/2+i\mu)(5/2+i\mu)\left(\frac{k_1}{k_3}\right)^{-1/2-i\mu}+(\mu\rightarrow-\mu).
	\end{equation}
	
	\subsubsection{Calculation of $I_{--}$}
	
	$I_{--}$ diagram is in general complicated since it does not factorize into $\eta$ and $\eta^\prime$ integrals. But we can still calculate the the nonanalytic terms in $k_3$ in the squeezed limit. This is because in the squeezed limit, $\eta^\prime$ integral contributes when $-\eta^\prime \sim {\mathcal{O}(\frac{1}{k_3}})$, whereas the contribution of $\eta$ integral is dominant when $-\eta\sim {\mathcal{O}(\frac{1}{k_{12}})}$. Thus when $k_3\ll k_{12}$, one of the step functions in $G_{--}$ \eqref{++} drops out and the integral approximately factorizes. The $\eta^\prime$ integral then is identical to what we had for $I_{+-}$; whereas the only change for $\eta$ integral is that $k_{12}\rightarrow -k_{12}$. Thus we have
	\begin{multline}
	I_{--}+I_{+-} \\
	= \frac{\rho_2^2}{\dot{\phi}_0}\frac{1}{64 k_1^2 k_2^2 k_3^2}\Gamma\left(\frac{1}{2}+i\mu\right)^2 \Gamma(-2i\mu)\left(\frac{3}{2}+i\mu\right)\left(\frac{5}{2}+i\mu\right)\left(\frac{k_1}{k_3}\right)^{-\frac{1}{2}-i\mu}(1-e^{i\pi(-\frac{1}{2}-i\mu)})\\+(\mu\rightarrow-\mu).
	\end{multline}
	
	\subsubsection{Three-Point Function}
	
	The full three point function can be written as a sum of $I_{++},I_{+-},I_{-+},I_{--}$. This gives
	\begin{multline}
	\langle\xi(\vec{k}_1) \xi(\vec{k}_2) \xi(\vec{k}_3)\rangle = \frac{\rho_2^2}{\dot{\phi}_0}\frac{1}{16 k_1^2 k_2^2 k_3^2}\times \\
	\left(\Gamma\left(\frac{1}{2}+i\mu\right)^2 \Gamma\left(-2i\mu\right)\left(\frac{3}{2}+i\mu\right)\left(\frac{5}{2}+i\mu\right)(1+i\sinh(\pi\mu))\left(\frac{k_3}{k_1}\right)^{\frac{1}{2}+i\mu}+(\mu\rightarrow-\mu)\right).
	\end{multline}
	This gives $F$  as a function of $\left(\frac{k_3}{k_1}\right)$ as defined in \eqref{Ffunction} to be
	\begin{multline}\label{singlehsr}
	F^{\text{single}}_{h} = - \frac{1}{4}\times\rho_2^2\\
	\left(\Gamma\left(\frac{1}{2}+i\mu\right)^2 \Gamma\left(-2i\mu\right)\left(\frac{3}{2}+i\mu\right)\left(\frac{5}{2}+i\mu\right)(1+i\sinh(\pi\mu))\left(\frac{k_3}{k_1}\right)^{\frac{3}{2}+i\mu}+(\mu\rightarrow-\mu)\right).
	\end{multline}
	where, $\rho_2 = \frac{2 c_2 v \dot{\phi}_0 }{\Lambda^2}; \hspace{2em} \alpha =- \frac{c_2 \dot{\phi}_0^2}{\Lambda^2}$.
	
	For the case of the Goldstone effective description of inflation, with $\Lambda \ll f_\pi$, the relevant terms in the lagrangian are given by \eqref{higgscouplingeft}. The calculation of $F^{\text{single}}_{h}$ follows identical steps as above with the difference that the operator $\mathcal{D}= k_1^2 k_2^2\partial_{k_{12}}^2$. This results into, the replacement $\left(\frac{3}{2}+i\mu\right)\left(\frac{5}{2}+i\mu\right)\rightarrow \frac{1}{2}\left(\frac{1}{2}+i\mu\right)\left(\frac{3}{2}+i\mu\right)$. Hence the final answer reads (taking $d_2=1$),
	\begin{multline}\label{singleheft}
	F^{\text{single}}_{h} = - \frac{1}{8}\times\lambda_2\left(\frac{v f_\pi}{\Lambda}\right)^2\times\\
	\left(\Gamma\left(\frac{1}{2}+i\mu\right)^2 \Gamma\left(-2i\mu\right)\left(\frac{1}{2}+i\mu\right)\left(\frac{3}{2}+i\mu\right)(1+i\sinh(\pi\mu))\left(\frac{k_3}{k_1}\right)^{\frac{3}{2}+i\mu}+(\mu\rightarrow-\mu)\right).
	\end{multline}

	\subsection{Calculation of Double Exchange Diagram}
	
	We see from the Fig. \ref{fig:three-diagrams} (a) that there is a quadratic mixing between inflaton and Higgs field $h$. For numerical simplification one can define a mixed propagator which captures this mixing \cite{Chen:2017ryl}. Using this mixed propagator, we then calculate double exchange diagram numerically. We first focus on single field slow roll inflation.
	
	\subsubsection{Mixed Propagators} 
	
	A mixed propagator is characterized by the $3-$momentum (say $k$) flowing through the line and bulk time coordinate $\eta$. We have to sum over all time instants $\eta^\prime$ where the mixing occurs. The bulk time coordinate $\eta$ can be part of either time or anti-time ordering. Let us first consider the case where $\eta$ comes from time ordering, and denote that mixed propagator by, $D_{+}(\eta,k)$. Then we have two possibilities, (a) $\eta^\prime$ comes also from time ordering, in which case the we can write the contribution of the mixing part of the entire diagram as, 
	\begin{equation}
	\mathcal{I}_+ = -\int_{-\infty}^{0}\frac{d\eta^\prime}{\eta^{\prime 4}} g_{+}(k,\eta^\prime)\left(\theta(\eta-\eta^\prime)\bar{f}(\eta)f(\eta^\prime)+\theta(\eta^\prime-\eta)\bar{f}(\eta^\prime)f(\eta)\right),
	\end{equation}
	and, (b) $\eta^\prime$ comes from anti-time ordering, for which we get,
	\begin{equation}
	\mathcal{I}_- = +\int_{-\infty}^{0}\frac{d\eta^\prime}{\eta^{\prime 4}} g_{-}(k,\eta^\prime)\bar{f}(\eta^\prime)f(\eta).
	\end{equation}
	The overall signs are due to again $e^{\pm i\int dt H}$ and we have omitted a factor of $+i$ for simplicity which we will restore in the final expression for the mixed propagator. We can then rewrite $\mathcal{I}_+$ as
	\begin{equation}
	\mathcal{I}_+ = -(\mathcal{I}_{-})^*+\bar{f}(\eta)\int_{\eta}^{0}\frac{d\eta^\prime}{\eta^{\prime 4}} g_{+}(k,\eta^\prime)f(\eta^\prime)-f(\eta)\int_{\eta}^{0}\frac{d\eta^\prime}{\eta^{\prime 4}} g_{+}(k,\eta^\prime)\bar{f}(\eta^\prime).
	\end{equation}
	So the whole contribution of the mixed propagator when $\eta$ comes from time ordering is
	\begin{multline}
	D_{+}(\eta,k)=(+i)(-\rho_2)(\mathcal{I}_++\mathcal{I}_-) \\
	= (+i)(-\rho_2) \left(2 i \text{Im}(\mathcal{I}_-)+\bar{f}(\eta)\int_{\eta}^{0}\frac{d\eta^\prime}{\eta^{\prime 4}}g_{+}(k,\eta^\prime) f(\eta^\prime)-f(\eta)\int_{\eta}^{0}\frac{d\eta^\prime}{\eta^{\prime 4}}g_{+}(k,\eta^\prime) \bar{f}(\eta^\prime)\right).
	\end{multline}
	We have restored the factor of $+i$ and also put the mixing vertex $-\rho_2$ from \eqref{higgsexpansion}. 
	
	Using the mode functions for inflatons \eqref{infmodes} and massive scalars \eqref{mode1} and \eqref{mode2}, and also the relations, \eqref{h1},\eqref{h2}, we can evaluate the mixed propagator analytically. For convenience, we define the function $J_{+}(-\eta k)=\frac{8 k^3}{\pi\rho}\mathcal{D}_+(\eta,k)$:
	\begin{multline}\label{Jplus}
	J_+(x)= x^{\frac{3}{2}}\left[\sqrt{2\pi}\text{sech}(\pi\mu)e^{\pi\mu/2}\left(H_{i\mu}^{(2)}(x)e^{i\pi/4}+H_{-i\mu}^{(1)}(x)e^{-i\pi/4}\right) \right. \\
	\left. -2\sqrt{x}H_{i\mu}^{(1)}(x)\left(\text{csch}(\pi\mu)\mathsf{F}(\mu,x)+(1-\text{coth}(\pi\mu))\mathsf{F}(-\mu,x)\right)  \right. \\
	\left. -2\sqrt{x}H_{i\mu}^{(2)}(x)\left(\text{csch}(\pi\mu)\mathsf{F}(\mu,x)-(1+\text{coth}(\pi\mu))\mathsf{F}(-\mu,x)\right)\right],
	\end{multline}
	where $x=-k\eta$, and $\mathsf{F}(\mu,x)$ is given in terms of hypergeometric function ${}_2F_2$, 
	\begin{equation}
	\mathsf{F}(\mu,x) = \frac{1}{\Gamma(1-i\mu)}\frac{1}{2\mu+i}\left(\frac{x}{2}\right)^{-i\mu}{}_2 F_2(\frac{1}{2}-i\mu,\frac{1}{2}-i\mu;\frac{3}{2}-i\mu,1-2i\mu;-2ix).
	\end{equation}
	We will be using the small argument limit of $J_+(x)$,
	\begin{equation}
	J_+(x)=A(\mu)x^{\frac{3}{2}}(x/2)^{i\mu}+A(-\mu)x^{\frac{3}{2}}(x/2)^{-i\mu},
	\end{equation}
	where
	$A(\mu)=-2\sqrt{2/\pi}\text{sech}(\pi\mu)\Gamma(-i\mu)\sin(\frac{\pi}{4}+\frac{i\pi\mu}{2})$.
	
	\subsubsection{Three-Point Function}
	
	After we incorporate appropriate internal lines into mixed propagators, the double exchange diagram effectively contains a single vertex. Thus we only have two diagrams to calculate, with one of them being the conjugate of the other. Let us start with the time ordered contribution. Since the lagrangian \eqref{higgsexpansion} contains a term of the form $\frac{\alpha}{\dot{\phi}_0}\dot{\xi}h^2$ then we get
	\begin{equation}
	+i \frac{\alpha}{\dot{\phi}_0}\int\frac{d\eta}{\eta^4}\left(-\frac{k_1^2\eta^2}{2k_1^3}\right)e^{ik_1\eta}D_{+}(\eta,k_2)D_{+}(\eta,k_3).
	\end{equation}
	
	Once again we will be interested in the limit $k_3\ll k_1$, so using the small argument expansion of $J_{+}$,
	\begin{multline}
	+i \frac{\alpha}{\dot{\phi}_0} \left(-\frac{1}{2k_1}\right)\int\frac{d\eta}{\eta^2}e^{ik_1\eta}\mathcal{D}_{+}(\eta,k_1)\mathcal{D}_{+}(\eta,k_3) \\ =
	-\frac{i}{2}\frac{\alpha}{\dot{\phi}_0} \frac{\pi^2\rho_2^2}{64k_1^3k_3^3}\left(A(\mu)s(\mu)\left(\frac{k_3}{k_1}\right)^{3/2}\left(\frac{k_3}{2k_1}\right)^{i\mu}+A(-\mu)s(-\mu)\left(\frac{k_3}{k_1}\right)^{3/2}\left(\frac{k_3}{2k_1}\right)^{-i\mu}\right),
	\end{multline}
	where we have defined $s(\mu)=\int_{0}^{\infty}\frac{dx}{x^2}e^{-ix}J_{+}(x)x^{3/2+i\mu}$.
	The full three point function then becomes, after adding the anti-time ordered contribution, 
	\begin{equation}
	\frac{\pi^2\rho_2^2}{64k_1^3k_3^3}(-\frac{i}{2} \frac{\alpha}{\dot{\phi}_0})\left(A(\mu)s(\mu)-A^*(-\mu)s^*(-\mu)\right)\left(\frac{k_3}{k_1}\right)^{3/2}\left(\frac{k_3}{2k_1}\right)^{i\mu}+(\mu\rightarrow-\mu).
	\end{equation}
	This gives $F$ as a function of $\left(\frac{k_3}{k_1}\right)$ after summing over permutation $k_1\leftrightarrow k_2$,
	\begin{equation}\label{doublehsr}
	F^{\text{double}}_{h} =  \alpha\rho_2^2 \frac{i\pi^2}{16}\left(A(\mu)s(\mu)-A^*(-\mu)s^*(-\mu)\right)\left(\frac{k_3}{k_1}\right)^{3/2}\left(\frac{k_3}{2k_1}\right)^{i\mu}+(\mu\rightarrow-\mu).
	\end{equation}

	For $m<\frac{3H}{2}$ we get an appropriately modified version of the above expression. In terms of the function, $\bar{s}(\nu)=\int_{0}^{\infty}\frac{dx}{x^2}e^{-ix}I_{+}(x)x^{3/2-\nu}$
	we have,
	\begin{equation}
	F^{\text{double}}_{h} =  -\alpha\rho_2^2 \frac{\pi^2}{8}B(\nu)\text{Im}(\bar{s}(\nu))\left(\frac{k_3}{k_1}\right)^{\frac{3}{2}-\nu},
	\end{equation}
	where $ B(\nu)\equiv-2^{\nu+1}\Gamma(\nu)\sqrt{2/\pi}\sec(\pi\nu)\sin(\pi/4-\pi\nu/2)$.
	In deriving the above we have not kept a contribution of the form $\left(\frac{k_3}{k_1}\right)^{3/2+\nu}$. 
	
	For the Goldstone description we see from \eqref{higgscouplingeft} that the functional form for $F^{\text{double}}_{h}$ is identical to above. For the overall coefficient we change, $\rho_2^2\alpha\rightarrow \frac{1}{2}\lambda_2(\lambda_2 v)^2f_\pi^2$, and hence,
	\begin{equation}\label{doubleheft}
	F^{\text{double}}_{h} =  \lambda_2(\lambda_2 v f_\pi)^2 \frac{i\pi^2}{32}\left(A(\mu)s(\mu)-A^*(-\mu)s^*(-\mu)\right)\left(\frac{k_3}{k_1}\right)^{3/2}\left(\frac{k_3}{2k_1}\right)^{i\mu}+(\mu\rightarrow-\mu).
	\end{equation}
	\subsection{Calculation of Triple Exchange Diagram}
	Using the mixed propagator, the triple exchange diagram can also be calculated in an identical manner \cite{Chen:2017ryl}. Using the cubic Higgs vertex, $\frac{\lambda_h v}{2}h^3$ we can write the time ordered diagram as,
	\begin{equation}
	(-i)\frac{\lambda_h v}{2}\int\frac{d\eta}{\eta^4}D_{+}(\eta,k_1)D_{+}(\eta,k_2)D_{+}(\eta,k_3).
	\end{equation}
	Again in the limit $k_3\gg k_1$ we can use the small argument expansion of $J_{+}(x)$,
	\begin{multline}
	(-i)\frac{\lambda_h v}{2}\int\frac{d\eta}{\eta^4}D_{+}(\eta,k_1)D_{+}(\eta,k_2)D_{+}(\eta,k_3)\\
	=(-i)\frac{\lambda_h v}{2}\times \frac{\pi^3\rho_2^3}{8^3k_1^3k_3^3}\left(A(\mu)t(\mu)\left(\frac{k_3}{k_1}\right)^{\frac{3}{2}}\left(\frac{k_3}{2k_1}\right)^{i\mu}+A(-\mu)t(-\mu)\left(\frac{k_3}{k_1}\right)^{\frac{3}{2}}\left(\frac{k_3}{2k_1}\right)^{-i\mu}\right),
	\end{multline}
	where $A(\mu)=-2\sqrt{2/\pi}\text{sech}(\pi\mu)\Gamma(-i\mu)\sin(\frac{\pi}{4}+\frac{i\pi\mu}{2})$ and $t(\mu)=\int_{0}^{\infty}\frac{dx}{x^4}J_{+}(x)^2x^{\frac{3}{2}+i\mu}$.
	After adding the anti-time ordered contribution and permutation $k_1\leftrightarrow k_2$ we get
	\begin{equation}\label{triplehsr}
	F^{\text{triple}}_{h}=\frac{\pi^3\rho_2^3\dot{\phi}_0\lambda_h v}{128}(+i)\left(A(\mu)t(\mu)-A^*(-\mu)t^*(-\mu)\right)\left(\frac{k_3}{k_1}\right)^{\frac{3}{2}}\left(\frac{k_3}{2k_1}\right)^{i\mu}+(\mu\rightarrow-\mu).
	\end{equation}
	For the Goldstone description we see from \eqref{higgscouplingeft} that the functional form for $F^{\text{triple}}_{h}$ is identical to above. For the overall coefficient we change, $\rho_2^3\rightarrow (\lambda_2 v)^3$, and hence,
	\begin{equation}\label{tripleheft}
	F^{\text{triple}}_{h}=\frac{\pi^3\lambda_2^3v^3f_\pi^2\lambda_h v}{128}(+i)\left(A(\mu)t(\mu)-A^*(-\mu)t^*(-\mu)\right)\left(\frac{k_3}{k_1}\right)^{\frac{3}{2}}\left(\frac{k_3}{2k_1}\right)^{i\mu}+(\mu\rightarrow-\mu).
	\end{equation}
	\section{Massive Vector Fields in dS Space}\label{appvectormode} 
	
	Here we will derive mode functions for massive spin-1 fields \cite{Lee:2016vti}, which will be useful in the next appendix in computing NG mediated by $Z$-type particles.

	\subsection{Mode Functions in Momentum Space}

	We start with the lagrangian,
	\begin{equation}
	\int d^4x \sqrt{-g}\left(-\frac{1}{4}F_{\mu\nu}^2-\frac{1}{2}m^2 Z_\mu^2\right),
	\end{equation}
	where $F_{\mu\nu}=\nabla_\mu Z_\nu-\nabla_\nu Z_\mu$. Variation of the action yields the equation of motion,
	\begin{equation}\label{eomforA}
	\nabla_\mu F^{\mu\nu}=m^2 Z^\nu.
	\end{equation}
	Taking the divergence of both sides and also using the fact that $\nabla_\mu\nabla_\nu F^{\mu\nu}\propto R_{\mu\nu}F^{\mu\nu}=0$, we get
	\begin{equation}\label{constraintforA}
	\nabla_\mu Z^{\mu}=0.
	\end{equation}
	Mode functions for $Z^\mu$ are then obtained by solving \eqref{eomforA} and \eqref{constraintforA}. 
	The NG correlators involve mixing the inflaton with the $Z$, which is constrained by the spatial rotation and translation invariance. Therefore only the longitudinal $Z$ polarization, which is a spatial scalar, can appear since the inflaton is obviously scalar. We are interested in the mode functions for this degree of freedom.
	It is shared between the timelike component $Z_\eta$ and the longitudinal spatial component $Z_{\text{long}} = \vec{Z} \cdot \hat{k}$ with $\hat{k}$ being a unit vector pointing in the direction of propagation. 
	
	Fourier transforming from $(\eta,\vec{x})$ to $(\eta,\vec{k})$ coordinates, the constraint equation reads,
	\begin{equation}\label{constraintforAc}
	\eta^2\partial_\eta Z_\eta-2\eta Z_\eta = i \eta^2 k Z_{\text{long}}.
	\end{equation} 
	From \eqref{eomforA} and \eqref{constraintforAc} we get the equation of motion for the component $Z_\eta$,
	\begin{equation}
	\partial_\eta^2 Z_\eta-\frac{2}{\eta}\partial_\eta Z_\eta-\partial_i^2Z_\eta+\frac{(m^2+2)}{\eta^2}Z_\eta=0.
	\end{equation}
	This is almost identical to the equation of motion for the scalar \eqref{eomforscalar}. The solutions are again given in terms of Hankel functions, but with $\mu^2 = m^2+2-\frac{9}{4}=m^2-\frac{1}{4}$. After we obtain the mode function for $Z_\eta$, that for $Z_{\text{long}}$ is simply obtained from the constraint equation \eqref{constraintforAc}. 
	In parallel with the case of scalars, the quantum field is obtained by elevating the free superposition coefficients in the general classical solution to linear combinations of creation and destruction operators on the Bunch-Davies vacuum, 
	
	\begin{equation}
	\begin{aligned}
	Z_\eta(\eta,\vec{k})&=h_{k,0}(\eta)b_{\vec{k}}^\dagger+\bar{h}_{k,0}(\eta)b_{-\vec{k}},\\
	Z_{\text{long}}(\eta,\vec{k})&=h_{k,l}(\eta)b_{\vec{k}}^\dagger+\bar{h}_{k,l}(\eta)b_{-\vec{k}},
	\end{aligned}	
	\end{equation}
	where the mode functions are
	\begin{gather}
	\bar{h}_{k,0}(\eta)=N e^{-\frac{\pi\mu}{2}}(-\eta)^{\frac{3}{2}}H_{i\mu}^{(1)}(-k\eta) \\
	\bar{h}_{k,l}(\eta)=Ne^{-\frac{\pi\mu}{2}}\left(+\frac{i}{2k}\right)(-\eta)^{\frac{1}{2}}\left(-H_{i\mu}^{(1)}(-k\eta)+k\eta H_{i\mu+1}^{(1)}(-k\eta)-k\eta H_{i\mu-1}^{(1)}(-k\eta)\right),
	\end{gather}
	with $N=e^{\frac{i\pi}{4}}\frac{\sqrt{\pi}}{2}\frac{k}{m}$. In the above $b$ and $b^\dagger$ are annihilation and creation operators for the longitudinal degree of freedom for spin-1.
	
	\section{NG due to $Z$ Exchange}\label{appvectorng}
	
	For single-field slow-roll inflation, 
	the lagrangian involving the inflaton and spin-1 $Z$ relevant for single-exchange (diagram (a) in Fig. \ref{fig:three-diagrams}) is given by \eqref{heavyhiggs} to be 
	\begin{equation}
	\mathcal{L} = \rho \eta \dot{\xi}Z_\eta + \frac{\rho}{\dot{\phi}_0}\eta^2\dot{\xi}\partial_i\xi Z_i,
	\end{equation}
	where $\rho=\frac{\rho_{1,Z}\rho_2}{m_h^2}$. As discussed earlier, we defer the computation of the double-exchange contribution to future work, although we have estimated its strength and it seems readily detectable in future measurements. 
	We have also shown earlier that the triple-exchange contribution is suppressed and can be neglected.
	
	We now  parallel the steps taken in the calculation of NG for the case of single-exchange of a scalar $h$. We start with $I_{-+}$ diagram. In this case the time ordered and anti-time ordered contribution factorize, and we can evaluate them separately.
	
	\paragraph{Time Ordered Contribution}
	
	\begin{equation}
	(+i\rho)\int_{-\infty}^{0}\frac{d\eta^\prime}{\eta^{\prime 4}}\eta^\prime \left(-\frac{k_3^2\eta^{\prime 2}}{2k_3^3}\right)e^{ik_3\eta^\prime}h_{k,0}(\eta)=
	\rho\frac{1}{4\sqrt{2}}\frac{1}{m_Z k_3^{\frac{3}{2}}}\Gamma\left(\frac{3}{2}+i\mu\right)\Gamma\left(\frac{3}{2}-i\mu\right).
	\end{equation}
	
	\paragraph{Anti-time Ordered Contribution}
	
	Defining $p=\frac{k_1}{k_3}$,
	\begin{gather}
	(-i\frac{\rho}{\dot{\phi}_0})\int_{-\infty}^{0}\frac{d\eta}{\eta^4}\eta^2\left(\frac{-k_2^2\eta^2}{2k_2^3}\right)e^{-ik_2\eta}\left(\frac{-ik_{1i}}{2k_1^3}\right)e^{-ik_1\eta}(1+ik_1\eta)\hat{k}_{3i} \bar{h}_{k,l}(\eta)\\
	=\frac{i\rho}{\dot{\phi}_0}\frac{1}{16k_2k_1^2k_3^{\frac{3}{2}}}\frac{\sqrt{\pi}}{m_Z}e^{\frac{i\pi}{4}}(\hat{k}_1\cdot\hat{k}_3)\times\\
	\left(-f_1(\frac{1}{2},2p)+ipf_1(\frac{3}{2},2p)+f_2(\frac{3}{2},2p)-ipf_2(\frac{5}{2},2p)-f_3(\frac{3}{2},2p)+ip f_3(\frac{5}{2},2p)\right),
	\end{gather}
	where different integrals involving Hankel functions have been evaluated using \eqref{h1},
	
	\begin{equation}
	\begin{aligned}
	f_1(n,p)=\frac{(+i/2)^n}{\Gamma(n+3/2)}\frac{1}{\sqrt{\pi}}\Gamma(n+1-i\mu)\Gamma(n+1+i\mu)\\\times{}_2F_1(n+1+i\mu,n+1-i\mu,n+3/2,(1-p)/2)\\
	f_2(n,p)=(+i)\frac{(+i/2)^n}{\Gamma(n+3/2)}\frac{1}{\sqrt{\pi}}\Gamma(n+2-i\mu)\Gamma(n+i\mu)\\\times{}_2F_1(n+i\mu,n+2-i\mu,n+3/2,(1-p)/2)\\
	f_3(n,p)=(-i)\frac{(+i/2)^n}{\Gamma(n+3/2)}\frac{1}{\sqrt{\pi}}\Gamma(n+2+i\mu)\Gamma(n-i\mu)\\\times{}_2F_1(n+2+i\mu,n-i\mu,n+3/2,(1-p)/2)
	\end{aligned}
	\end{equation}

	We multiply the above two contributions and also sum over $I_{++},I_{--},I_{+-}$ diagrams. Finally we sum over permutations $\vec{k}_1\leftrightarrow \vec{k}_2$ to get,
	\begin{gather}\label{singlezsr}
	F^{\text{single}}_{Z}=\frac{\rho^2}{16\pi m_Z^2}\sin^2\theta\Gamma(\frac{3}{2}+i\mu)\Gamma(\frac{3}{2}-i\mu)\cosh(\pi\mu) \times\\
	\left((7-5i\mu+16\mu^2+4i\mu^3)\Gamma(\frac{3}{2}+i\mu)^2\Gamma(-2-2i\mu)(1-i\sinh(\pi\mu))\left(\frac{k_3}{k_1}\right)^{\frac{5}{2}+i\mu}+(\mu\rightarrow-\mu)\right),
	\end{gather}
	where $\theta=\hat{k}_1\cdot\hat{k}_3$ and we have used large negative argument expansion of hypergeometric function. 
	
	For the Goldstone description, we see from  \eqref{inf-Z-goldstone} that the functional form of $F^{\text{single}}_{Z}$ is identical to the above. The overall coefficient is changed to $\rho\rightarrow \frac{vm_Z}{2\Lambda}$(taking $\text{Im}(d_{1})=1$). Hence we get
	\begin{gather}\label{singlezeft}
	F^{\text{single}}_{Z}=\left(\frac{v}{2\Lambda}\right)^2\frac{1}{16\pi}\sin^2\theta\Gamma(\frac{3}{2}+i\mu)\Gamma(\frac{3}{2}-i\mu)\cosh(\pi\mu) \times\\
	\left((7-5i\mu+16\mu^2+4i\mu^3)\Gamma(\frac{3}{2}+i\mu)^2\Gamma(-2-2i\mu)(1-i\sinh(\pi\mu))\left(\frac{k_3}{k_1}\right)^{\frac{5}{2}+i\mu}+(\mu\rightarrow-\mu)\right).
	\end{gather}
	
\end{document}